\magnification=1200
\hoffset=.0cm
\voffset=.0cm
\baselineskip=.57cm plus .57mm minus .57mm

%
%
%
%
\def\ref#1{\lbrack#1\rbrack}
%
%
%
%
\input amssym.def
\input amssym.tex
%
%
\font\teneusm=eusm10              
\font\seveneusm=eusm7             
\font\fiveeusm=eusm5                 
%
%

%
%

%
%
\newfam\eusmfam
\textfont\eusmfam=\teneusm
\scriptfont\eusmfam=\seveneusm
\scriptscriptfont\eusmfam=\fiveeusm

\def\proclaim #1. #2\par{\medbreak{\bf #1.\enspace}{\it #2}\par\medbreak}
%
%
%
%
%
\def\dim{{\rm dim}\hskip 1pt}
\def\ind{{\rm ind}\hskip 1pt}
\def\sgn{{\rm sgn}\hskip 1pt}
\def\deg{{\rm deg}\hskip 1pt}

\def\dom{{\rm dom}\hskip 1pt}

\def\supp{{\rm supp}\hskip 2pt}
\def\pfaff{{\rm pfaff}\hskip 2pt}
\def\tr{{\rm tr}\hskip 1pt}

\def\Hom{{\rm Hom}\hskip 1pt}

\def\hst1{\hskip 1pt}

%
%
%
%
%

\hrule\vskip.5cm
\hbox to 16.5 truecm{October 2000  \hfil DFUB 00--16}
\hbox to 16.5 truecm{Version 3  \hfil hep-th/0010110}
\vskip.5cm\hrule
\vskip.9cm
\centerline{\bf RELATIVE TOPOLOGICAL INTEGRALS AND }   
\centerline{\bf RELATIVE CHEEGER--SIMONS DIFFERENTIAL CHARACTERS}   
\vskip.4cm
\centerline{by}
\vskip.4cm
\centerline{\bf Roberto Zucchini}
\centerline{\it Dipartimento di Fisica, Universit\`a degli Studi di Bologna}
\centerline{\it V. Irnerio 46, I-40126 Bologna, Italy}
\centerline{\it and }
\centerline{\it INFN, sezione di Bologna}
\vskip.9cm
\hrule
\vskip.6cm
\centerline{\bf Abstract} 
\vskip.4cm
\par\noindent
Topological integrals appear frequently in Lagrangian field theories.
On manifolds without boundary, they can be treated in the framework of 
(absolute) (co)homology using the formalism of Cheeger--Simons differential 
characters. String and D--brane theory involve field theoretic models on 
worldvolumes with boundary. On manifolds with boundary, the proper
treatment of topological integrals requires a generalization of 
the usual differential topological set up and leads naturally
to relative (co)homology and relative Cheeger--Simons differential 
characters. In this paper, we present a construction of relative 
Cheeger--Simons differential characters which is computable in principle
and which contains the ordinary Cheeger--Simons differential characters
as a particular case.
\vfill\eject

\par\vskip.6cm
\item{\bf 0.} {\bf Introduction}
\vskip.4cm
\par

Topological integrals appear frequently in Lagrangian field theories
such as Chern--Simons model, Wess--Zumino--Witten model, gauge theory and  
D--brane theory, to mention only the most popular and best known.
They are formal integrals on topologically non trivial manifolds
of differential forms which are only locally defined.
The integrand thus suffers ambiguities on overlapping coordinate patches, 
making the definition of integration problematic. 
In physics, the problem of the proper definition of topological integrals
has been studied by several authors since the mid eighties \ref{1,2,3}
and also recently it has been the object of a number of studies
\ref{4,5,6,7,8}. 
In mathematics, the interest in this topic dates back at least to the 
early seventies when it was attempted to frame the Chern--Simons forms 
associated to connections on a principal bundle in appropriate global 
differential topological structures on its base space. 
It resulted in the theory of Cheeger--Simons differential characters
\ref {9,10,11} whose apparent 
relation with the smooth versionsof Deligne cohomology \ref{12} 
and Deligne--Beilinson cohomology \ref{13,14} developed a decade later 
was noticed in the early ninenties and has been reconsidered recently  
\ref{15}.

Virtually all the above studies deal with absolute cohomology
and differential characters. A  generalization of the formalism 
appropriate for relative cohomology and differential characters
has not been fully worked out to the best of our knowledge.
This is attempted in the present paper.

The reason why this is an interesting problem and not a mere
academic exercise is shown by the physical examples illustrated below
in which the relevance of relative cohomology and differential characters
should be apparent. Since we have physical applications in mind,
we want to provide a constructive treatment, i. e. one computable 
at least in principle. For this reason, we opt for a formulation closer in 
spirit to Cheeger's and Simons', which is somewhat more concrete and thus 
more suitable for the physicists' computational needs.  We shall do this using 
the machinery of \v Cech (co)homology as in \ref{1,2,3}. We shall not use 
partitions of unity as in \ref{8}, since these are required by distribution 
valued quantized fields, while the fields relevant in our examples  
are background semiclassical fields. Though we work mostly in the framework 
of relative integer cohomology, our formulation presumably
might be extended to more general relative cohomology theories,
in particular K--theory. 

Consider a spacetime $X$ and a D--brane occupying a submanifold 
$Y\subseteq X$ in type II string theory. The background is characterized by 
the NS NS field $B_2$. Further, the D--brane carries a $U(1)$ gauge field 
$A_1$. For a string with world sheet $\Sigma_2\subseteq X$ such that
$\partial\Sigma_2\subseteq Y$, the path integral measure
contains a factor
$$
\pfaff(D_\Sigma)\exp\bigg(i\int_{\Sigma_2} B_2
-i \int_{\partial\Sigma_2} A_1\bigg),
\eqno(0.1)
$$
where $D_\Sigma$ is the Dirac operator on $\Sigma_2$ and $\pfaff(D_\Sigma)$ 
is its pfaffian \ref{16}. If $\partial\Sigma_2=\emptyset$, the sign of 
$\pfaff(D_\Sigma)$ is uniquely defined. In order for the path integral 
measure to be well--defined $H_3=dB_2$ is required to be a globally 
defined closed 3--form with quantized fluxes through any closed 3--folds 
$P_3\subseteq X$:
$$
\int_{P_3} H_3\in 2\pi\Bbb Z.
\eqno(0.2)
$$
If $\partial\Sigma_2\not=\emptyset$, the sign of
$\pfaff(D_\Sigma)$ is not uniquely defined in general, signaling 
a global world--sheet anomaly. 
Consistency requires that this anomaly be canceled by an equal 
and opposite anomaly of the exponential factor of (0.1). 
In order for this to be possible, the 2--form $B_{A2}=B_2-dA_1$ must be 
globally defined on $Y$ so that the restriction of $H_3$ on $Y$ is exact.  
The quantization condition (0.2) gets generalized as
$$
\int_{P_3} H_3-\int_{Q_2}B_{A2}-\pi\int_{Q_2}w_2\in 2\pi\Bbb Z, 
\eqno(0.3)
$$
for $P_3\subseteq X$, $Q_2\subseteq Y$ with $\partial P_3=Q_2$,
where $w_2$ is a closed 2--form on $Y$ representing the second 
Stiefel--Whitney class of $Y$ modulo 2. In the simple case where $B_2=0$, 
$Y$ turns out to be a Spin$_c$ manifold and $A_1$ a Spin$_c$ connection. 
See \ref{16} for more details and \ref{17} for a related analysis.

The problem of D--branes in group manifolds has received a great deal of 
attention recently \ref{18,19,20,21,22,23}. The central issue here
is the proper definition of D0 charge and its quantization.
Consider a D--brane located in a submanifold $K$ of a compact simple Lie
group $G$. The background is characterized by a closed 3-form 
$H_3$ on $G$, the trace of the third wedge power of the left invariant 
Maurer--Cartan form of $G$.
According to \ref{24}, the D0 charge $Q$ of a D2--brane contained in 
the D--brane is defined if $H_3=dL_2$ on $K$ for some 2--form $L_2$ 
globally defined on $K$ and is given by
$$
Q=\int_{V_3} H_3-\int_{Z_2}L_2, 
\eqno(0.4)
$$
for $V_3\subseteq G$, $Z_2\subseteq K$ with $\partial V_3=Z_2$. 
When $H_3$ is cohomologically trivial, $Q$ is quantized as $Q\in 2\pi\Bbb Z$
in the usual way. When $H_3$ has a fundamental period (level) $k$,
$Q$ is quantized as $Q\in 2\pi\Bbb Z_k$. These quantization rules  
have to be compared with (0.3).

Consider $N$ coinciding D--branes of type II string theory spanning 
a world--volume $W$ in the space time $X$. The background fields are 
the spin connection $\omega_1$, the NS NS B--field $B_2$ and the 
R R field $C$. Further, the set of branes 
carries a $U(N)$ Chan--Paton gauge field $A_1$ \ref{25}. 
Here, we assume that $B_2=0$. $C$ is an odd/even degree 
form field for type IIA/IIB strings. $C$ is not globally defined in $X$ in 
general. Only its field strength $G=dC$ is. The D--brane is carries 
R R charges and thus couples to the R R field $C$ via the Wess--Zumino term. 
Thus, the path integral contains a factor of the form
$$
\pfaff(D_W)\exp\bigg(i\int_W\tr\exp_\wedge(iF_2/2\pi)
\wedge\widehat A^{1\over 2}(R_{TW2})\wedge\widehat A^{-{1\over 2}}(R_{NW2})
\wedge C\bigg),
\eqno(0.5)
$$
where $\pfaff(D_W)$ is the pfaffian of the Dirac operator on $W$
and $R_{TW2}$, $R_{NW2}$ and $F_2$ are 
the curvatures of the tangent and normal bundles $TW$, $NW$ of $W$ and 
the gauge field strength, respectively. $\widehat A$ denotes 
the A--roof genus. This factor is required and explicitly determined
by gauge and gravitational anomaly cancellation \ref{26,27,28}. 
As before, the sign of $\pfaff(D_W)$ suffers in general
an ambiguity which signals a global anomaly.
The proper definition of the path integral measure requires
some kind of quantization condition for the R R curvature $G$.
This reads
$$
\int_U\widehat A^{1\over 2}(R_{TU2})\wedge
\widehat A^{-{1\over 2}}(R_{NU2})\wedge G-\pi\int_U\nu
\in 2\pi\Bbb Z,
\eqno(0.6)
$$
for any closed submanifold $U$ of $X$ of dimension one unit larger than $W$,
where $R_{TW2}$, $R_{NW2}$ are the curvatures of the tangent and normal 
bundles $TU$, $NU$ of $U$ nd $\nu$ is a closed form representing the 
pfaffian anomaly modulo 2.
In the last three years it has become clear that a realistic theory of 
D--brane R R charges and R R fields in type II string theory requires 
K theory when $B_2=0$ and some sort of twisted generalization thereof 
when $B_2\not=0$ \ref{28,29,30,31,32}.
In any case, a form of generalized cohomology is involved which maps to a 
full lattice in ordinary real cohomology as is apparent from (0.6). 

A generalization for open membranes is still to be worked out \ref{33}. 
It presumably involves adding in the exponential in the right hand side of 
(0.5) a suitable integral on $Z=\partial Y$ leading to a structure similar 
to (0.1). This is however just a speculation for the time being.

The above examples show clearly that the geometrical framework 
suitable for the analysis of these matters is provided by 
relative singular homology and (some generalization of) 
integral cohomology. To make this clearer and also to render the rest 
of the paper more easily readable, we recall briefly some of the basic 
definitions. (See ref. \ref{34,35,36,37,38,39} for background material.)

Let $X$, $Y$ be smooth manifolds with $Y\subseteq X$.
Denote by $i:Y\rightarrow X$ the smooth inclusion map.
A relative singular $p-1$--cycle $(S_{p-1},T_{p-2})$ of $X$ mod $Y$
is a pair of singular chains of $X$, $Y$, respectively, satisfying
$$
\partial S_{p-1}-i_*T_{p-2}=0, \quad
-\partial T_{p-2}=0.
\eqno(0.7)
$$
A relative de Rham $p$--cocycle $(\Xi_p,\Upsilon_{p-1})$ of $X$ mod $Y$
is a pair of forms of $X$, $Y$, respectively, satisfying
$$
d\Xi_p=0,\quad i^*\Xi_p-d\Upsilon_{p-1}=0.
\eqno(0.8)
$$
Locally, there are forms $\tilde\Xi_{p-1}$, $\tilde\Upsilon_{p-2}$ in 
$X$, $Y$, respectively, such that
$$
\Xi_p=d\tilde\Xi_{p-1},\quad 
\Upsilon_{p-1}=i^*\tilde\Xi_{p-1}-d\tilde\Upsilon_{p-2}.
\eqno(0.9)
$$
The associated relative topological integral is the formal integral 
$$
\int_{S_{p-1}}\tilde\Xi_{p-1}-\int_{T_{p-2}}\tilde\Upsilon_{p-2}.
\eqno(0.10)
$$
In general, its value is determined only up to a quantized ambiguity. 
In the simplest case, the ambiguity is just integer valued
\footnote{}{}
\footnote{${}^1$}{Here and in the following, we neglect an inessential factor 
$2\pi$ appearing in the physical quantization conditions.}.
This translates into a quantization condition for the relative 
de Rham $p$--cocycle $(\Xi_p,\Upsilon_{p-1})$
of the form
$$
\int_{s_p}\Xi_p-\int_{t_{p-1}}\Upsilon_{p-1}\in \Bbb Z,
\eqno(0.11)
$$
for any relative singular $p$--cycle $(s_p,t_{p-1})$. 
For more general quantized ambiguities, we have totally analogous
generalized quantization conditions.

In the first example illustrated above, $(\Sigma_2,\partial\Sigma_2)$
is a relative singular 2--cycle and $(H_3,B_{A2})$ is a relative
3--cocycle. The argument of the exponential in (0.1) is the 
associated topological integral. The quantization condition 
(0.3) holds for every relative singular 3--cycle $(P_3,Q_2)$. 
Similarly, in the second example, $(V_3,Z_2)$ is a relative singular 
3--cycle, $(H_3,L_2)$ is a relative 3--cocycle and $Q$ expresses the canonical
pairing of relative singular 3 homology and relative de Rham 3 cohomology.
Quantization selects a sublattice of the latter. Similar considerations 
might apply to an open membrane generalization of the third example.

Since $\tilde\Xi_{p-1}$, $\tilde\Upsilon_{p-2}$ are only local forms 
in general, the proper definition of the topological integral (0.10) is not
a straightforward matter. However, any reasonable definition  
should satisfy the following a priori requirements 
up to the usual quantized ambiguity.
To begin with, we expect the topological integral to depend linearly on the 
relative cycle $(S_{p-1},T_{p-2})$ and the relative cocycle 
$(\Xi_p,\Upsilon_{p-1})$.
Further, we expect some kind of Stokes' theorem to hold. 
So, when the relative singular $p-1$--cycle $(S_{p-1},T_{p-2})$ 
is a relative boundary, 
$$
S_{p-1}=\partial s_p-i_*t_{p-1},\quad
T_{p-2}=-\partial t_{p-1},
\eqno(0.12)
$$
for some singular chains $s_p$, $t_{p-1}$ in $X$, $Y$, respectively, then 
$$
\int_{S_{p-1}}\tilde\Xi_{p-1}-\int_{T_{p-2}}\tilde\Upsilon_{p-2}
=\int_{s_p}\Xi_p-\int_{t_{p-1}}\Upsilon_{p-1},
\eqno(0.13)
$$
where the integrals in the right hand side are computed according to the 
ordinary differential geometric prescription.
Finally, we would like the topological integral to reduce to an ordinary 
integral when the forms $\tilde\Xi_{p-1}$, $\tilde\Upsilon_{p-2}$ are 
globally defined in $X$, $Y$, respectively.
So, when the relative de Rham $p$--cocycle $(\Xi_p,\Upsilon_{p-1})$
is a relative coboundary,
$$
\Xi_p=d\xi_{p-1},\quad 
\Upsilon_{p-1}=i^*\xi_{p-1}-d\upsilon_{p-2},
\eqno(0.14)
$$
for some globally defined forms $\xi_{p-1}$, $\upsilon_{p-2}$ on $X$, $Y$, 
respectively, then 
$$
\int_{S_{p-1}}\tilde\Xi_{p-1}-\int_{T_{p-2}}\tilde\Upsilon_{p-2}
=\int_{S_{p-1}}\xi_{p-1}-\int_{T_{p-2}}\upsilon_{p-2},
\eqno(0.15)
$$
where again the integrals in the right hand side are computed according to the 
ordinary differential geometric prescription.

The plan of the paper is as follows. In sect. 1, we introduce the  basic 
notions of relative homology and cohomology. In sect. 2, we provide an 
explicit construction of the family of relative Cheeger--Simons differential 
characters and show independence form covering choices. 
In sect. 3, we analyze in detail its formal properties.
Finally, sect. 4 contains a few concluding remarks.

\eject
\par\vskip.6cm
\item{\bf 1.} {\bf Relative singular, de Rham and \v Cech (co)homology}
\vskip.4cm

This is a review of some basic material on relative singular, de Rham and 
\v Cech (co)homology. The reader interested in a more thorough
treatment is suggested to consult standard textbooks such as \ref{34,35,36}.

\vskip.6cm
\item{\it 1.} {\it Basic definitions and facts}
\vskip.4cm
\par

Let $M$ be a smooth manifold.
Let ${\cal O}=\{O_\alpha|\alpha\in A\}$ be an open covering of $M$.
Here, $A$ is a countable index set. We set, for $k\geq 0$,
$$
O_{\alpha_0,\ldots,\alpha_k}=O_{\alpha_0}\cap\cdots\cap O_{\alpha_k}.
\eqno(1.1.1)
$$
The $k$--th nerve of ${\cal O}$ is
$$
N({\cal O},k)=\{(\alpha_0,\ldots,\alpha_k)\in A^{k+1}|
O_{\alpha_0,\ldots,\alpha_k}\not= \emptyset\}.
\eqno(1.1.2)
$$
${\cal O}$ is a good covering if all the non empty 
$O_{\alpha_0,\ldots,\alpha_k}$ are contractible.

For $r\in\Bbb Z$, we denote by ${\cal S}_r(M)$ the group of (generalized)
dimension $r$ singular chains of $M$: ${\cal S}_r(M)=0$, for $r\leq -2$,  
${\cal S}_{-1}(M)=\Bbb Z$ and ${\cal S}_r(M)$ is the group of ordinary 
smooth, finite singular chains of $M$ of dimension $r$, 
for $0\leq r$. A dimension $r$ singular chain $U_r$ is characterized by 
its support $\supp U_r\subseteq M$. By convention, $\supp U_r=\emptyset$ for
$r\leq -1$. For any non empty open subset $O$ of $M$, we denote by 
${\cal S}^O_r(M)$ the group of all dimension $r$ chains 
$U_r$ such that $\supp U_r\subseteq O$. Clearly, ${\cal S}^O_r(M)$ is a 
subgroup of ${\cal S}_r(M)$. 

We define a homomorphism $b:{\cal S}_r(M)\rightarrow{\cal S}_{r-1}(M)$ by
$$
bU_r=\partial_s U_r.
\eqno(1.1.3)
$$
Here, for $1\leq r$, $\partial_s$ is the customary simplicial boundary 
operator, while, for $r=0$, $\partial_s U_0=\ind U_0$, where 
$\ind\sum_P n_P P=\sum_P n_P$ for a dimension $0$ chain $\sum_P n_P P$
\footnote{}{}
\footnote{${}^1$}{In dimension $0$, the definition of the 
boundary operator $b$ given here differs from the customary one of singular 
homology, where $b$ vanishes. As a consequence, the $0$ dimensional 
homology groups corresponding to the two definitions of $b$ are also 
different. Our definition ensures that the statement above (1.1.7) holds.}. 
$b$ is nilpotent
$$
b^2=0.
\eqno(1.1.4)
$$

Let ${\cal O}$ be an open covering of $M$.
For $r\in\Bbb Z$, we denote by ${\cal S}^{\cal O}_r(M)$
the subgroup of ${\cal O}$--small elements of ${\cal S}_r(M)$:
${\cal S}^{\cal O}_r(M)={\cal S}_r(M)$, for $r\leq 0$, and 
${\cal S}^{\cal O}_r(M)$ is the subgroup of ${\cal S}_r(M)$
formed by the singular chains made up of simplices the support of 
each of which is contained in some open set of ${\cal O}$, for
$1\leq r$.
There exists a homomorphism $q:{\cal S}_r(M)\rightarrow{\cal S}_r(M)$,
called barycentric subdivision operator, with the following properties.
$q$ is a chain map
$$
qb-bq=0.
\eqno(1.1.5)
$$
$q$ is homotopic to the identity,  i. e. there is a 
homomorphism $c:{\cal S}_r(M)\rightarrow{\cal S}_{r+1}(M)$ such that
$$
bc+cb=q-1.
\eqno(1.1.6)
$$
Most importantly, for any $U_r\in {\cal S}_r(M)$ there is an integer
$k(U_r,{\cal O})\geq 0$ such that $q^kU_r\in {\cal S}^{\cal O}_r(M)$
for $k\geq k(U_r,{\cal O})$. $q$ and $c$ preserve ${\cal O}$--smallness:
for any $U_r\in {\cal S}^{\cal O}_r(M)$, $qU_r\in {\cal S}^{\cal O}_r(M)$ and 
$cU_r\in {\cal S}^{\cal O}_{r+1}(M)$. Further, for any $U_r\in {\cal S}_r(M)$,
$cU_r$ is degenerate, i. e. it is made up of simplices each of which, 
considered as a smooth map of the standard $r+1$ simplex into $M$, has rank 
smaller than $r+1$. An explicit construction of $q$ and $c$ can be found 
in \ref{36}.

Let ${\cal O}=\{O_\alpha|\alpha\in A\}$ be an open covering of $M$.
For $k,r\in\Bbb Z$, we denote by ${\cal C}_{k,r}(M,{\cal O})$  
the group of finite \v Cech singular chains of $\cal O$ in $M$ of \v Cech 
degree $k$ and dimension $r$: ${\cal C}_{k,r}(M,{\cal O})=0$, for $k\leq -2$,  
${\cal C}_{-1,r}(M,{\cal O})={\cal S}^{\cal O}_r(M)$ and 
${\cal C}_{k,r}(M,{\cal O})$ is the group of alternating maps 
$U_{k,r}:A^{k+1}\rightarrow {\cal S}^{\cal O}_r(M)$ such that
$(U_{k,r})_{\alpha_0,\ldots,\alpha_k}=0$ 
for $(\alpha_0,\ldots,\alpha_k)\not\in N({\cal O},k)$,
$(U_{k,r})_{\alpha_0,\ldots,\alpha_k}\in 
{\cal S}^{O_{\alpha_0,\ldots,\alpha_k}}_r(M)$ for 
$(\alpha_0,\ldots,\alpha_k)\in N({\cal O},k)$ 
and $(U_{k,r})_{\alpha_0,\ldots,\alpha_k}\not=0$ only for a finite number
of $(\alpha_0,\ldots,\alpha_k)$, for $0\leq k$.
Note that the \v Cech singular chains are automatically ${\cal O}$--small.
The \v Cech singular chains of ${\cal C}_{-1,r}(M,{\cal O})$ are called 
simply singular chains, on account of the definition given above.
The \v Cech singular chains of ${\cal C}_{k,-1}(M,{\cal O})$ 
are called integer \v Cech chains, since they are integer valued.

The operator $b$ yields a homomorphism
$b:{\cal C}_{k,r}(M,{\cal O})\rightarrow{\cal C}_{k,r-1}(M,{\cal O})$ 
in obvious fashion.
It is known that the homology of $({\cal C}_{k,*}(M,{\cal O}),b)$ 
vanishes for $k>-1$, if $\cal O$ is a good covering.

We define a homomorphism $\beta:{\cal C}_{k,r}(M,{\cal O})\rightarrow
{\cal C}_{k-1,r}(M,{\cal O})$ by
$$
(\beta U_{k,r})_{\alpha_0,\ldots,\alpha_{k-1}}
=\sum_{\alpha\in A}
(U_{k,r})_{\alpha,\alpha_0,\ldots,\alpha_{k-1}}.
\eqno(1.1.7)
$$
$\beta$ is a differential
$$
\beta^2=0.
\eqno(1.1.8)
$$
The homology of $({\cal C}_{*,r}(M,{\cal O}),\beta)$ is known to vanish
for $r>-1$ for any covering $\cal O$.

$b$ and $\beta$ commute
$$
b\beta-\beta b=0.
\eqno(1.1.9)
$$

For $r\in\Bbb Z$, we denote by ${\cal D}^r(M)$ the vector space of 
(generalized) degree $r$ differential forms of $M$: ${\cal D}^r(M)=0$, 
for $r\leq -2$, 
${\cal D}^{-1}(M)=\Bbb R$ and ${\cal D}^r(M)$ is the vector space of 
ordinary smooth differential forms of $M$ of de Rham degree $r$, for 
$0\leq r$. More generally, one may consider degree $r$ differential 
forms $\Xi^r$ which are defined only on a domain $\dom\Xi^r\subseteq M$.
By convention, $\dom\Xi^r=M$ for $r\leq -1$. 
For any non empty open subset $O$ of $M$, we denote by ${\cal D}_O^r(M)$
the vector space of all degree $r$ forms 
$\Xi^r$ such that $\dom \Xi^r\supseteq O$. Clearly,
${\cal D}^r(M)$ is a subspace of ${\cal D}_O^r(M)$.

We define a homomorphism $d:{\cal D}^r(M)\rightarrow{\cal D}^{r+1}(M)$ by
$$
d\Xi^r= d_{dR}\Xi^r.
\eqno(1.1.10)
$$
Here, for $0\leq r$, $d_{dR}$ is the usual de Rham differential
while, for $r=-1$, $d_{dR}\Xi^{-1}$ is the constant $0$--form
corresponding to the constant $\Xi^{-1}$
\footnote{}{}
\footnote{${}^2$}{The definition given here of coboundary operator
$d$ in degree $-1$ is a rather natural extension of the usual
de Rham differential which allows the treatment of degree $-1$
on the same footing as non negative degree. It further 
ensures that the statement above (1.1.12) holds.}.
$d$ is a differential
$$
d^2=0.
\eqno(1.1.11)
$$

Let ${\cal O}=\{O_\alpha|\alpha\in A\}$ be an open covering of $M$.
For $k,r\in\Bbb Z$, we denote by ${\cal C}^{k,r}(M,{\cal O})$ the vector
space of \v Cech--de Rham cochains of $\cal O$ in $M$ of \v Cech degree $k$ 
and de Rham degree $r$: ${\cal C}^{k,r}(M,{\cal O})=0$, for $k\leq -2$, 
${\cal C}^{-1,r}(M,{\cal O})$ is the vector space of forms 
$\Xi^r\in{\cal D}^r(M)$ and 
${\cal C}^{k,r}(M,{\cal O})$ is the vector space of alternating maps 
$\Xi^{k,r}:A^{k+1}\rightarrow {\cal D}^r(M)$ such that
$(\Xi^{k,r})_{\alpha_0,\ldots,\alpha_k}=0$ 
for $(\alpha_0,\ldots,\alpha_k)\not\in N({\cal O},k)$ and 
$(\Xi^{k,r})_{\alpha_0,\ldots,\alpha_k}\in
{\cal D}_{O_{\alpha_0,\ldots,\alpha_k}}^r(M)$
for $(\alpha_0,\ldots,\alpha_k)\in N({\cal O},k)$, for $0\leq k$.
The \v Cech--de Rham cochains of ${\cal C}^{-1,r}(M,{\cal O})$ 
are called simply de Rham cochains, on account of the definition given above.
The \v Cech--de Rham cochains of ${\cal C}^{k,-1}(M,{\cal O})$ 
are called real \v Cech cochains, since they are real valued.

The operator $d$ yields a homomorphism
$d:{\cal C}^{k,r}(M,{\cal O})\rightarrow{\cal C}^{k,r+1}(M,{\cal O})$. 
By Poincar\'e's lemma, the cohomology of $({\cal C}^{k,*}(M,{\cal O}),d)$ 
vanishes for $k>-1$, if $\cal O$ is a good covering.

We define a homomorphism $\delta:{\cal C}^{k,r}(M,{\cal O})\rightarrow
{\cal C}^{k+1,r}(M,{\cal O})$ by
$$
(\delta \Xi^{k,r})_{\alpha_0,\ldots,\alpha_{k+1}}
=\sum_{l=0}^{k+1}(-1)^l
(\Xi^{k,r})_{\alpha_0,\ldots,\alpha_{l-1},\alpha_{l+1},\ldots,\alpha_{k+1}}
\big|_{O_{\alpha_0,\ldots,\alpha_{k+1}}}
\eqno(1.1.12)
$$
$\delta$ is nilpotent
$$
\delta^2=0.
\eqno(1.1.13)
$$
The cohomology of $({\cal C}^{*,r}(M,{\cal O}),\delta)$ is known to vanish
for $r>-1$ for any covering $\cal O$.

$d$ and $\delta$ commute
$$
d\delta-\delta d=0.
\eqno(1.1.14)
$$

A degree $k$ real \v Cech cochain $\Xi^{k,-1}\in{\cal C}^{k,-1}(M,{\cal O})$ 
is called integer if $(\Xi^{k,-1})_{\alpha_0,\ldots,\alpha_k}\in\Bbb Z$
for all $(\alpha_0,\ldots,\alpha_k)\in A^{k+1}$. Such integer cochains form a 
lattice subgroup ${\cal C}_{\Bbb Z}^{k,-1}(M,{\cal O})$ of 
${\cal C}^{k,-1}(M,{\cal O})$.

Let $U_{k,r}\in{\cal C}_{k,r}(M,{\cal O})$, 
$\Xi^{k,r}\in{\cal C}^{k,r}(M,{\cal O})$. For $k\geq 0$, we set
$$
\langle U_{k,r},\Xi^{k,r}\rangle
=\cases{{1\over k!}\sum_{(\alpha_0,\ldots,\alpha_k)\in N({\cal O},k)}
\int_{(U_{k,r})_{\alpha_0,\ldots,\alpha_k}}
(\Xi^{k,r})_{\alpha_0,\ldots,\alpha_k}, &{if $r\geq 0$},\cr
&\cr
{1\over k!}\sum_{(\alpha_0,\ldots,\alpha_k)\in N({\cal O},k)}
(U_{k,-1})_{\alpha_0,\ldots,\alpha_k}
(\Xi^{k,-1})_{\alpha_0,\ldots,\alpha_k}, &{if $r=-1$},\cr
&\cr
\vphantom{\sum_{(\alpha_0,\ldots,\alpha_k)\in N({\cal O},k)}\int}
0,&{if $r\leq -2$}.\cr}
\eqno(1.1.15)
$$
For $k=-1$, similar expressions hold but with the sum over
the $k$--th nerve of the covering and the factor $1/k!$ omitted. 
The integrals in the right hand side are convergent, since 
all singular chains have compact support. 
The sum in the right hand side is convergent,
as all \v Cech singular chains are finite by definition.
One has
$$
\eqalignno{\vphantom{\bigg(}
\langle U_{k,r},d\Xi^{k,r-1}\rangle=& \ \langle bU_{k,r},\Xi^{k,r-1}\rangle,
&(1.1.16)\cr
\vphantom{\bigg(}
\langle U_{k,r},\delta\Xi^{k-1,r}\rangle
=& \ \langle \beta U_{k,r},\Xi^{k-1,r}\rangle.
&(1.1.17)\cr}
$$
These duality relations play a fundamental role in the following. 

Let ${\cal O}=\{O_\alpha|\alpha\in A\}$, ${\cal O}'=\{O'{}_{\alpha'}
|\alpha'\in A'\}$ be open coverings of the manifold M. ${\cal O}'$ is 
called a refinement of ${\cal O}$ if there is a map $f:A'\rightarrow A$ 
such that $O'{}_{\alpha'}\subseteq O_{f(\alpha')}$ for $\alpha'\in A'$. 
The refinement map $f$ defines a homomorphism $f^*:{\cal C}^{k,r}
(M,{\cal O})\rightarrow {\cal C}^{k,r}(M,{\cal O}')$  
of the corresponding spaces of \v Cech--de Rham cochains by 
$$
f^*\Xi^{k,r}{}_{\alpha'{}_0\ldots\alpha'{}_l}=
\Xi^{k,r}{}_{f(\alpha'{}_0)\ldots f(\alpha'{}_l)}
\big|_{O'{}_{\alpha'{}_0\ldots \alpha'{}_l}}.
\eqno(1.1.18)
$$ 
$f^*$ is a cochain map,  i. e. 
$$
f^*\delta=\delta'f^*.
\eqno(1.1.19)
$$
The resulting homomorphism of cohomology depends only on the coverings
${\cal O}$, ${\cal O}'$ but not on the refinement map $f$.

In the rest of this section, we shall describe briefly the main
versions of $Y$ relative homology and cohomology of $X$
for a pair of manifolds $X$, $Y$ such that $Y\subseteq X$. 

Let $p\in\Bbb N$, $p\geq 2$. 
Let $X$, $Y$ be smooth manifolds with $\dim X\geq p$,
$\dim Y\geq p-1$ and such that $Y\subseteq X$. 
Let $i:Y\rightarrow X$ be the smooth inclusion map.

Let $\cal O$ be an open covering of $X$ and let ${\cal O}\cap Y$
be the open covering of $Y$ induced by $\cal O$. 

\vskip.6cm
\item{\it 2.} {\it Relative homology and cohomology}
\vskip.4cm
\par

A $Y$ relative singular $p-1$--chain 
$(S_{p-1},T_{p-2})$ of $X$ is a pair of singular chains 
$S_{p-1}\in{\cal S}_{p-1}(X)$, 
$T_{p-2}\in{\cal S}_{p-2}(Y)$.
A $Y$ relative singular $p-1$--chain $(S_{p-1},T_{p-2})$
of $X$ is a cycle if 
$$
\eqalignno{\vphantom{\bigg(}
&bS_{p-1}-i_*T_{p-2}=0,&(1.2.1a)\cr
\vphantom{\bigg(}
&-bT_{p-2}=0.&(1.2.1b)\cr}
$$
A $Y$ relative singular $p-1$--cycle $(S_{p-1},T_{p-2})$
of $X$ is a boundary if it is of the form
$$
\eqalignno{\vphantom{\bigg(}
& S_{p-1}=bs_p-i_*t_{p-1},&(1.2.2a)\cr
\vphantom{\bigg(}
& T_{p-2}=-bt_{p-1},&(1.2.2b)\cr}
$$
where $(s_p, t_{p-1})$ is an arbitrary $Y$ relative singular 
$p$--chain of $X$. 
We denote by $C^s_{p-1}(X,Y)$, $Z^s_{p-1}(X,Y)$, 
$B^s_{p-1}(X,Y)$
the groups of $Y$ relative singular $p-1$--chains,
cycles and boundaries of $X$, respectively.
Two relative $p-1$--cycles are equivalent 
if their difference is a relative $p-1$--boundary.
The equivalence classes of $Y$ relative singular 
$p-1$--cycles of $X$ form the $p-1$--th relative singular homology 
group $H^s_{p-1}(X,Y)$.

A $Y$ relative singular $p-1$--chain (respectively a cycle, a boundary)
$(S_{p-1},T_{p-2})$ is said $\cal O$--small if $S_{p-1}$ is 
$\cal O$--small and $T_{p-2}$ is ${\cal O}\cap Y$--small in the sense 
defined in the previous subsection. 
We denote by $C^{s{\cal O}}_{p-1}(X,Y)$, $Z^{s{\cal O}}_{p-1}(X,Y)$, 
$B^{s{\cal O}}_{p-1}(X,Y)$
the groups of ${\cal O}$--small $Y$ relative singular $p-1$--chains,
cycles and boundaries of $X$, respectively.
Two ${\cal O}$--small relative $p-1$--cycles are equivalent 
if their difference is an ${\cal O}$--small relative $p-1$--boundary.
The equivalence classes of ${\cal O}$--small $Y$ relative singular 
$p-1$--cycles of $X$ form the $p-1$--th ${\cal O}$--small 
relative singular homology group $H^{s{\cal O}}_{p-1}(X,Y)$.

An ${\cal O}$--small $Y$ relative singular $p-1$--chain $(S_{p-1},T_{p-2})$ 
can be viewed as a pair of \v Cech singular chains
$(S_{-1,p-1},T_{-1,p-2})$ with $S_{-1,p-1}\in{\cal C}_{-1,p-1}(X,{\cal O})$, 
$T_{-1,p-2}\in{\cal C}_{-1,p-2}(Y,{\cal O}\cap Y)$. 
We shall use both notations interchangeably depending on context.

A $Y$ relative integer \v Cech $p-1$--chain $(S_{p-1,-1},T_{p-2,-1})$
of $X$ is a pair of integer \v Cech chains 
$S_{p-1,-1}\in{\cal C}_{p-1,-1}(X,{\cal O})$, 
$T_{p-2,-1}\in{\cal C}_{p-2,-1}(Y,{\cal O}\cap Y)$.
A $Y$ relative integer \v Cech $p-1$--chain $(S_{p-1,-1},T_{p-2,-1})$
of $X$ is a cycle if 
$$
\eqalignno{\vphantom{\bigg(}
&\beta S_{p-1,-1}-i_*T_{p-2,-1}=0&(1.2.3a)\cr
\vphantom{\bigg(}
&-\beta T_{p-2,-1}=0.&(1.2.3b)\cr}
$$
A $Y$ relative integer \v Cech $p-1$--cycle $(S_{p-1,-1},T_{p-2,-1})$
of $X$ is a boundary if it is of the form
$$
\eqalignno{\vphantom{\bigg(}
& S_{p-1,-1}=\beta s_{p,-1}-i_*t_{p-1,-1},&(1.2.4a)\cr
\vphantom{\bigg(}
& T_{p-2,-1}=-\beta t_{p-1,-1},&(1.2.4b)\cr}
$$
where $(s_{p,-1},t_{p-1,-1})$ is an arbitrary $Y$ relative integer \v Cech 
$p$--chain of $X$. 
We denote by $C^C_{p-1}(X,Y,{\cal O})$, $Z^C_{p-1}(X,Y,{\cal O})$, 
$B^C_{p-1}(X,Y,{\cal O})$
the groups of $Y$ relative integer \v Cech $p-1$--chains,
cycles and boundaries of $X$, respectively.
Two relative $p-1$--cycles are equivalent 
if their difference is a relative $p-1$--boundary.
The equivalence classes of $Y$ relative integer \v Cech 
$p-1$--cycles of $X$ form the $p-1$--th relative integer \v Cech homology 
group $H^C_{p-1}(X,Y,{\cal O})$. 

For $r\in \Bbb N$, set $I_r=\{0,1,2,\ldots,r\}$.
A $Y$ relative \v Cech singular $p-1$--intertwiner of $X$ is a sequence
$(S_{-1,p-1},T_{-1,p-2};$ $\{V_{k,p-1-k}|k\in I_{p-1}\}, 
\{Z_{k,p-2-k}|k\in I_{p-2}\};S_{p-1,-1},T_{p-2,-1})$ 
with $S_{-1,p-1}\in{\cal C}_{-1,p-1}(X,$ ${\cal O})$, 
$T_{-1,p-2}\in{\cal C}_{-1,p-2}(Y,{\cal O}\cap Y)$,
$V_{k,p-1-k}\in{\cal C}_{k,p-1-k}(X,{\cal O})$,
$Z_{k,p-2-k}\in{\cal C}_{k,p-2-k}(Y,{\cal O}\cap Y)$,
$S_{p-1,-1}\in{\cal C}_{p-1,-1}(X,{\cal O})$, 
$T_{p-2,-1}\in{\cal C}_{p-2,-1}(Y,{\cal O}\cap Y)$
satisfying 
$$
\eqalignno{\vphantom{\bigg(}&
S_{-1,p-1}=\beta V_{0,p-1},&(1.2.5a)\cr  
\vphantom{\bigg(}&
T_{-1,p-2}=\beta Z_{0,p-2}, &(1.2.5b)\cr 
\vphantom{\bigg(}&
b V_{k,p-1-k} = \beta V_{k+1,p-2-k}+(-1)^k i_*Z_{k,p-2-k},
\quad 0\leq k\leq p-2, &(1.2.6a)\cr
\vphantom{\bigg(}&
b Z_{k,p-2-k} = \beta Z_{k+1,p-3-k},
\quad 0\leq k\leq p-3, &(1.2.6b)\cr
\vphantom{\bigg(}&
S_{p-1,-1}=b V_{p-1,0},&(1.2.7a)\cr
\vphantom{\bigg(}&
T_{p-2,-1}=-(-1)^{p-2} b Z_{p-2,0}.
&(1.2.7b)\cr}
$$
Note that $(S_{p-1},T_{p-2})\in Z^{s{\cal O}}_{p-1}(X,Y)$ (cfr. eq. 
(1.2.1)) and $(S_{p-1,-1},T_{p-2,-1})\in Z^C_{p-1}(X,$ 
$Y,{\cal O})$ (cfr. eq. (1.2.3)). A $Y$ relative singular 
\v Cech $p-1$--intertwiner
$(S_{-1,p-1},T_{-1,p-2};$ $\{V_{k,p-1-k}|k\in I_{p-1}\}, 
\{Z_{k,p-2-k}|k\in I_{p-2}\};S_{p-1,-1},T_{p-2,-1})$ 
of $X$ is said trivial if
$$
\eqalignno{\vphantom{\bigg(}&
 S_{-1,p-1}=bs_{-1,p}-i_*t_{-1,p-1}, &(1.2.8a)\cr  
\vphantom{\bigg(}&
 T_{-1,p-2}=-bt_{-1,p-1}, &(1.2.8b)\cr
\vphantom{\bigg(}&
V_{k,p-1-k} = b v_{k,p-k}+\beta v_{k+1,p-1-k}+(-1)^k i_*z_{k,p-1-k},
\quad 0\leq k\leq p-1,&(1.2.9a)\cr
\vphantom{\bigg(}&
Z_{k,p-2-k} = b z_{k,p-1-k}+\beta z_{k+1,p-2-k},
\quad 0\leq k\leq p-2, &(1.2.9b)\cr 
\vphantom{\bigg(}&
S_{p-1,-1}=\beta s_{p,-1}-i_*t_{p-1,-1},&(1.2.10a)\cr
\vphantom{\bigg(}&
 T_{p-2,-1}=-\beta t_{p-1,-1},&(1.2.10b)\cr}
$$ 
where $s_{-1,p}\in{\cal C}_{-1,p}(X,{\cal O})$, $t_{-1,p-1}\in
{\cal C}_{-1,p-1}(Y,{\cal O}\cap Y)$, $v_{k,p-k}\in
{\cal C}_{k,p-k}(X,{\cal O})$, for $0\leq k\leq p$,
$z_{k,p-1-k}\in{\cal C}_{k,p-1-k}(Y,{\cal O}\cap Y)$, for $0\leq k\leq p-1$,
$s_{p,-1}\in{\cal C}_{p,-1}(X,{\cal O})$, $t_{p-1,-1}\in
{\cal C}_{p-1,-1}(Y,{\cal O}\cap Y)$ are such that
$$
\eqalignno{\vphantom{\bigg(}&
s_{-1,p}=\beta v_{0,p},&(1.2.11a)\cr  
\vphantom{\bigg(}&
t_{-1,p-1}=-\beta z_{0,p-1},&(1.2.11b)\cr  
\vphantom{\bigg(}&
s_{p,-1}=b v_{p,0},&(1.2.12a)\cr
\vphantom{\bigg(}&
t_{p-1,-1}=-(-1)^{p-1} b z_{p-1,0}.&(1.2.12b)\cr}
$$
We denote by $ZI^{Cs}_{p-1}(X,Y,{\cal O})$, $BI^{Cs}_{p-1}(X,Y,{\cal O})$
the groups of $Y$ relative \v Cech singular $p-1$--intertwiners,
and trivial intertwiners of $X$, respectively.
Two relative $p-1$--intertwiners are equivalent 
if their difference is trivial.
The equivalence classes of $Y$ relative \v Cech singular 
$p-1$--intertwiners of $X$ form a group $HI^{Cs}_{p-1}(X,Y,{\cal O})$.
The notion of intertwiner given here is the generalization of 
that of `element' of \ref{39} suitable for relative homology.

A $Y$ relative de Rham $p$--cochain $(\Xi^p,\Upsilon^{p-1})$
of $X$ is a pair of de Rham cochains 
$\Xi^p\in{\cal D}^p(X)$ , 
$\Upsilon^{p-1}\in{\cal D}^{p-1}(Y)$.
A $Y$ relative de Rham $p$--cochain $(\Xi^p,\Upsilon^{p-1})$
of $X$ is a cocycle if 
$$
\eqalignno{\vphantom{\bigg(}
&d\Xi^p=0,&(1.2.13a)\cr
\vphantom{\bigg(}
&i^*\Xi^p-d\Upsilon^{p-1}=0.&(1.2.13b)\cr}
$$
A $Y$ relative de Rham $p$--cocycle $(\Xi^p,\Upsilon^{p-1})$
of $X$ is a coboundary if it is of the form
$$
\eqalignno{\vphantom{\bigg(}
& \Xi^p=d\xi^{p-1},&(1.2.14a)\cr
\vphantom{\bigg(}
&  \Upsilon^{p-1}= i^*\xi^{p-1}-d\upsilon^{p-2},&(1.2.14b)\cr}
$$
where $(\xi^{p-1},\upsilon^{p-2})$ is an arbitrary $Y$ relative 
de Rham $p-1$--cochain of $X$. 
We denote by $C_{dR}^p(X,Y)$, $Z_{dR}^p(X,Y)$, $B_{dR}^p(X,Y)$
the vector spaces of $Y$ relative de Rham $p$--cochains,
cocycles and coboundaries of $X$, respectively.
Two relative $p$--cocycles are equivalent 
if their difference is a $p$--coboundary.
The equivalence classes of $Y$ relative de Rham $p$--cocycles of $X$
span the $p$--th relative de Rham cohomology space $H_{dR}^p(X,Y)$.

A $Y$ relative de Rham $p-1$--cochain $(\Xi^p,\Upsilon^{p-1})$
can be viewed as a pair of \v Cech--de Rham cochains 
$(\Xi^{-1,p},\Upsilon^{-1,p-1})$ with 
$\Xi^{-1,p}\in{\cal C}^{-1,p}(X,{\cal O})$, 
$\Upsilon^{-1,p-1}\in{\cal C}^{-1,p-1}(Y,{\cal O}\cap Y)$.
We shall use both notations interchangeably depending on context.

A $Y$ relative real \v Cech $p$--cochain $(\Xi^{p,-1},\Upsilon^{p-1,-1})$
of $X$ is a pair of real \v Cech cochains 
$\Xi^{p,-1}\in{\cal C}^{p,-1}(X,{\cal O})$, 
$\Upsilon^{p-1,-1}\in{\cal C}^{p-1,-1}(Y,{\cal O}\cap Y)$.
A $Y$ relative real \v Cech $p$--cochain $(\Xi^{p,-1},\Upsilon^{p-1,-1})$
of $X$ is a cocycle if 
$$
\eqalignno{\vphantom{\bigg(}
&\delta \Xi^{p,-1}=0,&(1.2.15a)\cr
\vphantom{\bigg(}
&i^*\Xi^{p,-1}-\delta \Upsilon^{p-1,-1}=0.&(1.2.15b)\cr}
$$
A $Y$ relative real \v Cech $p$--cocycle $(\Xi^{p,-1},\Upsilon^{p-1,-1})$
of $X$ is a coboundary if it is of the form
$$
\eqalignno{\vphantom{\bigg(}
& \Xi^{p,-1}=\delta \xi^{p-1,-1},&(1.2.16a)\cr
\vphantom{\bigg(}
& \Upsilon^{p-1,-1}=i^*\xi^{p-1,-1}-\delta\upsilon^{p-2,-1},&(1.2.16b)\cr}
$$
where $(\xi^{p-1,-1},\upsilon^{p-2,-1})$ is an arbitrary $Y$ relative 
real \v Cech $p-1$--cochain of $X$.
We denote by $C_C^p(X,Y,{\cal O})$, $Z_C^p(X,Y,{\cal O})$, 
$B_C^p(X,Y,{\cal O})$ the vector spaces of $Y$ relative real 
\v Cech $p$--cochains, cocycles and coboundaries of $X$, respectively.
Two relative $p$--cocycles are equivalent if their difference is a 
$p$--coboundary. The equivalence classes of $Y$ relative real \v Cech 
$p$--cocycles of $X$ form the $p$--th relative real \v Cech cohomology 
space $H_C^p(X,Y,{\cal O})$.

For $r\in \Bbb N$, set $I^r=\{0,1,2,\ldots,r\}$.
A $Y$ relative \v Cech--de Rham $p$--cointertwiner of $X$ is a sequence
$(\Xi^{-1,p},\Upsilon^{-1,p-1};$ $\{\Omega^{k,p-1-k}|k\in I^{p-1}\}, 
\{\Theta^{k,p-2-k}|k\in I^{p-2}\};\Xi^{p,-1},\Upsilon^{p-1,-1})$, 
where $\Xi^{-1,p}\in{\cal C}^{-1,p}(X,{\cal O})$, 
$\Upsilon^{-1,p-1}\in{\cal C}^{-1,p-1}(Y,{\cal O}\cap Y)$,
$\Omega^{k,p-1-k}\in{\cal C}^{k,p-1-k}(X,{\cal O})$,
$\Theta^{k,p-2-k}\in{\cal C}^{k,p-2-k}(Y,{\cal O}\cap Y)$,
$\Xi^{p,-1}\in{\cal C}^{p,-1}(X,{\cal O})$, 
$\Upsilon^{p-1,-1}\in{\cal C}^{p-1,-1}(Y,{\cal O}\cap Y)$
satisfy
$$
\eqalignno{\vphantom{\bigg(}&
\delta \Xi^{-1,p}= d\Omega^{0,p-1},&(1.2.17a)\cr
\vphantom{\bigg(}&
\delta \Upsilon^{-1,p-1}=-d\Theta^{0,p-2}+i^*\Omega^{0,p-1},&(1.2.17b)\cr
\vphantom{\bigg(}&
d \Omega^{k,p-1-k} = \delta \Omega^{k-1,p-k},
\quad 1\leq k\leq p-1, &(1.2.18a)\cr
\vphantom{\bigg(}&
d \Theta^{k,p-2-k} = \delta \Theta^{k-1,p-1-k}+(-1)^k i^*\Omega^{k,p-1-k},
\quad 1\leq k\leq p-2, &(1.2.18b)\cr
\vphantom{\bigg(}&
d\Xi^{p,-1} =\delta \Omega^{p-1,0},&(1.2.19a)\cr
\vphantom{\bigg(}&
d\Upsilon^{p-1,-1} =(-1)^{p-1}\big(\delta\Theta^{p-2,0}
+(-1)^{p-1}i^*\Omega^{p-1,0}\big).&(1.2.19b)\cr}
$$
Note that $(\Xi^p,\Upsilon^{p-1})\in Z_{dR}^p(X,Y)$ (cfr. eq. 
(1.2.13)) and $(\Xi^{p,-1},\Upsilon^{p-1,-1})\in Z_C^p(X,Y,{\cal O})$ 
(cfr. eq. (1.2.15)). We call a $Y$ relative \v Cech--de Rham 
$p$--cointertwiner $(\Xi^{-1,p},\Upsilon^{-1,p-1};$ 
$\{\Omega^{k,p-1-k}|k\in I^{p-1}\},$  
$\{\Theta^{k,p-2-k}|k\in I^{p-2}\};\Xi^{p,-1},\Upsilon^{p-1,-1})$
of $X$ trivial if
$$
\eqalignno{\vphantom{\bigg(}&
\Xi^{-1,p}=d\xi^{-1,p-1}, &(1.2.20a)\cr  
\vphantom{\bigg(}&
\Upsilon^{-1,p-1}=i^*\xi^{-1,p-1}- d\upsilon^{-1,p-2}, 
&(1.2.20b)\cr  
\vphantom{\bigg(}&
\Omega^{k,p-1-k} = d\omega^{k,p-2-k}+\delta\omega^{k-1,p-1-k},
\quad 0\leq k\leq p-1,&(1.2.21a)\cr
\vphantom{\bigg(}&
\Theta^{k,p-2-k} = d\theta^{k,p-3-k}+\delta\theta^{k-1,p-2-k}
+(-1)^ki^*\omega^{k,p-2-k},\quad 0\leq k\leq p-2,~~~~~ &(1.2.21b)\cr 
\vphantom{\bigg(}&
\Xi^{p,-1}=\delta\xi^{p-1,-1},&(1.2.22a)\cr
\vphantom{\bigg(}&
\Upsilon^{p-1,-1}=i^*\xi^{p-1,-1}-\delta\upsilon^{p-2,-1},&(1.2.22b)\cr}
$$
where $\xi^{-1,p-1}\in{\cal C}^{-1,p-1}(X,{\cal O})$,
$\upsilon^{-1,p-2}\in{\cal C}^{-1,p-2}(Y,{\cal O}\cap Y)$,
$\omega^{k,p-2-k}\in{\cal C}^{k,p-2-k}(X,{\cal O})$,
for $-1\leq k\leq p-1$, $\theta^{k,p-3-k}\in
{\cal C}^{k,p-3-k}(Y,{\cal O}\cap Y)$, for $-1\leq k\leq p-2$, 
$\xi^{p-1,-1}\in{\cal C}^{p-1,-1}(X,{\cal O})$,
$\upsilon^{p-2,-1}\in{\cal C}^{p-2,-1}(Y,{\cal O}\cap Y)$
with
$$
\eqalignno{\vphantom{\bigg(}&
\xi^{-1,p-1}=\omega^{-1,p-1},&(1.2.23a)\cr  
\vphantom{\bigg(}&
\upsilon^{-1,p-2}=\theta^{-1,p-2},&(1.2.23b)\cr  
\vphantom{\bigg(}&
\xi^{p-1,-1}=\omega^{p-1,-1},&(1.2.24a)\cr
\vphantom{\bigg(}&
\upsilon^{p-2,-1}=(-1)^{p-2}\theta^{p-2,-1}.&(1.2.24b)\cr}
$$
We denote by $ZI_{CdR}^p(X,Y,{\cal O})$, $BI_{CdR}^p(X,Y,$ ${\cal O})$ 
the spaces of $Y$ relative \v Cech--de Rham $p$--cointertwiners,
and trivial cointertwiners of $X$, respectively.
Two relative $p$--cointertwiners are equivalent 
if their difference is trivial.
The equivalence classes of $Y$ relative \v Cech--de Rham 
$p$--cointertwiners of $X$ form a space $HI_{CdR}^p(X,Y,{\cal O})$.
The notion of cointertwiner given here is the generalization of 
that of `coelement' of \ref{39} suitable for relative cohomology.

\par\vskip.6cm
\item{\it 3.} {\it Integral relative \v Cech cohomology and 
relative differential cocycles}
\vskip.4cm
\par

A $Y$ relative integer \v Cech $p$--cochain $(\hat\Xi^{p,-1},
\hat\Upsilon^{p-1,-1})$ of $X$ is a pair of integer \v Cech cochains 
$\hat\Xi^{p,-1}\in{\cal C}_{\Bbb Z}^{p,-1}(X,{\cal O})$, 
$\hat\Upsilon^{p-1,-1}\in{\cal C}_{\Bbb Z}^{p-1,-1}(Y,{\cal O}\cap Y)$.
Clearly, a relative integer \v Cech cochain is also a relative 
real \v Cech cochain.
A $Y$ relative integer \v Cech $p$--cochain 
$(\hat\Xi^{p,-1},\hat\Upsilon^{p-1,-1})$ of $X$ is a cocycle 
if it satisfies eq. (1.2.15) with $(\Xi^{p,-1},\Upsilon^{p-1,-1})$
replaced by $(\hat\Xi^{p,-1},\hat\Upsilon^{p-1,-1})$, so that
it is a cocycle also when seen as a relative real \v Cech cochain.
A $Y$ relative integer \v Cech $p$--cocycle  
$(\hat\Xi^{p,-1},\hat\Upsilon^{p-1,-1})$ of $X$ is a coboundary
if it satisfies eq. (1.2.16) with $(\xi^{p-1,-1},\upsilon^{p-2,-1})$
replaced by any $Y$ relative integer \v Cech $p-1$--cochain 
$(\hat\xi^{p-1,-1},\hat\upsilon^{p-2,-1})$, 
so that it is a coboundary also when seen as a relative real \v Cech cochain.
We denote by $C_{C\Bbb Z}^p(X,Y,{\cal O})$, $Z_{C\Bbb Z}^p(X,Y,{\cal O})$, 
$B_{C\Bbb Z}^p(X,Y,{\cal O})$ the groups of $Y$ relative integer 
\v Cech $p$--cochains, cocycles and coboundaries of $X$, respectively.
Two relative integer $p$--cocycles are equivalent if their difference is a 
integer $p$--coboundary. The equivalence classes of $Y$ relative integer  
\v Cech $p$--cocycles of $X$ form the $p$--th relative integer \v Cech 
cohomology group $H_{C\Bbb Z}^p(X,Y,{\cal O})$.

A $Y$ relative differential $p$--cocycle of $X$ is a \v Cech six--tuple
$(\Xi^{p,-1},\Upsilon^{p-1,-1};\Xi^{*p-1,-1},$ $\Upsilon^{*p-2,-1};
\hat\Xi^{p,-1},\hat\Upsilon^{p-1,-1})$, where
$\Xi^{p,-1}\in{\cal C}^{p,-1}(X,{\cal O})$, 
$\Upsilon^{p-1,-1}\in{\cal C}^{p-1,-1}(Y,{\cal O}\cap Y)$,
$\Xi^{*p-1,-1}\in{\cal C}^{p-1,-1}(X,{\cal O})$,
$\Upsilon^{*p-2,-1}\in{\cal C}^{p-2,-1}(Y,{\cal O}\cap Y)$,
$\hat\Xi^{p,-1}\in{\cal C}_{\Bbb Z}^{p,-1}(X,{\cal O})$, 
$\hat\Upsilon^{p-1,-1}$ $\in{\cal C}_{\Bbb Z}^{p-1,-1}(Y,{\cal O}\cap Y)$,
satisfying 
$$
\eqalignno{\vphantom{\bigg(}&
\delta \Xi^{p,-1}=0,&(1.3.1a)\cr
\vphantom{\bigg(}&
i^*\Xi^{p,-1}-\delta \Upsilon^{p-1,-1}=0,&(1.3.1b)\cr
\vphantom{\bigg(}&
\delta \Xi^{*p-1,-1}=\hat\Xi^{p,-1}-\Xi^{p,-1},&(1.3.2a)\cr
\vphantom{\bigg(}&
i^*\Xi^{*p-1,-1}-\delta \Upsilon^{*p-2,-1}=
\hat\Upsilon^{p-1,-1}-\Upsilon^{p-1,-1},&(1.3.2b)\cr
\vphantom{\bigg(}&
\delta \hat\Xi^{p,-1}=0,&(1.3.3a)\cr
\vphantom{\bigg(}&
i^*\hat\Xi^{p,-1}-\delta \hat\Upsilon^{p-1,-1}=0.&(1.3.3b)\cr}
$$
Note that $(\Xi^{p,-1},\Upsilon^{p-1,-1})\in Z_C^p(X,Y,{\cal O})$ 
and $(\hat\Xi^{p,-1},\hat\Upsilon^{p-1,-1})
\in Z_{C\Bbb Z}^p(X,Y,{\cal O})$ (cfr. eq. (1.2.15)). 
A $Y$ relative differential $p$--cocycle 
$(\Xi^{p,-1},\Upsilon^{p-1,-1};\Xi^{*p-1,-1},\Upsilon^{*p-2,-1};
\hat\Xi^{p,-1},$ $\hat\Upsilon^{p-1,-1})$ of $X$ is a differential 
coboundary if 
$$
\eqalignno{\vphantom{\bigg(}&
\Xi^{p,-1}=\delta\xi^{p-1,-1}, &(1.3.4a)\cr  
\vphantom{\bigg(}&
 \Upsilon^{p-1,-1}=i^*\xi^{p-1,-1}- \delta\upsilon^{p-2,-1}, 
&(1.3.4b)\cr
\vphantom{\bigg(}&
\Xi^{*p-1,-1} =\hat\xi^{p-1,-1}-\xi^{p-1,-1},&(1.3.5a)\cr
\vphantom{\bigg(}&
\Upsilon^{*p-2,-1} =\hat\upsilon^{p-2,-1}-\upsilon^{p-2,-1}, &(1.3.5b)\cr 
\vphantom{\bigg(}&
\hat \Xi^{p,-1}=\delta\hat \xi^{p-1,-1},&(1.3.6a)\cr 
\vphantom{\bigg(}&
\hat \Upsilon^{p-1,-1}= i^*\hat \xi^{p-1,-1}-\delta\hat \upsilon^{p-2,-1},
&(1.3.6b)\cr}
$$
where $\xi^{p-1,-1}\in{\cal C}^{p-1,-1}(X,{\cal O})$,
$\upsilon^{p-2,-1}\in{\cal C}^{p-2,-1}(Y,{\cal O}\cap Y)$,
$\hat \xi^{p-1,-1}\in{\cal C}_{\Bbb Z}^{p-1,-1}(X,{\cal O})$,
$\hat \upsilon^{p-2,-1}\in{\cal C}_{\Bbb Z}^{p-2,-1}(Y,{\cal O}\cap Y)$.
We denote by $ZD_C^p(X,Y,{\cal O})$, $BD_C^p(X,Y,{\cal O})$
the groups of $Y$ relative differential $p$--cocycles
and coboundaries of $X$, respectively.
Two relative differential $p$--cocycles are equivalent 
if their difference is a differential coboundary.
The equivalence classes of $Y$ relative differential 
$p$--cocycle of $X$ form a group $HD_C^p(X,Y,{\cal O})$.
An analogous notion of differential cocycle has been introduced 
for the absolute case in \ref{15}.

A $Y$ relative differential $p$-cocycle 
$(\Xi^{p,-1},\Upsilon^{p-1,-1};\Xi^{*p-1,-1},\Upsilon^{*p-2,-1};
\hat\Xi^{p,-1},\hat\Upsilon^{p-1,-1})$ of $X$ is torsion if
it is of the form (1.3.4)--(1.3.6) 
with $\hat \xi^{p-1,-1}\in{\cal C}^{p-1,-1}(X,{\cal O})$,
$\hat \upsilon^{p-2,-1}\in{\cal C}^{p-2,-1}(Y,{\cal O}\cap Y)$ 
subject to the condition $\delta\hat \xi^{p-1,-1}
\in{\cal C}_{\Bbb Z}^{p,-1}(X,{\cal O})$,
$i^*\hat \xi^{p-1,-1}-\delta\hat \upsilon^{p-2,-1}
\in{\cal C}_{\Bbb Z}^{p-1,-1}(Y,{\cal O}\cap Y)$.
Torsion differential cocycles form a subgroup 
$ZD_{Ct}^p(X,Y,{\cal O})$ of $ZD_C^p(X,$ $Y,{\cal O})$.
Being invariant under translation by $BD_C^p(X,Y,{\cal O})$,
$ZD_{Ct}^p(X,Y,{\cal O})$ projects to a subgroup 
$HD_{Ct}^p(X,Y,{\cal O})$ of $HD_C^p(X,Y,{\cal O})$.

\par\vskip.6cm
\item{\it 4.} {\it Relative homology and $\cal O$--small homology isomorphism
and the relative \v Cech singular/ \v Cech--de Rham isomorphisms}
\vskip.4cm
\par
The barycentric subdivision operator $q$ (cfr. subsect. 1.1)
acts on relative chains in obvious fashion. For any relative chain 
$(S_{p-1},T_{p-2})\in C^s_{p-1}(X,Y)$, there is an integer 
$k(S_{p-1},T_{p-2},{\cal O})\geq 0$ such that 
$(q^kS_{p-1},q^kT_{p-2})\in C^{s{\cal O}}_{p-1}(X,Y)$ is $\cal O$--small for 
$k\geq k(S_{p-1},T_{p-2},{\cal O})$. By the chain relation (1.1.5),
if $(S_{p-1},T_{p-2})\in Z^s_{p-1}(X,Y)$ is a relative 
cycle, then $(q^kS_{p-1},q^kT_{p-2})\in Z^{s{\cal O}}_{p-1}(X,Y)$ 
also is. If $(S_{p-1},T_{p-2})\in B^s_{p-1}(X,Y)$ is a relative boundary, 
then $(q^kS_{p-1},q^kT_{p-2})\in B^{s{\cal O}}_{p-1}(X,Y)$ 
also is and the corresponding relative chain $(q^ks_p,q^kt_{p-1})$ 
is $\cal O$--small for $k$ large enough (cfr. eq. (1.2.2)). 
Using the chain relation (1.1.5) and the homotopy relation (1.1.6), 
it is possible to construct 
a chain equivalence of the complex of $Y$ relative singular chains
and that of $\cal O$--small $Y$ relative singular chains for any open covering 
${\cal O}$ of $X$ \ref{36}. Hence, the corresponding homologies are isomorphic
$$
H^s_{p-1}(X,Y) \cong H^{s{\cal O}}_{p-1}(X,Y).
\eqno(1.4.1)
$$

We say that the open covering $\cal O$ of $X$ is 
a good covering of the pair $X$, $Y$, if $\cal O$ is a good covering of $X$ 
and ${\cal O}\cap Y$  is good a covering of $Y$. (See appendix A1.)

An $\cal O$--small $Y$ relative singular $p-1$--cycle $(S_{p-1},T_{p-2})
\in Z^{s{\cal O}}_{p-1}(X,Y)$ and a $Y$ relative integer \v Cech $p$--cycle
$(S_{p-1,-1},T_{p-2,-1})\in Z^C_{p-1}(X,Y,{\cal O})$
are said compatible if they fit into some $Y$ relative 
\v Cech singular $p$--intertwiner $(S_{-1,p-1},T_{-1,p-2};
\{V_{k,p-1-k}\},\{Z_{k,p-2-k}\};$ $S_{p-1,-1},T_{p-2,-1})
\in ZI^{Cs}_{p-1}(X,Y,{\cal O})$ (cfr. eqs. (1.2.1), (1.2.3), 
(1.2.5)--(1.2.7)). 
From eqs. (1.2.2), (1.2.4), (1.2.8)--(1.2.10), 
it follows that any $\cal O$--small $Y$ relative singular $p-1$--boundary
$(S_{p-1},T_{p-2})\in B^{s{\cal O}}_{p-1}(X,Y)$
is always compatible with any $Y$ relative integer 
\v Cech $p$--boundary $(S_{p-1,-1},T_{p-2,-1})\in B^C_{p-1}(X,Y,{\cal O})$
through a trivial intertwiner in $BI^{Cs}_{p-1}(X,Y,{\cal O})$. 
Therefore, the compatibility relation in $Z^{s{\cal O}}_{p-1}(X,Y)\times 
Z^C_{p-1}(X,Y,{\cal O})$ defined above induces a compatibility relation in 
$H^{s{\cal O}}_{p-1}(X,Y)\times H^C_{p-1}(X,Y,{\cal O})$ at the level of 
relative homology. A fundamental theorem states that, when $\cal O$ is 
a good covering of the pair $X$, $Y$, this relation
is actually an isomorphism
$$
H^{s{\cal O}}_{p-1}(X,Y) \cong H^C_{p-1}(X,Y,{\cal O}).
\eqno(1.4.2)
$$
Its proof is analogous to that of the absolute case \ref{34}.
On account of the isomorphism (1.4.1), we find out that,  
for such coverings, $H^C_{p-1}(X,Y,{\cal O})$ does not depend
on $\cal O$ up to isomorphism.
 
A $Y$ relative de Rham $p$--cocycle $(\Xi^p,\Upsilon^{p-1})\in 
Z_{dR}^p(X,Y)$ and a $Y$ relative real \v Cech $p$--cocycle
$(\Xi^{p,-1},\Upsilon^{p-1,-1})\in Z_C^p(X,Y,{\cal O})$ 
are said compatible if they fit into some $Y$ relative 
\v Cech--de Rham $p$--cointertwiner $(\Xi^{-1,p},\Upsilon^{-1,p-1};
\{\Omega^{k,p-1-k}\},\{\Theta^{k,p-2-k}\};\Xi^{p,-1},\Upsilon^{p-1,-1})$ 
$\in ZI_{CdR}^p(X,Y,{\cal O})$ (cfr. eqs. (1.2.13), (1.2.15), 
(1.2.17)--(1.2.19)). 
From eqs. (1.2.14), (1.2.16), (1.2.20)--(1.2.22), 
it follows that a $Y$ relative de Rham $p$--coboundary
$(\Xi^p,\Upsilon^{p-1})\in B_{dR}^p(X,Y)$ is compatible with any 
$Y$ relative real \v Cech $p$--coboundary $(\Xi^{p,-1},\Upsilon^{p-1,-1})$
$\in B_C^p(X,Y,{\cal O})$ through a trivial cointertwiner in
$BI_{CdR}^p(X,Y,{\cal O})$. Therefore, the compatibility 
relation in $Z_{dR}^p(X,Y)\times Z_C^p(X,Y,{\cal O})$ defined above induces 
a compatibility relation in $H_{dR}^p(X,Y)\times H_C^p(X,Y,{\cal O})$ 
at the level of relative cohomology. A fundamental theorem states that,
when $\cal O$ is a good covering of the pair $X$, $Y$, this
relation is actually an isomorphism
$$
H_{dR}^p(X,Y) \cong H_C^p(X,Y,{\cal O}),
\eqno(1.4.3)
$$
so that for such coverings $H_C^p(X,Y,{\cal O})$ does not depend 
on $\cal O$ up to isomorphism. Again, the proof 
is analogous to that of the absolute case \ref{34}. 

\par\vskip.6cm
\item{\it 5.} {\it Integrality in relative cohomology}
\vskip.4cm
\par

As is well--known, given any Abelian group $G$, by dualization via the functor 
$\Hom_{\Bbb Z}(\cdot,$ $G)$ of the singular chain complex of a manifold $M$, 
one can construct the singular cochain complex of $M$ with coefficients 
in $G$. When an open covering ${\cal O}$ of $M$ is given, one can 
similarly define ${\cal O}$--small singular cochains and \v Cech singular 
cochains with coefficients in $G$. This allows one to set up a 
cohomological framework that parallels completely the original homological 
one. (See refs. \ref{34,36} for background material.)
The generalization to the relative case is straightforward.

Proceeding as outlined above, it is possible to introduce the 
relative real singular cohomology space $H^p_s(X,Y)$ 
and the relative integer singular cohomology group 
$H^p_{s\Bbb Z}(X,Y)$. 

The natural inclusion of the group of relative integer singular cochains
into the space of relative real singular cochains is a cochain map. 
Thus, there is a canonical homomorphism $H^p_{s\Bbb Z}(X,Y)
\rightarrow H^p_s(X,Y)$ of the relative singular cohomology.
Its kernel $Tor_s^p(X,Y)$ is the relative singular torsion subgroup 
of $H^p_{s\Bbb Z}(X,Y)$.
Its range $\tilde H_{s\Bbb Z}^p(X,Y)$ is the relative integer 
singular cohomology lattice of $H^p_s(X,Y)$. 

The above setting has  faithful translation in relative \v Cech 
cohomology. Let $\cal O$ be a covering of $X$. 
The inclusion $C_{C\Bbb Z}^p(X,Y,{\cal O})\rightarrow C_C^p(X,Y,{\cal O})$ 
is a cochain map (cfr. subsect. 1.3). Thus, it induces a
homomorphism $H_{C\Bbb Z}^p(X,Y,{\cal O})\rightarrow H_C^p(X,Y,{\cal O})$ 
of the relative integer \v Cech cohomology group 
into the relative real \v Cech cohomology space.
Its kernel $Tor_C^p(X,Y,{\cal O})$ is the relative \v Cech torsion subgroup 
of $H_{C\Bbb Z}^p(X,Y,{\cal O})$.
Its range $\tilde H_{C\Bbb Z}^p(X,Y,{\cal O})$ is the relative integer 
\v Cech cohomology lattice of $H_C^p(X,Y,{\cal O})$. 

If ${\cal O}$ is restricted to be a good open covering of $X$, $Y$
(cfr. subsect. 1.4), then relative singular cohomology and  
relative \v Cech cohomology are completely isomorphic:
$$
\eqalignno{
H^p_{s\Bbb Z}(X,Y)&\cong H_{C\Bbb Z}^p(X,Y,{\cal O}),&(1.5.1)\cr
Tor_s^p(X,Y)&\cong Tor_C^p(X,Y,{\cal O}),&(1.5.2)\cr
H^p_s(X,Y)&\cong H_C^p(X,Y,{\cal O}),&(1.5.3)\cr
\tilde H_{s\Bbb Z}^p(X,Y)&\cong\tilde H_{C\Bbb Z}^p(X,Y,{\cal O}).&(1.5.4)\cr}
$$
The above isomorphisms are consistent: the isomorphisms
(1.5.2), (1.5.4) are the restriction the isomorphisms
(1.5.1), (1.5.3), respectively. Further, the homomorphism 
$H_{C\Bbb Z}^p(X,Y,{\cal O})$ $\rightarrow H_C^p(X,Y,{\cal O})$ 
is obtained by the composition of the homomorphism 
$H^p_{s\Bbb Z}(X,Y)\rightarrow H^p_s(X,Y)$ 
with the isomorphisms (1.5.1), (1.5.3).
The proofs are formally analogous to that of the isomorphism (1.4.3), 
though extra work must be done to show the isomorphism of
$\cal O$--small relative singular cohomology and relative 
singular cohomology. Note that, by (1.5.1)--(1.5.4), 
$H_{C\Bbb Z}^p(X,Y,{\cal O})$, $Tor_C^p(X,Y,{\cal O})$,
$H_C^p(X,Y,{\cal O})$, $\tilde H_{C\Bbb Z}^p(X,Y,{\cal O})$ are all
independent from the good open covering ${\cal O}$ up to isomorphism.

Let ${\cal O}$ be a good open covering of $X$, $Y$.
A $Y$ relative real \v Cech $p$--cocycle $(\Xi^{p,-1},$ $\Upsilon^{p-1,-1})
\in Z_C^p(X,Y,{\cal O})$ 
is said cohomologically integer if it fits into some $Y$ relative 
differential $p$--cocycle $(\Xi^{p,-1},\Upsilon^{p-1,-1};\Xi^{*p-1,-1},$
$\Upsilon^{*p-2,-1};\hat\Xi^{p,-1},\hat\Upsilon^{p-1,-1})
\in ZD_C^p(X,Y,{\cal O})$ (cfr. eqs. (1.2.15), (1.3.1)--(1.3.3)). 
From eqs. (1.2.16), (1.3.4)--(1.3.6), it follows that any $Y$ relative 
real \v Cech $p$--coboundary $(\Xi^{p,-1},\Upsilon^{p-1,-1})
\in B_C^p(X,Y,{\cal O})$ is 
always cohomologically integer, being part of a  
differential coboundary in  $BD_C^p(X,Y,{\cal O})$. 
We denote by $\tilde Z_{C\Bbb Z}^p(X,Y,{\cal O})$ the subgroup of
$Z_C^p(X,Y,{\cal O})$ formed by the cohomologically integer relative 
real \v Cech $p$--cocycles. Being invariant under translation by 
$B_C^p(X,Y,{\cal O})$, $\tilde Z_{C\Bbb Z}^p(X,Y,{\cal O})$ projects 
to a lattice of $H_C^p(X,Y,{\cal O})$. Clearly, $Z_{C\Bbb Z}^p(X,Y,{\cal O})
\subseteq \tilde Z_{C\Bbb Z}^p(X,Y,{\cal O})$ (cfr. subsect. 1.3)
and the lattice mentioned 
is precisely the relative integer \v Cech cohomology lattice 
$\tilde H_{C\Bbb Z}^p(X,Y,{\cal O})$ introduced above.

A $Y$ relative de Rham $p$--cocycle $(\Xi^p,\Upsilon^{p-1})\in 
Z_{dR}^p(X,Y)$ is said cohomologically integer, if it is compatible with
some cohomologically integer $Y$ relative real \v Cech $p$--cocycle
$(\Xi^{p,-1},\Upsilon^{p-1,-1})\in \tilde Z_{C\Bbb Z}^p(X,Y,{\cal O})$
for some good open covering ${\cal O}$ of $X$, $Y$ (cfr. subsect. 1.4). 
A $Y$ relative de Rham $p$--coboundary $(\Xi^p,\Upsilon^{p-1})\in 
B_{dR}^p(X,Y)$ is always cohomologically integer, since it is compatible with
a $Y$ relative real \v Cech $p$--coboundary $(\Xi^{p,-1},\Upsilon^{p-1,-1})
\in B_C^p(X,Y,{\cal O})$, which is necessarily cohomologically integer,
for any good open covering ${\cal O}$.
We denote by $Z_{dR\Bbb Z}^p(X,Y)$ be the subgroup of $Z_{dR}^p(X,Y)$
formed by the cohomologically integer relative de Rham $p$--cocycles. 
Since $Z_{dR\Bbb Z}^p(X,Y)$ is invariant under
translation by $B_{dR}^p(X,Y)$, $Z_{dR\Bbb Z}^p(X,Y)$ projects 
to a lattice subgroup $H_{dR\Bbb Z}^p(X,Y)$ of $H_{dR}^p(X,Y)$.
For any fixed good covering ${\cal O}$, every cohomologically integer
de Rham $p$--cocycle $(\Xi^p,\Upsilon^{p-1})\in Z_{dR\Bbb Z}^p(X,Y)$
is compatible with a cohomologically integer $Y$ relative real \v Cech 
$p$--cocycle $(\Xi^{p,-1},\Upsilon^{p-1,-1})\in \tilde 
Z_{C\Bbb Z}^p(X,Y,{\cal O})$. Further, $H_{dR\Bbb Z}^p(X,Y)$ corresponds 
precisely to the relative integer \v Cech cohomology lattice 
$\tilde H_{C\Bbb Z}^p(X,Y,{\cal O})$ under the \v Cech--de Rham 
isomorphism (1.4.3).

From (1.4.3), (1.5.3), (1.5.4) and the above discussion, one deduces the 
isomorphisms
$$
\eqalignno{
H_{dR}^p(X,Y)&\cong H_s^p(X,Y),&(1.5.5)\cr
H_{dR\Bbb Z}^p(X,Y) &\cong \tilde H_{s\Bbb Z}^p(X,Y), &(1.5.6)\cr}
$$
the isomorphism (1.5.6) being the restriction of that (1.5.5).
\eject

\par\vskip.6cm
\item{\bf 2.} {\bf The relative Cheeger--Simons differential characters}
\vskip.4cm
\par

Let $p$, $X$, $Y$ satisfy the assumptions stated at the end of subsect.
1.1 and let $\cal O$ be a good covering of $X$, $Y$ (cfr. subsect. 1.4).

\par\vskip.6cm
\item{\it 1.} {\it Construction of the maps $I_1^{\cal O}$ and $I_2^{\cal O}$}
\vskip.4cm
\par

We now define two basic realvalued functions, $I_1^{\cal O}$, $I_2^{\cal O}$,
of the appropriate relative data. In view of the construction
of relative Cheeger--Simons differential characters, we analyze 
in detail the properties of $I_1^{\cal O}$, $I_2^{\cal O}$, when the 
relative data are varied by trivial amounts.
Here, we systematically use the notation (1.1.15) for conciseness.

The first function, $I_1^{\cal O}$, depends on the following relative data: 
a  relative \v Cech singular $p-1$--intertwiner $(S_{-1,p-1},T_{-1,p-2};
\{V_{k,p-1-k}|k\in I_{p-1}\},\{Z_{k,p-2-k}|k\in I_{p-2}\};
S_{p-1,-1},$ $T_{p-2,-1})\in ZI^{Cs}_{p-1}(X,Y,{\cal O})$
(cfr. eqs. (1.2.5)--(1.2.7));
a  relative \v Cech--de Rham $p$--coin\-ter\-twiner 
$(\Xi^{-1,p},\Upsilon^{-1,p-1};\{\Omega^{k,p-1-k}|k\in I^{p-1}\}, 
\{\Theta^{k,p-2-k}|k\in I^{p-2}\};\Xi^{p,-1},\Upsilon^{p-1,-1})$
$\in ZI_{CdR}^p(X,Y,{\cal O})$ (cfr. eqs. (1.2.17)--(1.2.19)). 
It is given by
$$
I_1^{\cal O}
=\sum_{k=0}^{p-1}(-1)^k\langle V_{k,p-1-k},\Omega^{k,p-1-k}\rangle 
-\sum_{k=0}^{p-2}(-1)^k\langle Z_{k,p-2-k},\Theta^{k,p-2-k}\rangle.
\eqno(2.1.1)
$$

When the relative arguments are varied by arbitrary amounts
(generically denoted by $\Delta$) the variation $\Delta I_1^{\cal O}$ 
of $I_1^{\cal O}$ is given by 
$$
\eqalignno{\vphantom{\bigg(}
\Delta I_1^{\cal O}
=& \ \sum_{k=0}^{p-1}(-1)^k\langle V_{k,p-1-k},\Delta\Omega^{k,p-1-k}\rangle 
-\sum_{k=0}^{p-2}(-1)^k\langle Z_{k,p-2-k},\Delta\Theta^{k,p-2-k}\rangle &\cr
\vphantom{\bigg(} 
&+\sum_{k=0}^{p-1}(-1)^k\langle\Delta V_{k,p-1-k},\Omega^{k,p-1-k}\rangle 
-\sum_{k=0}^{p-2}(-1)^k\langle\Delta Z_{k,p-2-k},\Theta^{k,p-2-k}\rangle &\cr
\vphantom{\bigg(} 
&+\sum_{k=0}^{p-1}(-1)^k\langle\Delta V_{k,p-1-k},
\Delta\Omega^{k,p-1-k}\rangle 
-\sum_{k=0}^{p-2}(-1)^k\langle\Delta Z_{k,p-2-k},\Delta 
\Theta^{k,p-2-k}\rangle.~~~~~~~
&(2.1.2)\cr}
$$
If $(\Delta S_{-1,p-1},\Delta T_{-1,p-2};\{\Delta V_{k,p-1-k}\},
\{\Delta Z_{k,p-2-k}\};\Delta S_{p-1,-1},\Delta T_{p-2,-1})
\in BI^{Cs}_{p-1}(X,Y,$ ${\cal O})$ is a 
trivial relative intertwiner (cfr. eqs (1.2.8)--(1.2.10)) and   
$(\Delta \Xi^{-1,p},\Delta \Upsilon^{-1,p-1};$ $\{\Delta \Omega^{k,p-1-k}\}, 
\{\Delta \Theta^{k,p-2-k}\};\Delta \Xi^{p,-1},\Delta \Upsilon^{p-1,-1})
\in BI_{CdR}^p(X,Y,{\cal O})$ is a trivial relative co\-in\-tertwiner 
(cfr. eqs. (1.2.20)--(1.2.22)), one has 
$$
\eqalignno{\vphantom{\bigg(}
\Delta I_1^{\cal O}
=& \ \langle S_{-1,p-1}, \xi^{-1,p-1}\rangle
-\langle T_{-1,p-2}, \upsilon^{-1,p-2} \rangle &\cr
\vphantom{\bigg(}
&+\langle s_{-1,p}, \Xi^{-1,p}\rangle
-\langle t_{-1,p-1}, \Upsilon^{-1,p-1} \rangle &\cr
\vphantom{\bigg(}
&+\langle s_{-1,p}, d\xi^{-1,p-1}\rangle
-\langle t_{-1,p-1},i^*\xi^{-1,p-1}- d\upsilon^{-1,p-2}\rangle\cr
\vphantom{\bigg(}
&+(-1)^{p-1}\Big[\langle S_{p-1,-1}, \xi^{p-1,-1}\rangle
-\langle T_{p-2,-1}, \upsilon^{p-2,-1} \rangle &\cr
\vphantom{\bigg(}
&+\langle s_{p,-1}, \Xi^{p,-1}\rangle
-\langle t_{p-1,-1}, \Upsilon^{p-1,-1} \rangle &\cr
\vphantom{\bigg(}
&+\langle s_{p,-1},\delta\xi^{p-1,-1}\rangle
-\langle t_{p-1,-1},i^*\xi^{p-1,-1}-\delta\upsilon^{p-2,-1}\rangle\Big].
&(2.1.3)\cr}
$$

The second function, $I_2^{\cal O}$, depends on the following relative data:
a  relative integer \v Cech $p-1$--cycle
$(S_{p-1,-1},T_{p-2,-1})\in Z^C_{p-1}(X,Y,{\cal O})$
(cfr. eq. (1.2.3));
a  relative differential $p$--cocycle 
$(\Xi^{p,-1},\Upsilon^{p-1,-1}; \Xi^{*p-1,-1},\Upsilon^{*p-2,-1};
\hat\Xi^{p,-1},\hat\Upsilon^{p-1,-1})\in ZD_C^p(X,Y,{\cal O})$
(cfr. eqs. (1.3.1)--(1.3.3)). It is given by
$$
I_2^{\cal O}
=(-1)^{p-1}\Big[\langle S_{p-1,-1},\Xi^{*p-1,-1}\rangle 
-\langle T_{p-2,-1},\Upsilon^{*p-2,-1}\rangle\Big]. 
\eqno(2.1.4)
$$
When the relative arguments are varied by arbitrary amounts
(again generically denoted by $\Delta$), the variation $\Delta I_2^{\cal O}$ 
of $I_2^{\cal O}$ is given by 
$$
\eqalignno{\vphantom{\bigg(}
\vphantom{\bigg(}
\Delta I_2^{\cal O} =& \  (-1)^{p-1}\Big[
\langle S_{p-1,-1},\Delta\Xi^{* p-1,-1}\rangle
-\langle T_{p-2,-1},\Delta \Upsilon^{* p-2,-1}\rangle &\cr
\vphantom{\bigg(}
&+\langle \Delta S_{p-1,-1},\Xi^{* p-1,-1}\rangle 
-\langle \Delta T_{p-2,-1},\Upsilon^{* p-2,-1}\rangle&\cr
\vphantom{\bigg(}
&+\langle \Delta S_{p-1,-1},\Delta \Xi^{*p-1,-1}\rangle 
-\langle \Delta T_{p-2,-1}, \Delta \Upsilon^{*p-2,-1}\rangle\Big].~~~~~~~
&(2.1.5)\cr}
$$
If $(\Delta S_{p-1,-1},\Delta T_{p-2,-1})\in B^C_{p-1}(X,Y,{\cal O})$ is a 
relative \v Cech boundary (cfr. eq. (1.2.2)) and $(\Delta \Xi^{p,-1},
\Delta \Upsilon^{p-1,-1}; \Delta \Xi^{*p-1,-1},\Delta \Upsilon^{*p-2,-1};
\Delta \hat\Xi^{p,-1},\Delta \hat\Upsilon^{p-1,-1})\in BD_{C}^p(X,Y,{\cal O})$
or $ZD_{Ct}^p(X,Y,{\cal O})$ is either a relative differential coboundary or 
a torsion relative differential cocycle (cfr. eqs. (1.3.4)--(1.3.6)), then
$$
\eqalignno{\vphantom{\bigg(}
\Delta I_2^{\cal O}
=& \ -(-1)^{p-1}\Big[\langle S_{p-1,-1},\xi^{p-1,-1}\rangle
-\langle T_{p-2,-1}, \upsilon^{p-2,-1} \rangle &\cr
\vphantom{\bigg(}
&+\langle s_{p,-1}, \Xi^{p,-1}\rangle
-\langle t_{p-1,-1}, \Upsilon^{p-1,-1} \rangle &\cr
\vphantom{\bigg(}
&+\langle s_{p,-1},\delta\xi^{p-1,-1}\rangle
-\langle t_{p-1,-1},i^*\xi^{p-1,-1}-\delta\upsilon^{p-2,-1}\rangle\Big]&\cr
\vphantom{\bigg(}
&+(-1)^{p-1}\Big[\langle S_{p-1,-1},\hat\xi^{p-1,-1}\rangle
-\langle T_{p-2,-1}, \hat\upsilon^{p-2,-1} \rangle &\cr
\vphantom{\bigg(}
&+\langle s_{p,-1}, \hat\Xi^{p,-1}\rangle
-\langle t_{p-1,-1}, \hat\Upsilon^{p-1,-1} \rangle &\cr
\vphantom{\bigg(}
&+\langle s_{p,-1},\delta\hat\xi^{p-1,-1}\rangle
-\langle t_{p-1,-1},i^*\hat\xi^{p-1,-1}-\delta\hat\upsilon^{p-2,-1}
\rangle\Big].&(2.1.6)\cr}
$$

Let $(S_{p-1},T_{p-2})\in Z^{s{\cal O}}_{p-1}(X,Y)$,
$(\Xi^p,\Upsilon^{p-1})\in Z_{dR\Bbb Z}^p(X,Y)$ be respectively an
${\cal O}$--small relative singular $p-1$--cycle 
(cfr. eq. (1.2.1) and subsect. 1.2) and a
cohomologically integer relative de Rham $p$--cocycle,
(cfr. eq. (1.2.13) and subsect. 1.5). From the discussion of subsects. 1.4
and 1.5, we can carry out the following construction. 

$(S_{p-1},T_{p-2})$ can be extended to some  relative 
\v Cech singular $p-1$--intertwiner $(S_{-1,p-1},$ $T_{-1,p-2};
\{V_{k,p-1-k}\},\{Z_{k,p-2-k}\};S_{p-1,-1},T_{p-2,-1})\in 
ZI^{Cs}_{p-1}(X,Y,{\cal O})$
(cfr. eqs. (1.2.5)--(1.2.7) and subsect. 1.4).
By standard \v Cech singular techniques, one easily sees that the 
intertwiner $(S_{-1,p-1},T_{-1,p-2};\{V_{k,p-1-k}\},\{Z_{k,p-2-k}\};
S_{p-1,-1},T_{p-2,-1})$ is defined up to a trivial relative 
\v Cech singular intertwiner of the form (1.2.8)--(1.2.10) 
with $s_{-1,p}=0$, $t_{-1,p-1}=0$. 

$(\Xi^p,\Upsilon^{p-1})$ can be extended to some relative 
\v Cech--de Rham $p$--cointertwiner $(\Xi^{-1,p},$ $\Upsilon^{-1,p-1};
\{\Omega^{k,p-1-k}\},$ $\{\Theta^{k,p-2-k}\};\Xi^{p,-1},\Upsilon^{p-1,-1})
\in ZI_{CdR}^p(X,Y,{\cal O})$
(cfr. eqs. (1.2.17)--(1.2.19) and subsect. 1.4).
By the standard \v Cech--de Rham techniques, it is easy to see that the 
cointertwiner $(\Xi^{-1,p},\Upsilon^{-1,p-1};\{\Omega^{k,p-1-k}\},
\{\Theta^{k,p-2-k}\};$ $\Xi^{p,-1},\Upsilon^{p-1,-1})$
is defined up to a trivial relative \v Cech--de Rham cointertwiner of
the form (1.2.20)--(1.2.22) with $\xi^{-1,p-1}=0$, $\upsilon^{-1,p-2}=0$.

As $(\Xi^p,\Upsilon^{p-1})$ is 
cohomologically integer, the relative real \v Cech $p$--cocycle 
$(\Xi^{p,-1},$ $\Upsilon^{p-1,-1})\in \tilde Z_{C\Bbb Z}^p(X,Y,{\cal O})$ is  
cohomologically integer as well (cfr. subsects. 1.5).
Then, $(\Xi^{p,-1},\Upsilon^{p-1,-1})$ fits into some relative 
differential $p$--cocycle $(\Xi^{p,-1},\Upsilon^{p-1,-1};\Xi^{*p-1,-1},$
$\Upsilon^{*p-2,-1};\hat\Xi^{p,-1},\hat\Upsilon^{p-1,-1})\in 
ZD_C^p(X,Y,{\cal O})$ (cfr. eqs. (1.3.1)--(1.3.3) and subsect. 1.5). 
As $(\Xi^{p,-1},\Upsilon^{p-1,-1})$ is defined only up to a relative real 
\v Cech coboundary of the form (1.2.16),
the relative differential cocycle $(\Xi^{p,-1},\Upsilon^{p-1,-1};
\Xi^{*p-1,-1},$ $\Upsilon^{*p-2,-1};\hat\Xi^{p,-1},\hat\Upsilon^{p-1,-1})$ is 
determined up to a torsion relative differential cocycle of the form 
(1.3.4)--(1.3.6). Indeed, when the
relative \v Cech torsion $Tor_C^p(X,Y,{\cal O})$ is non vanishing,
the cohomology class of the relative integer \v Cech cocycle 
$(\hat\Xi^{p,-1},\hat\Upsilon^{p-1,-1})$ in $H_{C\Bbb Z}^p(X,Y,{\cal O})$
is not uniquely fixed by that of the relative real \v Cech cocycle 
$(\Xi^{p,-1},\Upsilon^{p-1,-1})$ in $H_C^p(X,Y,{\cal O})$ 
and, thus, the ambiguity of the relative differential cocycle is not 
in general a relative differential coboundary.

Using the relative homological and cohomological data obtained in this 
way from $(S_{p-1},T_{p-2})$ and $(\Xi^p,\Upsilon^{p-1})$, we set 
$$
I^{\cal O}=I_1^{\cal O}+I_2^{\cal O}.
\eqno(2.1.7)
$$
Since, however. those data are not determined by $(S_{p-1},T_{p-2})$ 
and $(\Xi^p,\Upsilon^{p-1})$ in unique fashion, as explained above, 
$I^{\cal O}$ is affected by an ambiguity $\Delta I^{\cal O}$ 
which we are now going to compute.

From the above discussion, by inspection of (2.1.3), (2.1.6), 
it appears that the relevant ambiguities of the definition of 
the relative intertwiner $(S_{-1,p-1},T_{-1,p-2};\{V_{k,p-1-k}\},$ 
$\{Z_{k,p-2-k}\};\,\,S_{p-1,-1},T_{p-2,-1})$, ~the relative cointertwiner
$(\Xi^{-1,p},\Upsilon^{-1,p-1};\,\,\{\Omega^{k,p-1-k}\},$ 
$\{\Theta^{k,p-2-k}\};\Xi^{p,-1},$ $\Upsilon^{p-1,-1})$ and the relative 
differential cocycle $(\Xi^{p,-1},\Upsilon^{p-1,-1};\Xi^{*p-1,-1},$ $
\Upsilon^{*p-2,-1};\hat\Xi^{p,-1},\hat\Upsilon^{p-1,-1})$ 
are those parameterized by the relative 
integer \v Cech chain $(s_{p,-1},$ $t_{p-1,-1})$, the relative real \v  Cech
cochain $(\xi^{p-1,-1},\upsilon^{p-2,-1})$
and the relative real \v Cech cochain $(\hat\xi^{p-1,-1},
\hat\upsilon^{p-2,-1})$ subject to the condition that the relative coboundary
$(\delta\hat \xi^{p-1,-1},$ $i^*\hat\xi^{p-1,-1}
-\delta\hat \upsilon^{p-2,-1})$ is integer 
(cfr. (1.2.8)--(1.2.10), (1.2.20)--(1.2.22)
and (1.3.4)--(1.3.6) and the previous discussion). 
The crucial point to be noted here is that the relative \v Cech cochain
$(\xi^{p-1,-1},\upsilon^{p-2,-1})$ parameterizing the ambiguity of the
relative \v Cech cocycle $(\Xi^{p,-1},$ $\Upsilon^{p-1,-1})$ 
is the same for both the relative cointertwiner and the relative 
differential cocycle.
Taking this into account, from (2.1.3), (2.1.6) with $s_{-1,p}=0$, 
$t_{-1,p-1}=0$, $\xi^{-1,p-1}=0$, $\upsilon^{-1,p-2}=0$, we find that 
$$
\eqalignno{\vphantom{\bigg(}
\Delta I^{\cal O}
=& \ (-1)^{p-1}\Big[\langle S_{p-1,-1},\hat\xi^{p-1,-1}\rangle
-\langle T_{p-2,-1}, \hat\upsilon^{p-2,-1} \rangle &\cr
\vphantom{\bigg(}
&+\langle s_{p,-1}, \hat\Xi^{p,-1}\rangle
-\langle t_{p-1,-1}, \hat\Upsilon^{p-1,-1} \rangle &\cr
\vphantom{\bigg(}
&+\langle s_{p,-1},\delta\hat\xi^{p-1,-1}\rangle
-\langle t_{p-1,-1},i^*\hat\xi^{p-1,-1}-\delta\hat\upsilon^{p-2,-1}
\rangle\Big]. &(2.1.8)\cr}
$$
$\Delta I^{\cal O}$ is clearly non zero in general. Thus, $I^{\cal O}$
is not unambiguously defined. However, the above expression  
suggests that, under certain conditions, $\Delta I^{\cal O}$ might be 
integer valued. In such a case, $I^{\cal O}$ would be 
 unambiguously defined modulo integers. 

If $(\hat\xi^{p-1,-1},\hat\upsilon^{p-2,-1})$ were
an relative integer \v Cech cochain, $\Delta I^{\cal O}$
would be integer. However, because of torsion, the relative \v Cech cochain
$(\hat\xi^{p-1,-1},\hat\upsilon^{p-2,-1})$ is real, being only 
subject to the condition that the relative coboundary
$(\delta\hat\xi^{p-1,-1},i^*\hat \xi^{p-1,-1}-\delta\hat\upsilon^{p-2,-1})$
is integer. So the first two terms of the right hand side of (2.1.8) and, 
thus, $\Delta I^{\cal O}$ are generally not integer valued.

If we insist that $\Delta I^{\cal O}$ be integer, we have to restrict 
the ambiguity inherent in the choice of the differential cocycle 
$(\Xi^{p,-1},\Upsilon^{p-1,-1};$ $\Xi^{*p-1,-1},\Upsilon^{*p-2,-1};
\hat\Xi^{p,-1},\hat\Upsilon^{p-1,-1})$ which is responsible 
for the non integrality of $(\hat\xi^{p-1,-1},\hat\upsilon^{p-2,-1})$. 
This can be achieved in two steps.

We first restrict the choice of the relative integer \v Cech cocycle 
$(\hat\Xi^{p,-1},\hat\Upsilon^{p-1,-1})$ by fixing its cohomology class in 
$H_{C\Bbb Z}^p(X,Y,{\cal O})$ among those classes of 
$H_{C\Bbb Z}^p(X,Y,{\cal O})$ whose image in $H_C^p(X,Y,{\cal O})$
is represented by the relative real \v Cech cocycle 
$(\Xi^{p,-1},\Upsilon^{p-1,-1})$ (cfr. subsect. 1.5). 
By inspecting (1.3.4)--(1.3.6) for given $(\xi^{p-1,-1},\upsilon^{p-2,-1})$, 
it is easy to see that the relative real \v Cech cochain 
$(\hat\xi^{p-1,-1},\hat\upsilon^{p-2,-1})$ is restricted in this way to 
be integer up to a relative real \v Cech cocycle. 

Such a cocycle parameterizes the set of the possible choices of the relative 
real \v Cech cochain $(\Xi^{*p-1,-1},\Upsilon^{*p-2,-1})$ for given 
$(\Xi^{p,-1},\Upsilon^{p-1,-1})$, $(\hat\Xi^{p,-1},\hat\Upsilon^{p-1,-1})$. 
It is natural to identify two choices of $(\Xi^{*p-1,-1},\Upsilon^{*p-2,-1})$
if they yield the same value of $I^{\cal O}$ modulo integers
for all $(S_{p-1},T_{p-2})\in Z^{s{\cal O}}_{p-1}(X,Y)$.
From (2.1.8), on account of (1.2.3), it is apparent that two choices of 
$(\Xi^{*p-1,-1},\Upsilon^{*p-2,-1})$ are equivalent when their difference is 
a cohomologically integer relative cocycle (cfr. subsect. 1.5). 
Thus, the set of equivalence classes of choices of 
$(\Xi^{*p-1,-1},\Upsilon^{*p-2,-1})$ is parameterized by the quotient
$Z_C^{p-1}(X,Y,{\cal O})/$ $\tilde Z_{C\Bbb Z}^{p-1}(X,Y,{\cal O})$
or, what is the same, by the relative \v Cech cohomology torus
$H_C^{p-1}(X,Y,{\cal O})/$ $\tilde H_{C\Bbb Z}^{p-1}(X,Y,{\cal O})$.
From its definition, it is clear that the parametrization is non canonical, 
depending on an arbitrary choice of a reference relative 
Cech cochain $(\Xi^{*p-1,-1},\Upsilon^{*p-2,-1})$ corresponding to the  
origin of the torus.

We next restrict the choice of the relative real \v Cech cochain
$(\Xi^{*p-1,-1},\Upsilon^{*p-2,-1})$ by fixing its image in 
$H_C^{p-1}(X,Y,{\cal O})/\tilde H_{C\Bbb Z}^{p-1}(X,Y,{\cal O})$.

The relative real \v Cech cochain $(\hat\xi^{p-1,-1},
\hat\upsilon^{p-2,-1})$ is finally restricted to be integer
up to a cohomologically integer relative cocycle. 
From (2.1.8), using (1.2.3) again, it follows then that, 
once the above choices are made, the ambiguity
$\Delta I^{\cal O}$ is integer valued.

Recalling the isomorphisms of singular, de Rham and \v Cech cohomology 
discussed in sect. 1.5, we conclude that we can unambiguously define a 
family of maps 
$\Phi^{\cal O}:Z^{s{\cal O}}_{p-1}(X,Y)\rightarrow \Bbb R/\Bbb Z$ by 
$$
\Phi^{\cal O}(S,T;\Xi,\Upsilon)=I^{\cal O}~~~\hbox{mod $\Bbb Z$},
\eqno(2.1.9)
$$
for $(S_{p-1},T_{p-2})\in Z^{s{\cal O}}_{p-1}(X,Y)$, depending on
a choice of a relative integer singular cohomology class in 
$H_{s\Bbb Z}^p(X,Y)$, a representative $(\Xi^p,
\Upsilon^{p-1})\in Z_{dR\Bbb Z}^p(X,Y)$ of the image 
of such class in $H_{dR\Bbb Z}^p(X,Y)$ shown explicitly
and a point in the relative de Rham cohomology torus  
$H_{dR}^{p-1}(X,Y)/H_{dR\Bbb Z}^{p-1}(X,Y)$.
From (2.1.1), (2.1.4), (2.1.7), (2.1.9), it appears that $\Phi^{\cal O}$
is $\Bbb Z$ linear in the first argument. 

When $(S_{p-1},T_{p-2})$, $(\Xi^p,\Upsilon^{p-1})$ are shifted by amounts 
given by the right hand sides of (1.2.2), with $s_{-1,p}$, $t_{-1,p-1}$
${\cal O}$--small, and (1.2.14), respectively, one has
$$
\eqalignno{\vphantom{\bigg(}
\Delta \Phi^{\cal O}(S,T;\Xi,\Upsilon)
=& \ \langle S_{-1,p-1}, \xi^{-1,p-1}\rangle
-\langle T_{-1,p-2}, \upsilon^{-1,p-2} \rangle &\cr
\vphantom{\bigg(}
&+\langle s_{-1,p}, \Xi^{-1,p}\rangle
-\langle t_{-1,p-1}, \Upsilon^{-1,p-1} \rangle &\cr
\vphantom{\bigg(}
&+\langle s_{-1,p}, d\xi^{-1,p-1}\rangle
-\langle t_{-1,p-1},i^*\xi^{-1,p-1}- d\upsilon^{-1,p-2}\rangle,
\quad \hbox{mod $\Bbb Z$},&\cr
&&(2.1.10)\cr}
$$
as follows readily from (2.1.3).

For reasons explained above, the $H_{dR}^{p-1}(X,Y)/H_{dR\Bbb Z}^{p-1}(X,Y)$
parametrization of the maps $\Phi^{\cal O}$ is not canonical. 
The changes of the parametrization are in one--to--one correspondence with 
the shifts in the  relative de Rham cohomology torus.
By the isomorphisms of de Rham and \v Cech cohomologies of subsects. 1.4, 1.5,
any such shift is represented equivalently by either a relative de Rham 
cocycle $(\Pi^{-1,p-1},\Sigma^{-1,p-2})$ defined up to cohomologically integer
relative de Rham cocycles or a relative real \v Cech cocycle
$(\Pi^{p-1,-1},\Sigma^{p-2,-1})$ defined modulo a cohomologically integer 
relative real \v Cech cocycle. The variation of 
$\Phi^{\cal O}(S,T;\Xi,\Upsilon)$ caused by the shift is given by 
$$
\eqalignno{\vphantom{\bigg(}
\Delta \Phi^{\cal O}(S,T;\Xi,\Upsilon)
=& \ (-1)^{p-1}\Big[\langle S_{p-1,-1},\Pi^{p-1,-1}\rangle
-\langle T_{p-2,-1},\Sigma^{p-2,-1} \rangle\Big] 
\quad \hbox{mod $\Bbb Z$},~~~~~~~~ &(2.1.11)\cr}
$$
as follows from the first two terms 
of (2.1.8) with $(\hat\xi^{p-1,-1},\hat\upsilon^{p-2,-1})$
replaced by $(\Pi^{p-1,-1},$ $\Sigma^{p-2,-1})$.
It is straightforward to show that
$$
\eqalignno{\vphantom{\bigg(}
\Delta \Phi^{\cal O}(S,T;\Xi,\Upsilon)
=& \ -\Big[\langle S_{-1,p-1},\Pi^{-1,p-1}\rangle
-\langle T_{-1,p-2},\Sigma^{-1,p-2} \rangle\Big] 
\quad \hbox{mod $\Bbb Z$}.~ &(2.1.12)\cr}
$$
Indeed, consider 
the function $I_1^{\cal O}$, eq. (2.1.1). 
If we vary the \v Cech--de Rham $p$--cointertwiner
$(\Xi^{-1,p},\Upsilon^{-1,p-1};\!\{\Omega^{k,p-1-k}\},$ 
$\{\Theta^{k,p-2-k}\};\Xi^{p,-1},\Upsilon^{p-1,-1})$
by a vanishing amount 
$(\Delta \Xi^{-1,p},$ $\Delta\Upsilon^{-1,p-1};~\{\Delta \Omega^{k,p-1-k}\},$ 
$\{\Delta \Theta^{k,p-2-k}\};\Delta \Xi^{p,-1},\Delta \Upsilon^{p-1,-1})$,
then $\Delta I_1^{\cal O}=0$ trivially. On the other hand,
the totally vanishing trivial  
\v Cech--de Rham $p$--cointertwiner can be written in the form 
(1.2.20)--(1.2.22) with $(\xi^{-1,p-1},\upsilon^{-1,p-2})=
(\Pi^{-1,p-1},\Sigma^{-1,p-2})$,
$(\xi^{p-1,-1},$ $\upsilon^{p-2,-1})=(\Pi^{p-1,-1},\Sigma^{p-2,-1})$
for suitable \v Cech--de Rham cochains $\omega^{k,p-2-k}$, 
$\theta^{k,p-3-k}$. So, by (2.1.3), $\Delta I_1^{\cal O}=0$
is given by the difference of the two above expressions.

\par\vskip.6cm
\item{\it 2.} {\it Dependence of $\Phi^{\cal O}$ on covering choices}
\vskip.4cm
\par

It is important to compare the result of the above construction for two 
choices of the underlying good open covering of $X$, $Y$, 
${\cal O}(1)$, ${\cal O}(2)$. 
The basic ideas consists in constructing suitable sequences of \v Cech 
(co)chains of the covering ${\cal O}(1)\cup{\cal O}(2)$ (disjoint union)
interpolating between the given \v Cech (co)chains of the individual coverings 
${\cal O}(1)$, ${\cal O}(2)$.

To this end, we explicitly indicate the \v Cech degree with respect 
the two coverings. 
So, $U_{k,l,n}$, say, is a \v Cech singular chain of \v Cech degree 
$k$, $l$ with respect to ${\cal O}(1)$, ${\cal O}(2)$, respectively, and 
dimension $n$. 
Similarly, $\Lambda^{k,l,n}$, say, is a \v Cech--de Rham cochain of 
\v Cech degree $k$, $l$ with respect to ${\cal O}(1)$, ${\cal O}(2)$, 
respectively, and form degree $n$.  
Accordingly, we have two operators $\beta_1$, $\beta_2$ 
defined as in (1.1.7) and obeying (1.1.8).
Similarly, we have two operators $\delta_1$, $\delta_2$ 
defined as in (1.1.12) and obeying (1.1.13).
Further, the pairs $\beta_1$, $\delta_1$ and $\beta_2$, $\delta_2$ 
independently satisfy the duality relations (1.1.17).
Conversely, we have just one operator $b$ and one operator $d$, 
which are the same as before and satisfy the duality relations (1.1.16).

A \v Cech singular chain $U_{k,l,n}$ is a \v Cech singular chain of
${\cal O}(1)\cup{\cal O}(2)$ of \v Cech degree $k+l+1$.
A \v Cech singular chain of the form $U_{k,-1,n}$ ($U_{-1,l,n}$) 
can be identified with a \v Cech singular chain 
$U^{(1)}_{k,-1,n}$ ($U^{(2)}_{-1,l,n}$) of ${\cal O}(1)$ 
(${\cal O}(2)$) of \v Cech degree $k$ ($l$) having the property of being
${\cal O}(2)$--small (${\cal O}(1)$--small).
The operator $\beta$ appropriate for the \v Cech singular chains 
of ${\cal O}(1)\cup{\cal O}(2)$ is the sum $\beta_1+(-1)^{\deg(1)+1}\beta_2$ 
while that for the \v Cech singular chains of ${\cal O}(1)$ (${\cal O}(2)$) 
is $\beta_1$ ($\beta_2$). 
Similarly, a \v Cech--de Rham cochain $\Lambda^{k,l,n}$ is 
a \v Cech--de Rham cochain of ${\cal O}(1)\cup{\cal O}(2)$ 
of \v Cech degree $k+l+1$. A \v Cech--de Rham cochain of the form 
$\Lambda^{k,-1,n}$ ($\Lambda^{-1,l,n}$) can be identified with a 
\v Cech--de Rham cochain $\Lambda_{(1)}^{k,-1,n}$ ($\Lambda_{(2)}^{-1,l,n}$) 
of ${\cal O}(1)$ (${\cal O}(2)$) of \v Cech degree $k$ ($l$).
The operator $\delta$ appropriate for the \v Cech--de Rham cochains 
of ${\cal O}(1)\cup{\cal O}(2)$ is the sum $\delta_1+(-1)^{\deg(1)+1}\delta_2$
while that for the \v Cech--de Rham cochains of ${\cal O}(1)$ 
(${\cal O}(2)$) is $\delta_1$ ($\delta_2$).

When stating that a sequence of (co)chains forms a relative (co)chain,
(co)cy\-cle, (co)boundary,  (trivial) (co)intertwiner etc.
it is necessary to specify the underlying covering and the relevant
$\beta$ or $\delta$ operators. If no label is attached to the (co)chains, 
it is understood that the covering is ${\cal O}(1)\cup{\cal O}(2)$ 
and the $\beta$ or $\delta$ operators are 
$\beta_1+(-1)^{\deg(1)+1}\beta_2$, $\delta_1+(-1)^{\deg(1)+1}\delta_2$.
If the label 1 (2) is attached to the (co)chains, 
it is understood that the covering is ${\cal O}(1)$ (${\cal O}(2)$) 
and the $\beta$ or $\delta$ operators are 
$\beta_1$ ($\beta_2$), $\delta_1$ ($\delta_2$). 

Set $J_r=\{(k,l)|k,l\in \Bbb Z, 0\leq k,l,~0\leq k+l\leq r\}$,
$K_r=\{k|k\in \Bbb Z,0\leq k\leq r\}$,  $r\in \Bbb N$.
We say that a sequence of chains $(S_{-1,-1,p-1},T_{-1,-1,p-2};
\{V_{k,l,p-1-k-l}|(k,l)\in J_{p-1}\},$ $\{Z_{k,l,p-2-k-l}|(k,l)\in J_{p-2}\};
\{S_{k,p-1-k,-1}|k\in K_{p-1}\},\{T_{k,p-2-k,-1}|k\in K_{p-2}\})$ 
interpolates two relative \v Cech singular $p-1$--intertwiners 
$(S^{(1)}_{-1,-1,p-1},T^{(1)}_{-1,-1,p-2};
\{V^{(1)}_{k,-1,p-1-k}|$ $k\in I_{p-1}\}, 
\{Z^{(1)}_{k,-1,p-2-k}|k\in I_{p-2}\};~
S^{(1)}_{p-1,-1,-1},~T^{(1)}_{p-2,-1,-1})$,~~ 
$(S^{(2)}_{-1,-1,p-1},~T^{(2)}_{-1,-1,p-2};$
$\{V^{(2)}_{-1,k,p-1-k}|k\in I_{p-1}\}, 
\{Z^{(2)}_{-1,k,p-2-k}|k\in I_{p-2}\};
S^{(2)}_{-1,p-1,-1},T^{(2)}_{-1,p-2,-1})$
(cfr. eqs. (1.2.5)--(1.2.7)),
if $S_{-1,-1,p-1}$, $V_{k,l,p-1-k-l}$, $S_{k,p-1-k,-1}$
are \v Cech singular chains of $X$,
$T_{-1,-1,p-2}$, $Z_{k,l,p-2-k-l}$, $T_{k,p-2-k,-1}$
are \v Cech singular chains in $Y$ and 
$$
\eqalignno{\vphantom{\bigg(}&
S_{-1,-1,p-1}=\beta_1\beta_2 V_{0,0,p-1},&(2.2.1a)\cr
\vphantom{\bigg(}&
T_{-1,-1,p-2}=-\beta_1\beta_2 Z_{0,0,p-2},&(2.2.1b)\cr
\vphantom{\bigg(}&
b V_{k,l,p-1-k-l} = \beta_1 V_{k+1,l,p-2-k-l}
+(-1)^{k+1}\beta_2 V_{k,l+1,p-2-k-l}&\cr
\vphantom{\bigg(}&
+(-1)^{k+l+1} i_*Z_{k,l,p-2-k-l},
\quad 0\leq k,l,\quad 0\leq k+l\leq p-2, &(2.2.2a)\cr
\vphantom{\bigg(}&
b Z_{k,l,p-2-k-l} = \beta_1 Z_{k+1,l,p-3-k-l}
+(-1)^{k+1}\beta_2 Z_{k,l+1,p-3-k-l},&\cr
\vphantom{\bigg(}&
\quad 0\leq k,l,\quad 0\leq k+l\leq p-3, &(2.2.2b)\cr
\vphantom{\bigg(}&
S_{k,p-1-k,-1}=bV_{k,p-1-k,0},
&(2.2.3a)\cr
\vphantom{\bigg(}&
T_{k,p-2-k,-1}=-(-1)^{p-1}bZ_{k,p-2-k,0},
&(2.2.3b)\cr}
$$
with
$$
\eqalignno{\vphantom{\bigg(}&
S^{(1)}_{-1,-1,p-1}=S_{-1,-1,p-1},\quad
S^{(2)}_{-1,-1,p-1}=S_{-1,-1,p-1},
&(2.2.4a)\cr
\vphantom{\bigg(}&
T^{(1)}_{-1,-1,p-2}=T_{-1,-1,p-2},\quad
T^{(2)}_{-1,-1,p-2}=T_{-1,-1,p-2},
&(2.2.4b)\cr
\vphantom{\bigg(}&
V^{(1)}_{k,-1,p-1-k} = \beta_2 V_{k,0,p-1-k},\quad
 V^{(2)}_{-1,k,p-1-k} = (-1)^k\beta_1 V_{0,k,p-1-k},&\cr
\vphantom{\bigg(}& 
\quad 0\leq k  \leq p-1, &(2.2.5a)\cr
\vphantom{\bigg(}&
Z^{(1)}_{k,-1,p-2-k} = -\beta_2 Z_{k,0,p-2-k},\quad 
 Z^{(2)}_{-1,k,p-2-k} = -(-1)^k\beta_1 Z_{0,k,p-2-k}, &\cr
\vphantom{\bigg(}&
\quad 0\leq k  \leq p-2, &(2.2.5b)\cr
\vphantom{\bigg(}&
S^{(1)}_{p-1,-1,-1} = \beta_2 S_{p-1,0,-1},\quad
S^{(2)}_{-1,p-1,-1} = (-1)^{p-1}\beta_1 S_{0,p-1,-1},&(2.2.6a)\cr
\vphantom{\bigg(}&
T^{(1)}_{p-2,-1,-1} = \beta_2 T_{p-2,0,-1},\quad 
T^{(2)}_{-1,p-2,-1} = (-1)^{p-2}\beta_1 T_{0,p-2,-1}. &(2.2.6b)\cr}
$$
It is straightforward to check that the above relations are 
compatible with the relations (1.2.5)--(1.2.7) obeyed by the relative 
\v Cech singular intertwiners.

For $r\in \Bbb N$, define 
$J^r=\{(k,l)|k,l\in \Bbb Z, -1\leq k,l,~-1\leq k+l\leq r\}$,
$K^r=\{k|k\in \Bbb Z,-1\leq k\leq r\}$.
We say that a sequence of cochains $(\Xi^{-1,-1,p},\Upsilon^{-1,-1,p-1};
\{\Omega^{k,l,p-2-k-l}|$ $(k,l)\in J^{p-2}\},
\{\Theta^{k,l,p-3-k-l}|(k,l)\in J^{p-3}\};
\{\Xi^{k,p-1-k,-1}|k\in K^p\},~\{\Upsilon^{k,p-2-k,-1}|k\in K^{p-1}\})$ 
interpolates two relative \v Cech--de Rham $p$--cointertwiners 
$(\Xi_{(1)}^{-1,-1,p},\Upsilon_{(1)}^{-1,-1,p-1};$
$\{\Omega_{(1)}^{k,-1,p-1-k}|k\in I^{p-1}\},~ 
\{\Theta_{(1)}^{k,-1,p-2-k}|k\in I^{p-2}\};~
\Xi_{(1)}^{p,-1,-1},~\Upsilon_{(1)}^{p-1,-1,-1})$,~~
$(\Xi_{(2)}^{-1,-1,p},$ $\Upsilon_{(2)}^{-1,-1,p-1};
\{\Omega_{(2)}^{-1,k,p-1-k}|k\in I^{p-1}\}, 
\{\Theta_{(2)}^{-1,k,p-2-k}|k\in I^{p-2}\};
\Xi_{(2)}^{-1,p,-1},\Upsilon_{(2)}^{-1,p-1,-1})$
(cfr. eqs. (1.2.17)--(1.2.19)), if 
$\Xi^{-1,-1,p-1}$, $\Omega^{k,l,p-1-k-l}$, $\Xi^{k,p-1-k,-1}$
are \v Cech--de Rham co\-chains of $X$,
$\Upsilon^{-1,-1,p-2}$, $\Theta^{k,l,p-2-k-l}$, $\Upsilon^{k,p-2-k,-1}$
are \v Cech--de Rham cochains in $Y$ and 
$$
\eqalignno{\vphantom{\bigg(}&
\delta_1\Xi^{-1,-1,p}=d\Omega^{0,-1,p-1},
&\cr
\vphantom{\bigg(}&
\delta_2\Xi^{-1,-1,p}=d\Omega^{-1,0,p-1}
&(2.2.7a)\cr
\vphantom{\bigg(}&
\delta_1\Upsilon^{-1,-1,p-1}=-d\Theta^{0,-1,p-2}+i^*\Omega^{0,-1,p-1},
&\cr
\vphantom{\bigg(}&
\delta_2\Upsilon^{-1,-1,p-1}=-d\Theta^{-1,0,p-2}+i^*\Omega^{-1,0,p-1},
&(2.2.7b)\cr
\vphantom{\bigg(}&
d \Omega^{k,l,p-2-k-l} = \delta_1 \Omega^{k-1,l,p-1-k-l}
+(-1)^{k+1}\delta_2 \Omega^{k,l-1,p-1-k-l},
&\cr
\vphantom{\bigg(}&
\quad -1\leq k,l,\quad 0\leq k+l\leq p-2, 
&(2.2.8a)\cr
\vphantom{\bigg(}&
d \Theta^{k,l,p-3-k-l} =\delta_1 \Theta^{k-1,l,p-2-k-l} 
+(-1)^{k+1}\delta_2 \Theta^{k,l-1,p-2-k-l}
&\cr
\vphantom{\bigg(}
&+(-1)^{k+l+1} i^*\Omega^{k,l,p-2-k-l},
\quad -1\leq k,l,\quad 0\leq k+l\leq p-3, &(2.2.8b)\cr
\vphantom{\bigg(}&
d\Xi^{k,p-1-k,-1}=\delta_1 \Omega^{k-1,p-1-k,0}
+(-1)^{k+1}\delta_2 \Omega^{k,p-2-k,0},
&(2.2.9a)\cr
\vphantom{\bigg(}&
d\Upsilon^{k,p-2-k,-1}=(-1)^{p-1}\Big(\delta_1\Theta^{k-1,p-2-k,0} 
+(-1)^{k+1}\delta_2 \Theta^{k,p-3-k,0}
&\cr
\vphantom{\bigg(}
&+(-1)^{p-1} i^*\Omega^{k,p-2-k,0}\Big),
&(2.2.9b)\cr}
$$
with
$$
\eqalignno{\vphantom{\bigg(}&
\Xi_{(1)}^{-1,-1,p}=\Xi^{-1,-1,p},\quad 
\Xi_{(2)}^{-1,-1,p}=\Xi^{-1,-1,p},
&(2.2.10a)\cr
\vphantom{\bigg(}&
\Upsilon_{(1)}^{-1,-1,p-1}=\Upsilon^{-1,-1,p-1},\quad
\Upsilon_{(2)}^{-1,-1,p-1}=\Upsilon^{-1,-1,p-1},
&(2.2.10b)\cr
\vphantom{\bigg(}&
\Omega_{(1)}^{k,-1,p-1-k} = \Omega^{k,-1,p-1-k},\quad
 \Omega_{(2)}^{-1,k,p-1-k} = \Omega^{-1,k,p-1-k}, 
\quad 0\leq k\leq p-1 &(2.2.11a)\cr
\vphantom{\bigg(}&
\Theta_{(1)}^{k,-1,p-2-k} = \Theta^{k,-1,p-2-k},\quad 
 \Theta_{(2)}^{-1,k,p-2-k} = \Theta^{-1,k,p-2-k}, 
\quad 0\leq k  \leq p-2, ~~~~~~~&(2.2.11b)\cr
\vphantom{\bigg(}&
\Xi_{(1)}^{p,-1,-1} = \Xi^{p,-1,-1},\quad
\Xi_{(2)}^{-1,p,-1} = \Xi^{-1,p,-1},&(2.2.12a)\cr
\vphantom{\bigg(}&
\Upsilon_{(1)}^{p-1,-1,-1} = \Upsilon^{p-1,-1,-1},\quad 
\Upsilon_{(2)}^{-1,p-1,-1} = \Upsilon^{-1,p-1,-1} &(2.2.12b)\cr}
$$
\footnote{}{}
\footnote{${}^1$}{In these formulae, it is assumed conventionally that any 
\v Cech--de Rham cochain $\lambda^{k,l,m}=0$ whenever $k$, $l$ do not
satisfy the restrictions listed at the beginning of this paragraph.}.
It is straightforward to check that the above relations are compatible with 
the relations (1.2.17)--(1.2.19) obeyed by the relative \v Cech--de Rham 
cointertwiners.

Using the interpolating sequences of (co)chains introduced above, 
one defines for $0\leq k\leq p-2$
$$
\eqalignno{\vphantom{\bigg(}
S_{1k}& = \sum_{l=0}^k\langle V_{l,k-l,p-1-k},
\delta_1 \Omega^{l-1,k-l,p-1-k}
+(-1)^{l+1} \delta_2 \Omega^{l,k-l-1,p-1-k}\rangle &\cr
\vphantom{\bigg(}
&+\sum_{l=0}^k\langle Z_{l,k-l,p-2-k},
\delta_1 \Theta^{l-1,k-l,p-2-k}
+(-1)^{l+1} \delta_2 \Theta^{l,k-l-1,p-2-k}\rangle.~~ &(2.2.13)\cr}
$$
Using the relations (1.1.16), (1.1.17), (2.2.2), (2.2.8), (2.2.5), (2.2.11), 
one finds that, for $1\leq k\leq p-2$, 
$$
\eqalignno{\vphantom{\bigg(}
&(-1)^k\langle V^{(1)}_{k,-1,p-1-k},\Omega_{(1)}^{k,-1,p-1-k}\rangle
-(-1)^k\langle V^{(2)}_{-1,k,p-1-k},\Omega_{(2)}^{-1,k,p-1-k}\rangle &\cr
\vphantom{\bigg(}
&-(-1)^k\langle Z^{(1)}_{k,-1,p-2-k},\Theta_{(1)}^{k,-1,p-2-k}\rangle
+(-1)^k\langle Z^{(2)}_{-1,k,p-2-k},\Theta_{(2)}^{-1,k,p-2-k}\rangle &\cr
\vphantom{\bigg(}
&+S_{1k}-S_{1k-1}=0.~~~~~&(2.2.14)\cr}
$$
Hence, 
$$
\eqalignno{\vphantom{\bigg(}
&\sum_{k=1}^{p-2}(-1)^k\langle V^{(1)}_{k,-1,p-1-k},
\Omega_{(1)}^{k,-1,p-1-k}\rangle
-\sum_{k=1}^{p-2}(-1)^k\langle Z^{(1)}_{k,-1,p-2-k},
\Theta_{(1)}^{k,-1,p-2-k}\rangle
&\cr
\vphantom{\bigg(}
-&\sum_{k=1}^{p-2}(-1)^k\langle V^{(2)}_{-1,k,p-1-k},
\Omega_{(2)}^{-1,k,p-1-k}\rangle
+\sum_{k=1}^{p-2}(-1)^k\langle Z^{(2)}_{-1,k,p-2-k},
\Theta_{(2)}^{-1,k,p-2-k}\rangle
&\cr
\vphantom{\bigg(}
+&S_{1p-2}-S_{10}=0.&(2.2.15)\cr}
$$
From the definition (2.2.13), using (1.1.17), (2.2.5), (2.2.11) with $k=0$,
one easily sees that
$$
\eqalignno{\vphantom{\bigg(}
S_{10}=& \ 
-\langle V^{(1)}_{0,-1,p-1},\Omega_{(1)}^{0,-1,p-1}\rangle
+\langle Z^{(1)}_{0,-1,p-2},\Theta_{(1)}^{0,-1,p-2}\rangle&\cr
\vphantom{\bigg(}
&+\langle V^{(2)}_{-1,0,p-1},\Omega_{(2)}^{-1,0,p-1}\rangle
-\langle Z^{(2)}_{-1,0,p-2},\Theta_{(2)}^{-1,0,p-2}\rangle.&(2.2.16)\cr}
$$
Further, from the definition (2.2.13), using (1.1.16), (1.1.17), (2.2.2a), 
(2.2.5a), (2.2.8), (2.2.11a), one finds
$$
\eqalignno{\vphantom{\bigg(}
S_{1p-2}=& \ \sum_{k=0}^{p-1}\langle V_{k,p-1-k,0},
\delta_1 \Omega^{k-1,p-1-k,0}+(-1)^{k+1} \delta_2 \Omega^{k,p-2-k,0}\rangle 
&\cr
\vphantom{\bigg(}
+&\sum_{k=0}^{p-2}\langle Z_{k,p-2-k,0},
\delta_1 \Theta^{k-1,p-2-k,0}
+(-1)^{k+1} \delta_2 \Theta^{k,p-3-k,0}
+(-1)^{p-1}i^*\Omega^{k,p-2-k,0}\rangle &\cr
\vphantom{\bigg(}
+&(-1)^{p-1}\langle V^{(1)}_{p-1,-1,0},\Omega_{(1)}^{p-1,-1,0}\rangle
-(-1)^{p-1}\langle V^{(2)}_{-1,p-1,0},\Omega_{(2)}^{-1,p-1,0}\rangle.
&(2.2.17)\cr}
$$
Let $I_1^{{\cal O}(1)}$ $(I_1^{{\cal O}(2)})$ be constructed 
according (2.1.1) using the above (co)intertwiners marked by the label 1 (2).
Substituting (2.2.16) and (2.2.17) into (2.2.15) and using 
(2.2.3), (2.2.9), one finds 
$$
I_1^{{\cal O}(2)}-I_1^{{\cal O}(1)}
=\sum_{k=0}^{p-1}\langle S_{k,p-1-k,-1},\Xi^{k,p-1-k,-1}\rangle
-\sum_{k=0}^{p-2}\langle T_{k,p-2-k,-1},\Upsilon^{k,p-2-k,-1}\rangle.
\eqno(2.2.18)
$$

We say that a sequence of chains 
$(\{S_{k,p-1-k,-1}|k\in K_{p-1}\},\{T_{k,p-2-k,-1}|k\in K_{p-2}\})$ 
interpolates two relative integer \v Cech $p-1$--cycles~ 
$(S^{(1)}_{p-1,-1,-1},~T^{(1)}_{p-2,-1,-1})$,~~ 
$(S^{(2)}_{-1,p-1,-1},$ $T^{(2)}_{-1,p-2,-1})$
(cfr. eqs. (1.2.3)), if 
$S_{k,p-1-k,-1}$ are integer \v Cech chains of $X$,
$T_{k,p-2-k,-1}$ are integer \v Cech chains in $Y$
and 
$$
\eqalignno{\vphantom{\bigg(}&
\beta_1 S_{k+1,p-2-k,-1}+(-1)^{k+1}\beta_2 S_{k,p-1-k,-1}
-i_*T_{k,p-2-k,-1}=0,&\cr
\vphantom{\bigg(}&
\quad 0\leq k\leq p-2,&(2.2.19a)\cr
\vphantom{\bigg(}&
-\beta_1 T_{k+1,p-3-k,-1}-(-1)^{k+1}\beta_2 T_{k,p-2-k,-1}
=0,\quad 0\leq k\leq p-3,&(2.2.19b)\cr}
$$
with
$$
\eqalignno{\vphantom{\bigg(}&
S^{(1)}_{p-1,-1,-1} = \beta_2 S_{p-1,0,-1},\quad
S^{(2)}_{-1,p-1,-1} = (-1)^{p-1}\beta_1 S_{0,p-1,-1},&(2.2.20a)\cr
\vphantom{\bigg(}&
T^{(1)}_{p-2,-1,-1} = \beta_2 T_{p-2,0,-1},\quad 
T^{(2)}_{-1,p-2,-1} = (-1)^{p-2}\beta_1 T_{0,p-2,-1}. &(2.2.20b)\cr}
$$
The above relations are compatible with the relations (1.2.3) 
obeyed by the integer \v Cech cycles.

We say that a sequence of cochains 
$(\{\Xi^{k,p-1-k,-1}|k\in K^p\},\{\Upsilon^{k,p-2-k,-1}|k\in K^{p-1}\};$
$\{\Xi^{*k,p-2-k,-1}|k\in K^{p-1}\},\{\Upsilon^{*k,p-3-k,-1}|k\in K^{p-2}\};
\{\hat\Xi^{k,p-1-k,-1}|k\in K^p\},\{\hat\Upsilon^{k,p-2-k,-1}|$ 
$k\in K^{p-1}\})$ interpolates two given relative differential $p$--cocycles
$(\Xi_{(1)}^{p,-1,-1},\Upsilon_{(1)}^{p-1,-1,-1};$
$\Xi_{(1)}^{*p-1,-1,-1},~\Upsilon_{(1)}^{*p-2,-1,-1};~
\hat\Xi_{(1)}^{p,-1,-1},~\hat\Upsilon_{(1)}^{p-1,-1,-1})$,~~ 
$(\Xi_{(2)}^{-1,p,-1},~\Upsilon_{(2)}^{-1,p-1,-1};~
\Xi_{(2)}^{*-1,p-1,-1},$ $\Upsilon_{(2)}^{*-1,p-2,-1};
\hat\Xi_{(2)}^{-1,p,-1},\hat\Upsilon_{(2)}^{-1,p-1,-1})$
(cfr. eqs. (1.3.1)--(1.3.3)), if
$\Xi^{k,p-1-k,-1}$, $\Xi^{*k,p-2-k,-1}$ are real \v Cech
cochains of $X$, $\hat\Xi^{k,p-1-k,-1}$ is an integer \v Cech cochain
of $X$, $\Upsilon^{k,p-2-k,-1}$, $\Upsilon^{*k,p-3-k,-1}$ are real \v Cech
cochains of $Y$, $\hat\Upsilon^{k,p-2-k,-1}$ is an integer \v Cech cochain
of $Y$ such that
$$
\eqalignno{\vphantom{\bigg(}&
\delta_1 \Xi^{k-1,p-k,-1}+(-1)^{k+1}\delta_2 \Xi^{k,p-1-k,-1}=0,
\quad -1\leq k \leq p+1, &(2.2.21a)\cr
\vphantom{\bigg(}&
i^*\Xi^{k,p-1-k,-1}-\delta_1 \Upsilon^{k-1,p-1-k,-1}
-(-1)^{k+1}\delta_2 \Upsilon^{k,p-2-k,-1}=0,&\cr
\vphantom{\bigg(}&
\quad -1\leq k\leq p,&(2.2.21b)\cr
\vphantom{\bigg(}&
\delta_1 \Xi^{*k-1,p-1-k,-1}+(-1)^{k+1}\delta_2 \Xi^{*k,p-2-k,-1}
=\hat\Xi^{k,p-1-k,-1}-\Xi^{k,p-1-k,-1},&\cr
\vphantom{\bigg(}&
\quad -1\leq k \leq p &(2.2.22a)\cr
\vphantom{\bigg(}&
i^*\Xi^{*k,p-2-k,-1}-\delta_1 \Upsilon^{*k-1,p-2-k,-1}
-(-1)^{k+1}\delta_2 \Upsilon^{*k,p-3-k,-1}&\cr
\vphantom{\bigg(}&
=\hat\Upsilon^{k,p-2-k,-1}-\Upsilon^{k,p-2-k,-1},
\quad -1\leq k\leq p-1,&(2.2.22b)\cr
\vphantom{\bigg(}&
\delta_1 \hat\Xi^{k-1,p-k,-1}+(-1)^{k+1}\delta_2 \hat\Xi^{k,p-1-k,-1}=0,
\quad -1\leq k \leq p+1, &(2.2.23a)\cr
\vphantom{\bigg(}&
i^*\hat\Xi^{k,p-1-k,-1}-\delta_1 \hat\Upsilon^{k-1,p-1-k,-1}
-(-1)^{k+1}\delta_2 \hat\Upsilon^{k,p-2-k,-1}=0,&\cr
\vphantom{\bigg(}&
\quad -1\leq k\leq p,&(2.2.23b)\cr}
$$
with
$$
\eqalignno{\vphantom{\bigg(}&
\Xi_{(1)}^{p,-1,-1}=\Xi^{p,-1,-1},\quad 
\Xi_{(2)}^{-1,p,-1}=\Xi^{-1,p,-1},
&(2.2.24a)\cr
\vphantom{\bigg(}&
\Upsilon_{(1)}^{p-1,-1,-1}=\Upsilon^{p-1,-1,-1},\quad
\Upsilon_{(2)}^{-1,p-1,-1}=\Upsilon^{-1,p-1,-1},
&(2.2.24b)\cr
\vphantom{\bigg(}&
\Xi_{(1)}^{*p-1,-1,-1}=\Xi^{*p-1,-1,-1},\quad 
\Xi_{(2)}^{*-1,p-1,-1}=\Xi^{*-1,p-1,-1},
&(2.2.25a)\cr
\vphantom{\bigg(}&
\Upsilon_{(1)}^{*p-2,-1,-1}=\Upsilon^{*p-2,-1,-1},\quad
\Upsilon_{(2)}^{*-1,p-2,-1}=\Upsilon^{*-1,p-2,-1},
&(2.2.25b)\cr
\vphantom{\bigg(}&
\hat\Xi_{(1)}^{p,-1,-1}=\hat\Xi^{p,-1,-1},\quad 
\hat\Xi_{(2)}^{-1,p,-1}=\hat\Xi^{-1,p,-1},
&(2.2.26a)\cr
\vphantom{\bigg(}&
\hat\Upsilon_{(1)}^{p-1,-1,-1}=\hat\Upsilon^{p-1,-1,-1},\quad
\hat\Upsilon_{(2)}^{-1,p-1,-1}=\hat\Upsilon^{-1,p-1,-1}
&(2.2.26b)\cr}
$$
(cfr. footnote 1).
It is straightforward to check that the above relations are compatible with 
the relations (1.3.1)--(1.3.3) obeyed by the relative differential cocycles.

Using the interpolating sequences of (co)chains just introduced, one defines
$$
\eqalignno{\vphantom{\bigg(}
S_2=& \ \sum_{k=0}^{p-1}\langle S_{k,p-1-k,-1},
\delta_1 \Xi^{*k-1,p-1-k,-1}+(-1)^{k+1}\delta_2 \Xi^{*k,p-2-k,-1}\rangle
&\cr
\vphantom{\bigg(}
-&\sum_{k=0}^{p-2}\langle T_{k,p-2-k,-1},
i^*\Xi^{*k,p-2-k,-1}-\delta_1 \Upsilon^{*k-1,p-2-k,-1}
-(-1)^{k+1}\delta_2 \Upsilon^{*k,p-3-k,-1}\rangle.&\cr
&&(2.2.27)\cr}
$$
Using (2.2.22), one has immediately
$$
\eqalignno{\vphantom{\bigg(}
S_2=& -\sum_{k=0}^{p-1}\langle S_{k,p-1-k,-1},\Xi^{k,p-1-k,-1}\rangle
+\sum_{k=0}^{p-2}\langle T_{k,p-2-k,-1},\Upsilon^{k,p-2-k,-1}\rangle
&\cr
\vphantom{\bigg(}
&+\sum_{k=0}^{p-1}\langle S_{k,p-1-k,-1},\hat\Xi^{k,p-1-k,-1}\rangle
-\sum_{k=0}^{p-2}\langle T_{k,p-2-k,-1},\hat\Upsilon^{k,p-2-k,-1}\rangle.
&(2.2.28)\cr}
$$
On the other hand, using (1.1.17), (2.2.19), (2.2.20), (2.2.25), one finds
$$
\eqalignno{\vphantom{\bigg(}
S_2=(-1)^{p-1}\Big[
&\langle S^{(2)}_{-1,p-1,-1},\Xi_{(2)}^{*-1,p-1,-1}\rangle 
-\langle T^{(2)}_{-1,p-2,-1},\Upsilon_{(2)}^{*-1,p-2,-1}\rangle &\cr
\vphantom{\bigg(}
-&\langle S^{(1)}_{p-1,-1,-1},\Xi_{(1)}^{*p-1,-1,-1}\rangle 
+\langle T^{(1)}_{p-2,-1,-1},\Upsilon_{(1)}^{*p-2,-1,-1}\rangle\Big].
&(2.2.29)\cr}
$$
Now, we note that
$$
\sum_{k=0}^{p-1}\langle S_{k,p-1-k,-1},\hat\Xi^{k,p-1-k,-1}\rangle
-\sum_{k=0}^{p-2}\langle T_{k,p-2-k,-1},\hat\Upsilon^{k,p-2-k,-1}\rangle
=0\quad\hbox{mod $\Bbb Z$}.
\eqno(2.2.30)
$$
Let $I_2^{{\cal O}(1)}$ $(I_2^{{\cal O}(2)})$ be constructed 
according (2.1.4) using the integer \v Cech cycle and the differential 
cocycle marked by the label 1 (2). From (2.2.28), (2.2.29), one has then
$$
I_2^{{\cal O}(2)}-I_2^{{\cal O}(1)}
=-\sum_{k=0}^{p-1}\langle S_{k,p-1-k,-1},\Xi^{k,p-1-k,-1}\rangle
+\sum_{k=0}^{p-2}\langle T_{k,p-2-k,-1},\Upsilon^{k,p-2-k,-1}\rangle
\quad\hbox{mod $\Bbb Z$}.
\eqno(2.2.31)
$$

Let $(S_{p-1},T_{p-2})\in Z^{s{\cal O}(i)}_{p-1}(X,Y)$, 
$(\Xi^p,\Upsilon^{p-1})\in Z_{dR\Bbb Z}^p(X,Y)$ be respectively 
an ${\cal O}(i)$--small relative singular cycle, $i=1,~2$, 
(cfr. eq. (1.2.1) and subsect. 1.2) and a cohomologically integer 
relative de Rham cocycle (cfr. eq. (1.2.13) and subsect. 1.5).
Let us now repeat the construction described in sect. 2.1 individually for 
each of the covering ${\cal O}(i)$.

Then, the relative singular cycles  
$(S^{(1)}_{-1,-1,p-1},T^{(1)}_{-1,-1,p-2})=(S_{-1,-1,p-1},T_{-1,-1,p-2})$, 
$(S^{(2)}_{-1,-1,p-1},T^{(2)}_{-1,-1,p-2})=(S_{-1,-1,p-1},T_{-1,-1,p-2})$
extend to \v Cech singular intertwiners 
$(S^{(1)}_{-1,-1,p-1},T^{(1)}_{-1,-1,p-2};\!\{V^{(1)}_{k,-1,p-1-k}\},\!
\{Z^{(1)}_{k,-1,p-2-k}\};\!S^{(1)}_{p-1,-1,-1},T^{(1)}_{p-2,-1,-1})$,
$(S^{(2)}_{-1,-1,p-1},$ $T^{(2)}_{-1,-1,p-2};\{V^{(2)}_{-1,k,p-1-k}\}, 
\{Z^{(2)}_{-1,k,p-2-k}\};S^{(2)}_{-1,p-1,-1},T^{(2)}_{-1,p-2,-1})$
defined up to shifts by trivial intertwiners leaving 
$(S^{(1)}_{-1,-1,p-1},T^{(1)}_{-1,-1,p-2})$, 
$(S^{(2)}_{-1,-1,p-1},T^{(2)}_{-1,-1,p-2})$
unchanged, respectively, (cfr. subsects. 1.4 and 2.1). 

In similar fashion, the relative de Rham cocycles 
$(\Xi_{(1)}^{-1,-1,p},\Upsilon_{(1)}^{-1,-1,p-1})
=(\Xi^{-1,-1,p},$ $\Upsilon^{-1,-1,p-1})$,
$(\Xi_{(2)}^{-1,-1,p},\Upsilon_{(2)}^{-1,-1,p-1})
=(\Xi^{-1,-1,p},\Upsilon^{-1,-1,p-1})$
extend to \v Cech--de Rham cointertwiners  
$(\Xi_{(1)}^{-1,-1,p},\Upsilon_{(1)}^{-1,-1,p-1};
\{\Omega_{(1)}^{k,-1,p-1-k}\},\{\Theta_{(1)}^{k,-1,p-2-k}\};
\Xi_{(1)}^{p,-1,-1},\Upsilon_{(1)}^{p-1,-1,-1})$,
$(\Xi_{(2)}^{-1,-1,p},\Upsilon_{(2)}^{-1,-1,p-1};~
\{\Omega_{(2)}^{-1,k,p-1-k}\},\!\{\Theta_{(2)}^{-1,k,p-2-k}\};~
\Xi_{(2)}^{-1,p,-1},\Upsilon_{(2)}^{-1,p-1,-1})$ 
defined up to shifts by trivial cointertwiners leaving 
$(\Xi_{(1)}^{-1,-1,p},\Upsilon_{(1)}^{-1,-1,p-1})$,
$(\Xi_{(2)}^{-1,-1,p},\Upsilon_{(2)}^{-1,-1,p-1})$
unchanged, respectively, (cfr. subsects. 1.4 and 2.1). 
In turn, the cohomologically integer real \v Cech 
cocycles $(\Xi_{(1)}^{p,-1,-1},\Upsilon_{(1)}^{p-1,-1,-1})$,
$(\Xi_{(2)}^{-1,p,-1},\Upsilon_{(2)}^{-1,p-1,-1})$ so obtained extend 
to relative differential cocycles
$(\Xi_{(1)}^{p,-1,-1},\Upsilon_{(1)}^{p-1,-1,-1};
\Xi_{(1)}^{*p-1,-1,-1},\Upsilon_{(1)}^{*p-2,-1,-1};
\hat\Xi_{(1)}^{p,-1,-1},\hat\Upsilon_{(1)}^{p-1,-1,-1})$,
$(\Xi_{(2)}^{-1,p,-1},\Upsilon_{(2)}^{-1,p-1,-1};$
$\Xi_{(2)}^{*-1,p-1,-1},\Upsilon_{(2)}^{*-1,p-2,-1};
\hat\Xi_{(2)}^{-1,p,-1},\hat\Upsilon_{(2)}^{-1,p-1,-1})$,
defined up to shifts by torsion differential cocycles leaving 
$(\Xi_{(1)}^{p,-1,-1},\Upsilon_{(1)}^{p-1,-1,-1})$,
$(\Xi_{(2)}^{-1,p,-1},\Upsilon_{(2)}^{-1,p-1,-1})$
unchanged, respectively, (cfr. subsects. 1.5 and 2.1). 

Using the above two sets of relative data, we can compute $I^{{\cal O}(i)}$,
$i=1,~2$, using (2.1.7). Since the choice of the relative data is not 
unique, $I^{{\cal O}(i)}$ is affected by an indetermination 
$\Delta I^{{\cal O}(i)}$ given by (2.1.8).

Next, our aim is to evaluate the difference $I^{{\cal O}(2)}-I^{{\cal O}(1)}$ 
modulo integers by constructing suitable interpolating sequences between
the relative data of the two coverings involved and exploiting the results
(2.2.18), (2.2.31). In order to do that, we have first to find out 
under which conditions such sequences do indeed exist.

Let us assume that $(S^{(1)}_{-1,-1,p-1},T^{(1)}_{-1,-1,p-2};
\{V^{(1)}_{k,-1,p-1-k}\},\{Z^{(1)}_{k,-1,p-2-k}\};
S^{(1)}_{p-1,-1,-1},$ $T^{(1)}_{p-2,-1,-1})$, $(S^{(2)}_{-1,-1,p-1},
T^{(2)}_{-1,-1,p-2};\{V^{(2)}_{-1,k,p-1-k}\},\!
\{Z^{(2)}_{-1,k,p-2-k}\};S^{(2)}_{-1,p-1,-1},T^{(2)}_{-1,p-2,-1})$ 
are two relative \v Cech singular $p-1$--intertwiners 
(cfr. eqs. (1.2.5)--(1.2.7)) such that
$(S^{(1)}_{-1,-1,p-1},T^{(1)}_{-1,-1,p-2})
=(S^{(2)}_{-1,-1,p-1},T^{(2)}_{-1,-1,p-2})$. 
Then, after possibly shifting the intertwiners by trivial intertwiners 
(cfr. eqs. (1.2.8)--(1.2.10))
preserving this condition, there exists a sequence of chains 
$(S_{-1,-1,p-1},T_{-1,-1,p-2};\{V_{k,l,p-1-k-l}\},\{Z_{k,l,p-2-k-l}\};$
$\{S_{k,p-1-k,-1}\},\{T_{k,p-2-k,-1}\})$ 
interpolating the intertwiners, i.e satisfying
(2.2.1)--(2.2.6). 

Here is a sketch of the proof. We begin with noting that, 
if $U_{k,l,n}$ is a \v Cech singular chain with $n>-1$ such that 
$\beta_1\beta_2U_{k,l,n}=0$, then there are 
\v Cech singular chains $U_{k+1,l,n}$, $U_{k,l+1,n}$ such that
$U_{k,l,n}=\beta_1U_{k+1,l,n}+(-1)^{k+1}\beta_2U_{k,l+1,n}$.
This follows from the triviality of the $\beta_1$, $\beta_2$
homology for ${\cal O}(1)$, ${\cal O}(2)$--small chains, respectively, 
when $n>-1$
and the fact that, if either $k<-1$ or $l<-1$, then $V_{k,l,n}=0$ 
for any \v Cech singular chains $V_{k,l,n}$. Set 
$$
\eqalignno{\vphantom{\bigg(}
&S_{-1,-1,p-1}=S^{(1)}_{-1,-1,p-1}=S^{(2)}_{-1,-1,p-1},&(2.2.32a)\cr
\vphantom{\bigg(}
&T_{-1,-1,p-2}=T^{(1)}_{-1,-1,p-2}=T^{(2)}_{-1,-1,p-2}.
&(2.2.32b)\cr}
$$ 
Then, $(S_{-1,-1,p-1},T_{-1,-1,p-2})$ is a relative singular $p-1$--cycle, 
$$
\eqalignno{\vphantom{\bigg(}
&bS_{-1,-1,p-1}-i_*T_{-1,-1,p-2}=0,&(2.2.33a)\cr
\vphantom{\bigg(}
&-bT_{-1,-1,p-2}=0.&(2.2.33b)\cr}
$$ 
Hence, there is a chain $V_{0,0,p-1}$ of $X$ and a
chain $Z_{0,0,p-2}$ of $Y$ satisfying (2.2.1).
By substituting (2.2.1) into (2.2.33), one finds that (2.2.2) holds 
for $k,~l=0$ for some chains $V_{1,0,p-2}$, $V_{0,1,p-2}$ of $X$ 
and $Z_{1,0,p-3}$, $Z_{0,1,p-3}$ of $Y$. The proof of (2.2.2) is 
completed by a straightforward induction on the value of $k+l$. 
$S_{k,p-1-k,-1}$, $T_{k,p-2-k,-1}$ are then defined according 
(2.2.3). Next, one verifies that 
relations (2.2.4)--(2.2.6) define two relative \v Cech singular 
$p-1$--intertwiners extending $(S^{(1)}_{-1,-1,p-1},T^{(1)}_{-1,-1,p-2})$, 
$(S^{(2)}_{-1,-1,p-1},T^{(2)}_{-1,-1,p-2})$. 
Thus, these intertwiners must equal the original intertwiners up to 
trivial shifts preserving $(S^{(1)}_{-1,-1,p-1},T^{(1)}_{-1,-1,p-2})$, 
$(S^{(2)}_{-1,-1,p-1},T^{(2)}_{-1,-1,p-2})$
(see the discussion of subsect. 1.4). 

The sequence of chains $(\{S_{k,p-1-k,-1}\},\{T_{k,p-2-k,-1}\})$ 
interpolates the integer \v Cech cycles 
$(S^{(1)}_{p-1,-1,-1},$ $T^{(1)}_{p-2,-1,-1})$,
$(S^{(2)}_{-1,p-1,-1},T^{(2)}_{-1,p-2,-1})$, i.e. 
it satisfies (2.2.19)--(2.2.20).
These statements are straightforwardly verified. 

Assume that ${\cal O}(1)\cup{\cal O}(2)$ is a good covering of the 
pair $X$, $Y$ and that $(\Xi_{(1)}^{-1,-1,p},$ $\Upsilon_{(1)}^{-1,-1,p-1};~
\{\Omega_{(1)}^{k,-1,p-1-k}\},\{\Theta_{(1)}^{k,-1,p-2-k}\};~
\Xi_{(1)}^{p,-1,-1},\Upsilon_{(1)}^{p-1,-1,-1})$, ~
$(\Xi_{(2)}^{-1,-1,p},\Upsilon_{(2)}^{-1,-1,p-1};$
$\{\Omega_{(2)}^{-1,k,p-1-k}\},\{\Theta_{(2)}^{-1,k,p-2-k}\};
\Xi_{(2)}^{-1,p,-1},\Upsilon_{(2)}^{-1,p-1,-1})$
are two relative \v Cech--de Rham $p$--co\-in\-ter\-twiners 
(cfr. eqs. (1.2.17)--(1.2.19)) such that
$(\Xi_{(1)}^{-1,-1,p},\Upsilon_{(1)}^{-1,-1,p-1})=
(\Xi_{(2)}^{-1,-1,p},$ $\Upsilon_{(2)}^{-1,-1,p-1})$.
Then, after possibly shifting the cointertwiners by trivial 
cointertwiners (cfr. eqs. (1.2.20)--(1.2.22))
preserving this condition, there exists a sequence of cochains 
$(\Xi^{-1,-1,p},\Upsilon^{-1,-1,p-1};~\{\Omega^{k,l,p-2-k-l}\},
\{\Theta^{k,l,p-3-k-l}\};$ $\{\Xi^{k,p-1-k,-1}\},\{\Upsilon^{k,p-2-k,-1}\})$ 
interpolating the cointertwiners, i.e. fulfilling (2.2.7)--(2.2.12).

Here is a sketch of the proof. As ${\cal O}(1)\cup{\cal O}(2)$ 
is a good covering of the pair $X$, $Y$, the cohomology isomorphism 
(1.4.3) holds true. Set
$$
\eqalignno{\vphantom{\bigg(}
&\Xi^{-1,-1,p}=\Xi_{(1)}^{-1,-1,p}=\Xi_{(2)}^{-1,-1,p},&(2.2.34a)\cr
\vphantom{\bigg(}
&\Upsilon^{-1,-1,p-1} =\Upsilon_{(1)}^{-1,-1,p-1}= \Upsilon_{(2)}^{-1,-1,p-1}.
&(2.2.34b)\cr}
$$
Then, $(\Xi^{-1,-1,p},\Upsilon^{-1,-1,p-1})$ is a relative de Rham 
$p$--cocycle, 
$$
\eqalignno{\vphantom{\bigg(}
&d\Xi^{-1,-1,p}=0,&(2.2.35a)\cr
\vphantom{\bigg(}
&i^*\Xi^{-1,-1,p}-d\Upsilon^{-1,-1,p-1}=0.&(2.2.35b)\cr}
$$ 
This can be extended to an ${\cal O}(1)\cup{\cal O}(2)$
relative cointertwiner, which is precisely the 
sequence of cochains interpolating the given relative 
\v Cech--de Rham cointertwiners we are looking for. Indeed, (2.2.7)--(2.2.9)
are nothing but the transcription of (1.2.17)--(1.2.19) for the 
covering ${\cal O}(1)\cup{\cal O}(2)$. 
One verifies that relations (2.2.10)--(2.2.12) 
define two relative \v Cech--de Rham $p$--cointertwiner extending 
$(\Xi_{(1)}^{-1,-1,p},\Upsilon_{(1)}^{-1,-1,p-1})$,
$(\Xi_{(2)}^{-1,-1,p},\Upsilon_{(2)}^{-1,-1,p-1})$.
Thus, these cointertwiners must equal the original cointertwiners up to 
trivial shifts preserving $(\Xi_{(1)}^{-1,-1,p},\Upsilon_{(1)}^{-1,-1,p-1})$,
$(\Xi_{(2)}^{-1,-1,p},$ $\Upsilon_{(2)}^{-1,-1,p-1})$
(see the discussion of subsect. 1.4). 

If the relative de Rham cocycles 
$(\Xi_{(1)}^{-1,-1,p},\Upsilon_{(1)}^{-1,-1,p-1})$,
$(\Xi_{(2)}^{-1,-1,p},\Upsilon_{(2)}^{-1,-1,p-1})$ 
are cohomologically integer, the relative real \v Cech cocycles
$(\Xi_{(1)}^{p,-1,-1},\Upsilon_{(1)}^{p-1,-1,-1})$,
$(\Xi_{(2)}^{-1,p,-1},$ $\Upsilon_{(2)}^{-1,p-1,-1})$ are also 
cohomologically integer (cfr. subsect. 1.5) and, therefore,  
fit into two relative differential $p$--cocycles
$(\Xi_{(1)}^{p,-1,-1},\Upsilon_{(1)}^{p-1,-1,-1};~
\Xi_{(1)}^{*p-1,-1,-1},\Upsilon_{(1)}^{*p-2,-1,-1};~
\hat\Xi_{(1)}^{p,-1,-1},$ $\hat\Upsilon_{(1)}^{p-1,-1,-1})$,
$(\Xi_{(2)}^{-1,p,-1},\Upsilon_{(2)}^{-1,p-1,-1};~\Xi_{(2)}^{*-1,p-1,-1},
\Upsilon_{(2)}^{*-1,p-2,-1};~\hat\Xi_{(2)}^{-1,p,-1},
\hat\Upsilon_{(2)}^{-1,p-1,-1})$ (cfr. eqs. (1.3.1)--(1.3.3)).
In that case, 
$(\{\Xi^{k,p-1-k,-1}\},\{\Upsilon^{k,p-2-k,-1}\})$ 
extends to a sequence of cochains 
$(\{\Xi^{k,p-1-k,-1}\},\{\Upsilon^{k,p-2-k,-1}\};~
\{\Xi^{*k,p-2-k,-1}\},\{\Upsilon^{*k,p-3-k,-1}\};~
\{\hat\Xi^{k,p-1-k,-1}\},$ $\{\hat\Upsilon^{k,p-2-k,-1}\})$
interpolating those cocycles, i.e. satisfying (2.2.21)--(2.2.26),
after possibly shifting the latter by torsion differential cocycles
(cfr. eqs. (1.3.4)--(1.3.6))
preserving $(\Xi_{(1)}^{p,-1,-1},\Upsilon_{(1)}^{p-1,-1,-1})$,
$(\Xi_{(2)}^{-1,p,-1},\Upsilon_{(2)}^{-1,p-1,-1})$.
Moreover, when the relative integer \v Cech cocycles
$(\hat\Xi_{(1)}^{p,-1,-1},\hat\Upsilon_{(1)}^{p-1,-1,-1})$,
$(\hat\Xi_{(2)}^{-1,p,-1},\hat\Upsilon_{(2)}^{-1,p-1,-1})$ 
are representatives of the same relative integer singular cohomology class 
via the isomorphism (1.5.1), the shifts by 
torsion differential cocycles preserve that cohomology class.

Indeed, the relative de Rham cocycle $(\Xi^{-1,-1,p},\Upsilon^{-1,-1,p-1})$ is 
cohomologically integer, so that the relative real \v Cech cocycle 
$(\{\Xi^{k,p-1-k,-1}\},\{\Upsilon^{k,p-2-k,-1}\})$ is similarly 
cohomologically integer. Thus, it can be extended to an
${\cal O}(1)\cup{\cal O}(2)$ relative differential cocycle, which is the 
desired interpolating sequence of cochains. (2.2.21)--(2.2.23) are indeed 
the transcription of (1.3.1)--(1.3.3) for the covering 
${\cal O}(1)\cup{\cal O}(2)$.
One then checks that relations (2.2.24)--(2.2.26) 
define two relative differential $p$--cocycles extending 
$(\Xi_{(1)}^{p,-1,-1},\Upsilon_{(1)}^{p-1,-1,-1})$,
$(\Xi_{(2)}^{-1,p,-1},\Upsilon_{(2)}^{-1,p-1,-1})$. Thus, they 
must equal the given relative differential cocycles up to a 
torsion differential cocycle preserving $(\Xi_{(1)}^{p,-1,-1},$
$\Upsilon_{(1)}^{p-1,-1,-1})$, $(\Xi_{(2)}^{-1,p,-1},
\Upsilon_{(2)}^{-1,p-1,-1})$. When the relative integer \v Cech cocycles
$(\hat\Xi_{(1)}^{p,-1,-1},\hat\Upsilon_{(1)}^{p-1,-1,-1})$,
$(\hat\Xi_{(2)}^{-1,p,-1},\hat\Upsilon_{(2)}^{-1,p-1,-1})$ 
are representatives of the same relative integer singular cohomology class,
the interpolating sequence of cochains can be chosen so that the 
relative integer \v Cech cocycle $(\{\hat\Xi^{k,p-1-k,-1}\},
\{\hat\Upsilon^{k,p-2-k,-1}\})$ is also a representative of that cohomology 
class. In that instance, relations (2.2.26) define two relative integer 
\v Cech cocycles representing again that cohomology class and thus 
equivalent to $(\hat\Xi_{(1)}^{p,-1,-1},\hat\Upsilon_{(1)}^{p-1,-1,-1})$,
$(\hat\Xi_{(2)}^{-1,p,-1},\hat\Upsilon_{(2)}^{-1,p-1,-1})$ in 
relative integer \v Cech cohomology.

The above statements remain true if one of the two coverings, say
${\cal O}(2)$, is substituted by a refinement ${\cal O}'(2)$
which is a good covering of $X$, $Y$ (cfr. subsects. 1.1, 1.4).  

Indeed, as ${\cal O}'(2)$ is a refinement of ${\cal O}(2)$, 
the associated refinement map $f_2$ induces 
a homomorphism $f_2{}^*$ of the space relative \v Cech--de Rham cochains 
of ${\cal O}(1)\cup{\cal O}(2)$ into that of ${\cal O}(1)\cup{\cal O}'(2)$,
which preserves the de Rham and \v Cech degrees, 
commutes with $d$ and is such that $f_2{}^*\delta_1=\delta_1f_2{}^*$, 
$f_2{}^*\delta_2=\delta'{}_2f_2{}^*$. 
Then, sequence of cochains obtained by applying $f_2{}^*$ to the interpolating 
sequence of cochains of ${\cal O}(1)$, ${\cal O}(2)$
is interpolating with respect to ${\cal O}(1)$, ${\cal O}'(2)$.

It is easy to see that the above conditions on the coverings
${\cal O}(1)$, ${\cal O}(2)$ are trivially satisfied for 
${\cal O}(1)={\cal O}(2)$, so that, in this special case,
interpolating sequences of cochains exist. 
Then, interpolating sequences exist also when ${\cal O}(2)$ is a 
refinement of ${\cal O}(1)$.

In summary, we have shown the following.

First, there indeed exists a sequence of chains 
$(S_{-1,-1,p-1},T_{-1,-1,p-2};\{V_{k,l,p-1-k-l}\},$ $\{Z_{k,l,p-2-k-l}\};
\{S_{k,p-1-k,-1}\},\{T_{k,p-2-k,-1}\})$ interpolating the intertwiners 
$(S^{(1)}_{-1,-1,p-1},$ $T^{(1)}_{-1,-1,p-2};\!\{V^{(1)}_{k,-1,p-1-k}\},
\{Z^{(1)}_{k,-1,p-2-k}\};S^{(1)}_{p-1,-1,-1},T^{(1)}_{p-2,-1,-1})$,
$(S^{(2)}_{-1,-1,p-1},T^{(2)}_{-1,-1,p-2};$ $\{V^{(2)}_{-1,k,p-1-k}\}, 
\{Z^{(2)}_{-1,k,p-2-k}\};~S^{(2)}_{-1,p-1,-1},T^{(2)}_{-1,p-2,-1})$
such that the sequence of chains 
$(\{S_{k,p-1-k,-1}\},\!\{T_{k,p-2-k,-1}\}\!)$ interpolates
the integer \v Cech cycles $(S^{(1)}_{p-1,-1,-1},\!T^{(1)}_{p-2,-1,-1}),$ 
$(S^{(2)}_{-1,p-1,-1},T^{(2)}_{-1,p-2,-1})$, possibly after shifting
the intertwiners by trivial intertwiners leaving 
$(S^{(1)}_{-1,-1,p-1},T^{(1)}_{-1,-1,p-2})$,
$(S^{(2)}_{-1,-1,p-1},T^{(2)}_{-1,-1,p-2})$
unchanged. 

Second, provided the good coverings ${\cal O}(1)$, ${\cal O}(2)$ 
satisfy the conditions illustrated above, 
there indeed exist a sequence of cochains  
$(\Xi^{-1,-1,p},\Upsilon^{-1,-1,p-1};\{\Omega^{k,l,p-2-k-l}\},$
$\{\Theta^{k,l,p-3-k-l}\};\{\Xi^{k,p-1-k,-1}\},\{\Upsilon^{k,p-2-k,-1}\})$
interpolating the cointertwiners $(\Xi_{(1)}^{-1,-1,p},$
$\Upsilon_{(1)}^{-1,-1,p-1};~\{\Omega_{(1)}^{k,-1,p-1-k}\},
\{\Theta_{(1)}^{k,-1,p-2-k}\};~\Xi_{(1)}^{p,-1,-1},
\Upsilon_{(1)}^{p-1,-1,-1})$, $(\Xi_{(2)}^{-1,-1,p},
\Upsilon_{(2)}^{-1,-1,p-1};$ $\{\Omega_{(2)}^{-1,k,p-1-k}\},\!
\{\Theta_{(2)}^{-1,k,p-2-k}\};\!\Xi_{(2)}^{-1,p,-1},
\Upsilon_{(2)}^{-1,p-1,-1})$ and a sequence of cochains
$(\{\Xi^{k,p-1-k,}$ ${}^{-1}\},\{\Upsilon^{k,p-2-k,-1}\};
\{\Xi^{*k,p-2-k,-1}\},\{\Upsilon^{*k,p-3-k,-1}\};
\{\hat\Xi^{k,p-1-k,-1}\},$ $\{\hat\Upsilon^{k,p-2-k,-1}\})$ 
interpolating the differential cocycles
$(\Xi_{(1)}^{p,-1,-1},\Upsilon_{(1)}^{p-1,-1,-1};
\Xi_{(1)}^{*p-1,-1,-1},\Upsilon_{(1)}^{*p-2,-1,-1};
\hat\Xi_{(1)}^{p,-1,-1},$ $ \hat\Upsilon_{(1)}^{p-1,-1,-1})$,
$(\Xi_{(2)}^{-1,p,-1},\Upsilon_{(2)}^{-1,p-1,-1};
\Xi_{(2)}^{*-1,p-1,-1},\Upsilon_{(2)}^{*-1,p-2,-1};
\hat\Xi_{(2)}^{-1,p,-1},\hat\Upsilon_{(2)}^{-1,p-1,-1})$, 
which are compatible in the sense that the end of the first interpolating 
sequence equals the beginning of the second, as shown by the notation,
possibly after shifting the cointertwiners and the differential cocycles
by trivial cointertwiners and torsion differential cocycles
leaving $(\Xi_{(1)}^{-1,-1,p},\Upsilon_{(1)}^{-1,-1,p-1})$, 
$(\Xi_{(2)}^{-1,-1,p},\Upsilon_{(2)}^{-1,-1,p-1})$ 
unchanged. 
Further, when the relative integer \v Cech cocycles
$(\hat\Xi_{(1)}^{p,-1,-1},\hat\Upsilon_{(1)}^{p-1,-1,-1})$,
$(\hat\Xi_{(2)}^{-1,p,-1},\hat\Upsilon_{(2)}^{-1,p-1,-1})$ 
represent of the same relative integer singular cohomology class, 
the shifts by torsion differential cocycles preserve that cohomology class.

Then, by (2.1.7), (2.2.18), (2.2.31), for given 
$(S_{p-1},T_{p-2})\in Z^{s{\cal O}(i)}_{p-1}(X,Y)$, $i=1,~2$, 
$(\Xi^p,\Upsilon^{p-1})$ $\in Z_{dR\Bbb Z}^p(X,Y)$, the difference
$I^{{\cal O}(2)}-I^{{\cal O}(1)}$ is integer, provided the relative data
employed in the construction of $I^{{\cal O}(i)}$ are suitably chosen. 
Since, however, this may not be the case, we see that a weaker result holds 
in general, namely
$$
I^{{\cal O}(2)}+\Delta I^{{\cal O}(2)}-I^{{\cal O}(1)}-\Delta I^{{\cal O}(1)}
=0\quad\hbox{mod $\Bbb Z$},
\eqno(2.2.36)
$$
where the indeterminations $\Delta I^{{\cal O}(i)}$, given by (2.1.8), 
account for the shifts relating the relative data used in $I^{{\cal O}(i)}$ 
and those for which the interpolating sequences exist. 

As discussed in subsect. 2.1, the indeterminations 
$\Delta I^{{\cal O}(i)}$ are not integer in general.
By demanding that $(\hat\Xi_{(1)}^{p,-1,-1},\hat\Upsilon_{(1)}^{p-1,-1,-1})$,
$(\hat\Xi_{(2)}^{-1,p,-1},\hat\Upsilon_{(2)}^{-1,p-1,-1})$
are representatives of a fixed cohomology class of 
$H_{s\Bbb Z}^p(X,Y)$ via (1.5.1), the $\Delta I^{{\cal O}(i)}$ are given 
modulo integers by expressions of the form of the right hand side of (2.1.11).

As we have seen in subsect. 2.1, for a given good covering ${\cal O}$, 
the $\Bbb Z$ linear functional $\Phi^{\cal O}:Z^{s{\cal O}}_{p-1}(X,Y)
\rightarrow\Bbb R/\Bbb Z$, eq. (2.1.9), depends on a choice 
of a relative integer singular cohomology class in $H_{s\Bbb Z}^p(X,Y)$, 
a representative $(\Xi^p,\Upsilon^{p-1})\in Z_{dR\Bbb Z}^p(X,Y)$ of the image 
of such class in $H_{dR\Bbb Z}^p(X,Y)$ and a point in the relative de Rham 
cohomology torus
$H_{dR}^{p-1}(X,Y)/H_{dR\Bbb Z}^{p-1}(X,Y)$. The parametrization of the 
family of maps $\Phi^{\cal O}$ in terms of $H_{dR}^{p-1}(X,Y)/
H_{dR\Bbb Z}^{p-1}(X,Y)$ is however not unique. A change of the 
parametrization changes $\Phi^{\cal O}$ by an amount given by (2.1.11),
(2.1.12). Thus, after fixing the cohomology class in $H_{s\Bbb Z}^p(X,Y)$ 
and its representative $(\Xi^p,\Upsilon^{p-1})\in Z_{dR\Bbb Z}^p(X,Y)$, 
there still is no natural way of comparing the maps $\Phi^{{\cal O}(1)}$, 
$\Phi^{{\cal O}(2)}$ for the good coverings ${\cal O}(1)$, ${\cal O}(2)$,
unless we have a mapping relating their 
$H_{dR}^{p-1}(X,Y)/H_{dR\Bbb Z}^{p-1}(X,Y)$ parametrizations.
This is precisely the origin of the residual indetrminations
$\Delta I^{{\cal O}(i)}$ of the previous paragraph.

Then, from (2.1.9) and (2.2.36), we can draw the following conclusions.
Let ${\cal O}(1)$, ${\cal O}(2)$ be two good coverings of $X$, $Y$
and ${\cal O}(12)$ be a common refinement of ${\cal O}(1)$, 
${\cal O}(2)$ which is also a good covering.
Let $(S_{p-1},T_{p-2})\in Z^{s{\cal O}(12)}_{p-1}(X,Y)$,
$(\Xi^p,\Upsilon^{p-1})\in Z_{dR\Bbb Z}^p(X,Y)$. 
The pairs of good coverings ${\cal O}(1)$, ${\cal O}(12)$ and 
${\cal O}(2)$, ${\cal O}(12)$ satisfy the requirements sufficient 
for the existence of interpolating sequences of cochains.
Then, 
$$
\Phi^{{\cal O}(i)}(S,T;\Xi,\Upsilon)=\Phi^{{\cal O}(12)}(S,T;\Xi,\Upsilon),
\quad i=1,~2, 
\eqno(2.2.37)
$$
provided the $H_{dR}^{p-1}(X,Y)/H_{dR\Bbb Z}^{p-1}(X,Y)$ parametrization
of $\Phi^{{\cal O}(i)}$, $\Phi^{{\cal O}(12)}$ is suitably chosen.
Thus, for $(S_{p-1},T_{p-2})\in Z^{s{\cal O}(12)}_{p-1}(X,Y)$,
$(\Xi^p,\Upsilon^{p-1})\in Z_{dR\Bbb Z}^p(X,Y)$, 
$$
\Phi^{{\cal O}(1)}(S,T;\Xi,\Upsilon)=\Phi^{{\cal O}(2)}(S,T;\Xi,\Upsilon),
\eqno(2.2.38)
$$
provided the $H_{dR}^{p-1}(X,Y)/H_{dR\Bbb Z}^{p-1}(X,Y)$ parametrizations
of $\Phi^{{\cal O}(1)}$, $\Phi^{{\cal O}(2)}$ are suitably chosen.

{\it Let us assume that the family of good open coverings of $X$, $Y$ 
is cofinal in the family of open coverings of $X$ (cfr. subsects. 1.4,
1.5).} The conditions under which this is the case will be analyzed separately.
in appendix A1.
Then, in the sense stated in (2.2.38), $\Phi^{\cal O}$ is 
independent from covering choices.

\vskip.6cm
\item{\it 3.} {\it Extension of $\Phi^{\cal O}$ to non $\cal O$--small 
relative cycles}
\vskip.4cm
\par

Since any dependence on a choice of open covering $\cal O$ is 
unnatural, we would like to extend the $\Bbb Z$ linear map 
$\Phi^{\cal O}:Z^{s{\cal O}}_{p-1}(X,Y)\rightarrow \Bbb R/\Bbb Z$ 
of subsect. 2.1 to a $\Bbb Z$ linear map   
$\Phi:Z^s_{p-1}(X,Y)\rightarrow \Bbb R/\Bbb Z$ independent from $\cal O$.
This can indeed be done using the barycentric subdivision operator $q$ 
introduced in subsect. 1.1 as follows.

Let us fix the cohomology class of $H_{s\Bbb Z}^p(X,Y)$, 
its representative $(\Xi^p,\Upsilon^{p-1})\in Z_{dR\Bbb Z}^p(X,$ $Y)$
and the point of the torus $H_{dR}^{p-1}(X,Y)/H_{dR\Bbb Z}^{p-1}(X,Y)$ 
involved in the definition of $\Phi^{\cal O}$. 
Let $(S_{p-1},T_{p-2})\in Z^s_{p-1}(X,Y)$ be a general relative singular 
$p-1$--cycle. Pick a good open covering ${\cal O}$ of the pair $X$, 
$Y$.
For a sufficiently large $k\geq 0$, $(q^kS_{p-1},q^kT_{p-2})
\in Z^{s{\cal O}}_{p-1}(X,Y)$ is ${\cal O}$--small. We then set 
$$
\Phi(S,T;\Xi,\Upsilon)
=\Phi^{\cal O}(q^kS,q^kT;\Xi,\Upsilon).
\eqno(2.3.1)
$$
Next, we shall show that the right hand side of (2.3.1)
does not depend on ${\cal O}$ and $k$, making the definition well--posed.

From (1.1.5), for $k,l\geq 0$, one has
$$
bc^{(k,l)}+c^{(k,l)}b=q^k-q^l, \quad c^{(k,l)}
=\sgn(k-l) \cdot c\sum_{r=\min(k,l)}^{\max(k,l)-1} q^r.
\eqno(2.3.2)
$$
So, by (1.2.1)
$$
\eqalignno{\vphantom{\bigg(}
&bc^{(k,l)}S_{p-1}+i_*c^{(k,l)}T_{p-2}=q^kS_{p-1}-q^lS_{p-1},&(2.3.3a)\cr
\vphantom{\bigg(}
&bc^{(k,l)}T_{p-2}=q^kT_{p-2}-q^lT_{p-2}.&(2.3.3b)\cr}
$$
Therefore, $(q^kS_{p-1}-q^lS_{p-1},q^kT_{p-2}-q^lT_{p-2})$ is the relative 
boundary of the relative chain $(c^{(k,l)}S_{p-1},-c^{(k,l)}T_{p-2})$.
Now, if $k$, $l$ are large enough, $q^rS_{p-1}$, $q^rT_{p-2}$ are both 
${\cal O}$--small for $r\geq\min(k,l)$. Since $c$ preserves
${\cal O}$--smallness and the range of $c$ contains only degenerate chains
(see subsect. 1.1), both $c^{(k,l)}S_{p-1}$ and $c^{(k,l)}T_{p-2}$ are  
${\cal O}$--small and degenerate, by (2.3.2). Recall that degenerate chains 
are invisible, that is the integral of any form on any such chain vanishes. 
So, recalling (1.1.15), (2.1.10)
$$
\eqalignno{\vphantom{\bigg(}
&\Phi^{\cal O}(q^kS,q^kT;\Xi,\Upsilon)
-\Phi^{\cal O}(q^lS,q^lT;\Xi,\Upsilon)&\cr
\vphantom{\bigg(}
=~&\Phi^{\cal O}(q^kS-q^lS,q^kT-q^lT;\Xi,\Upsilon)&\cr
\vphantom{\bigg(}
=~&\Phi^{\cal O}(bc^{(k,l)}S+i_*c^{(k,l)}T,bc^{(k,l)}T;\Xi,\Upsilon)&\cr
\vphantom{\bigg(}
=~&\langle c^{(k,l)}S_{p-1},\Xi^p\rangle
+\langle c^{(k,l)}T_{p-2},\Upsilon^{p-1}\rangle&\cr
\vphantom{\bigg(}
=~&0,\quad \hbox{mod $\Bbb Z$}. &(2.3.4)\cr}
$$
This shows that the right hand side of (2.3.1) is independent from $k$

Let ${\cal O}(1)$, ${\cal O}(2)$ be two good coverings. 
Let ${\cal O}(12)$ be a good 
covering refining both ${\cal O}(1)$, ${\cal O}(2)$ and let 
$k$ be large enough so that $(q^kS_{p-1},q^kT_{p-2})$ is 
${\cal O}(12)$--small. Then, 
$$
\Phi^{{\cal O}(1)}(q^kS,q^kT;\Xi,\Upsilon)
-\Phi^{{\cal O}(2)}(q^kS,q^kT;\Xi,\Upsilon)=0,
\eqno(2.3.5)
$$
by (2.2.38), provided the $H_{dR}^{p-1}(X,Y)/H_{dR\Bbb Z}^{p-1}(X,Y)$ 
parametrizations of $\Phi^{{\cal O}(1)}$, $\Phi^{{\cal O}(2)}$ are suitably 
chosen. This shows that the right hand side of (2.3.1) is independent 
from ${\cal O}$.

We have thus managed to define a mapping $\Phi:Z^s_{p-1}(X,Y)
\rightarrow \Bbb R/\Bbb Z$. It is easy to show that 
$\Phi$ is $\Bbb Z$ linear. For given $\Bbb Z$ linear 
combinations of relative cycles, one chooses a good
covering ${\cal O}$ and a subdivision 
degree $k$ large enough so that all the relative cycles involved are 
${\cal O}$--small. Then, the $\Bbb Z$ linearity of $\Phi$
follows trivially from that of $\Phi^{\cal O}$.

When the relative (co)cycles  $(S_{p-1},T_{p-2})$, $(\Xi^p,\Upsilon^{p-1})$ 
are shifted by the relative (co)boundaries given by the right hand sides of 
(1.2.2), (1.2.14), respectively, one has
$$
\eqalignno{\vphantom{\bigg(}
\Delta \Phi(S,T;\Xi,\Upsilon)
=& \ \langle S_{p-1}, \xi^{p-1}\rangle
-\langle T_{p-2}, \upsilon^{p-2} \rangle &\cr
\vphantom{\bigg(}
&+\langle s_p, \Xi^p\rangle
-\langle t_{p-1}, \Upsilon^{p-1} \rangle &\cr
\vphantom{\bigg(}
&+\langle s_p, d\xi^{p-1}\rangle
-\langle t_{p-1},i^*\xi^{p-1}- 
d\upsilon^{p-2}\rangle,\quad \hbox{mod $\Bbb Z$}.~~&
(2.3.6)\cr}
$$
Indeed, provided $k$ is large enough to make all chains involved
${\cal O}$--small, $\Delta \Phi(S,T;\Xi,\Upsilon)$ $=$
$\Delta \Phi^{\cal O}(q^kS,q^kT;\Xi,\Upsilon)$, which, on account of 
(2.1.10), is given by
$$
\eqalignno{\vphantom{\bigg(}
\Delta \Phi^{\cal O}(q^kS,q^kT;\Xi,\Upsilon)
=& \ \langle q^kS_{p-1}, \xi^{p-1}\rangle
-\langle q^kT_{p-2}, \upsilon^{p-2} \rangle &\cr
\vphantom{\bigg(}
&+\langle q^ks_p, \Xi^p\rangle
-\langle q^kt_{p-1}, \Upsilon^{p-1} \rangle &\cr
\vphantom{\bigg(}
&+\langle q^ks_p, d\xi^{p-1}\rangle
-\langle q^kt_{p-1},i^*\xi^{p-1}
- d\upsilon^{p-2}\rangle,\quad \hbox{mod $\Bbb Z$}.~~~~~~~ &
(2.3.7 )\cr}
$$
Using (2.3.2) and (1.2.1), one has
$$
\eqalignno{\vphantom{\bigg(}
&bc^{(k,0)}S_{p-1}+i_*c^{(k,0)}T_{p-2}=q^kS_{p-1}-S_{p-1},&(2.3. 8a)\cr
\vphantom{\bigg(}
&bc^{(k,0)}T_{p-2}=q^kT_{p-2}-T_{p-2}.&(2.3. 8b)\cr
\vphantom{\bigg(}
&bc^{(k,0)}s_p+c^{(k,0)}bs_p=q^ks_p-s_p,&(2.3. 8c)\cr
\vphantom{\bigg(}
&bc^{(k,0)}t_{p-1}+c^{(k,0)}bt_{p-1}=q^kt_{p-1}-t_{p-1}.&(2.3.8d)\cr}
$$
As the range of $c$ contains only degenerate chains, 
$c^{(k,0)}S_{p-1}$, $c^{(k,0)}T_{p-2}$, $c^{(k,0)}s_p$, 
$c^{(k,0)}t_{p-1}$ are all degenerate, hence invisible. Then, by (2.3.8), 
the chains $q^kS_{p-1}-S_{p-1}$, $q^kT_{p-2}-T_{p-2}$, $q^ks_p-s_p$, 
$q^kt_{p-1}-t_{p-1}$ are all invisible. It follows that 
the right hand side of (2.3.7) equals that of (2.3.6).

If we change the $H_{dR}^{p-1}(X,Y)/H_{dR\Bbb Z}^{p-1}(X,Y)$
parametrization, $\Phi(S,T;\Xi,\Upsilon)$ varies of an amount 
given by 
$$
\eqalignno{\vphantom{\bigg(}
\Delta \Phi(S,T;\Xi,\Upsilon)
=& \ -\Big[\langle S_{p-1},\Pi^{p-1}\rangle
-\langle T_{p-2},\Sigma^{p-2} \rangle\Big] 
\quad \hbox{mod $\Bbb Z$},~ &(2.3.9)\cr}
$$
for some relative de Rham cocycle $(\Pi^{p-1},\Sigma^{p-2})$  
defined up to cohomologically integer relative de Rham cocycles. 
Indeed, provided $k$ is so large that all chains involved are 
${\cal O}$--small, $\Delta \Phi(S,T;\Xi,\Upsilon)=
\Delta \Phi^{\cal O}(q^kS,q^kT;\Xi,\Upsilon)$, so that, by (2.1.12), 
$$
\eqalignno{\vphantom{\bigg(}
\Delta \Phi^{\cal O}(q^kS,q^kT;\Xi,\Upsilon)
=& \ -\Big[\langle q^kS_{p-1},\Pi^{p-1}\rangle
-\langle q^kT_{p-2},\Sigma^{p-2} \rangle\Big] 
\quad \hbox{mod $\Bbb Z$}.~ &(2.3.10)\cr}
$$
By (2.3.8), the chains $q^kS_{p-1}-S_{p-1}$, $q^kT_{p-2}-T_{p-2}$, 
are all invisible. It follows that 
the right hand side of (2.3.9) equals that of (2.3.10).

\vskip.6cm
\item{\it 4.} {\it The end product: the family ${\cal CS}^p_{X,Y}$ of 
relative Cheeger--Simons differential characters}
\vskip.4cm
\par

We have thus defined a family of $\Bbb Z$ linear mapping 
$\Phi:Z^s_{p-1}(X,Y)\rightarrow \Bbb R/\Bbb Z$
parame\-teri\-zed by a relative integer singular cohomology class in 
$H_{s\Bbb Z}^p(X,Y)$, a representative $(\Xi^p,\Upsilon^{p-1})\in 
Z_{dR\Bbb Z}^p(X,Y)$ of the image of such class in $H_{dR\Bbb Z}^p(X,Y)$ 
and a point in the relative de Rham cohomology torus
$H_{dR}^{p-1}(X,Y)/H_{dR\Bbb Z}^{p-1}(X,Y)$.
We claim that this is precisely the family of degree 
$p$ $Y$ relative Cheeger--Simons differential character of $X$, 
${\cal CS}^p_{X,Y}$. This will become clear in the next section. 
See also the heuristic discussion given in the introduction 
for comparison.
It is important to recall that the above construction works
provided the family of good open coverings of $X$, $Y$ is cofinal 
in the family of open coverings of $X$ (cfr. subsects. 1.4, 2.2).

\par\vskip.6cm
\item{\bf 3.} {\bf Formal properties of the relative Cheeger--Simons 
differential characters}
\vskip.4cm
\par

In this section, we shall define the relative Cheeger--Simons 
differential characters in abstract terms and study their main formal 
properties. This will lead us to identify the family ${\cal CS}^p_{X,Y}$
of these characters with the family of characters constructed 
in sect. 2.

Let $p$, $X$, $Y$ satisfy the same assumptions as in subsect. 1.2. 

\vskip.6cm
\item{\it 1.} {\it Basic properties of the relative Cheeger--Simons 
characters}
\vskip.4cm
\par

By definition, $\Phi\in{\cal CS}^p_{X,Y}$ if $\Phi:Z^s_{p-1}(X,Y)\rightarrow 
\Bbb R/\Bbb Z$ is a $\Bbb Z$ linear mapping and there is a relative 
de Rham $p$--cochain $(\Xi^p,\Upsilon^{p-1})\in C^p_{dR}(X,Y)$ such that
$$
\Phi(bs-i_*t,-bt)
=\langle s_p, \Xi^p\rangle-\langle t_{p-1}, \Upsilon^{p-1} \rangle 
\quad \hbox{mod $\Bbb Z$},
\eqno(3.1.1)
$$
for all relative singular chains $(s_p,t_{p-1})\in C_p^s(X,Y)$.
${\cal CS}^p_{X,Y}$ is clearly a group.

Let $\Phi\in{\cal CS}^p_{X,Y}$. If $(s_p,t_{p-1})\in Z_p^s(X,Y)$ 
is a relative singular $p$--cycle, then $\Phi(bs-i_*t,-bt)=0$,
by (1.2.1) (with $p$ replaced by $p+1$). From (3.1.1), we thus get the 
quantization condition
$$
\langle s_p,\Xi^p\rangle-\langle t_{p-1},\Upsilon^{p-1}\rangle\in \Bbb Z.
\eqno(3.1.2)
$$
Further, if $(s_p,t_{p-1})\in B_p^s(X,Y)$ is the boundary of a relative 
singular $p+1$--chain $(u_{p+1},v_p)$, one has from (3.1.2)
$$
\langle u_{p+1},d\Xi^p\rangle-\langle v_p,i^*\Xi^p-d\Upsilon^{p-1}\rangle
\in \Bbb Z,
\eqno(3.1.3)
$$
by (1.2.2) (with $p$ replaced by $p+1$) and (1.1.16), (1.1.17).
By (3.1.3), since $(u_{p+1},v_p)$ is arbitrary, $(\Xi^p,\Upsilon^{p-1})$
must satisfy (1.2.13) and is thus a relative de Rham cocycle.
From (3.1.2), $(\Xi^p,\Upsilon^{p-1})$ is cohomologically integer.
Therefore, for any $\Phi\in {\cal CS}^p_{X,Y}$, 
$(\Xi^p,\Upsilon^{p-1})\in Z_{dR\Bbb Z}^p(X,Y)$. 

To any $\Phi\in{\cal CS}^p_{X,Y}$ there is associated 
a well--defined relative integer singular cohomology class in 
$H_{s\Bbb Z}^p(X,Y)$ such that $(\Xi^p,\Upsilon^{p-1})$ is  
a representative of the image of such class in $H_{dR\Bbb Z}^p(X,Y)$.
Indeed, as $\Bbb R/\Bbb Z$ is a divisible group and $Z^s_{p-1}(X,Y)$
is a subgroup of the free group $C^s_{p-1}(X,Y)$, 
there is a $\Bbb Z$ linear mapping $\bar \Phi:C^s_{p-1}(X,Y)\rightarrow 
\Bbb R$ such that $\Phi=\bar\Phi\big|Z^s_{p-1}(X,Y)$ mod $\Bbb Z$. 
Then, by (3.1.1), 
$$
\Lambda^p(s_p)-\Gamma^{p-1}(t_{p-1})=\bar\Phi(bs-i_*t,-bt)
-\langle s_p,\Xi^p\rangle+\langle t_{p-1},\Upsilon^{p-1}\rangle,
\eqno(3.1.4)
$$
with $(s_p,t_{p-1})\in C_p^s(X,Y)$,
defines a relative integer singular cochain
$(\Lambda^p,\Gamma^{p-1})$, which is readily checked to be a cocycle 
cohomologically equivalent to $(\Xi^p,\Upsilon^{p-1})$. The choice 
of $\bar\Phi$ affects $(\Lambda^p,\Gamma^{p-1})$ at most by a relative
integer singular coboundary. Hence, the integer singular cohomology class of 
$(\Lambda^p,\Gamma^{p-1})$ 
is unambiguously determined by $\Phi$.

Let $(\Pi^{p-1},\Sigma^{p-2})\in C_{dR}^{p-1}(X,Y)$ be a relative de Rham 
$p-1$--cochain. Then, 
$$
\Phi(S,T)
=\langle S_{p-1},\Pi^{p-1}\rangle-\langle T_{p-2},\Sigma^{p-2} \rangle 
\quad \hbox{mod $\Bbb Z$},
\eqno(3.1.5)
$$
for $(S_{p-1},T_{p-2})\in Z_{p-1}^s(X,Y)$, defines a character 
$\Phi\in{\cal CS}^p_{X,Y}$. $\Phi$ depends only on the equivalence
class of $(\Pi^{p-1},\Sigma^{p-2})$ modulo the cohomologically integer 
relative de Rham $p-1$--cocycles of $Z_{dR\Bbb Z}^{p-1}(X,Y)$.
The class of $H_{s\Bbb Z}^p(X,Y)$ corresponding to $\Phi$ vanishes.
The relative de Rham cocycle $(\Xi^p,\Upsilon^{p-1})$ of $\Phi$
is the relative de Rham coboundary of $(\Pi^{p-1},\Sigma^{p-2})$
(cfr. eq. (1.2.14)) and $(\Xi^p,\Upsilon^{p-1})$ vanishes in the 
important case when $(\Pi^{p-1},\Sigma^{p-2})\in Z_{dR}^{p-1}(X,Y)$.

\vskip.6cm
\item{\it 2.} {\it The first relative Cheeger--Simons exact sequence}
\vskip.4cm
\par

From the above discussion, it follows that there is an exact sequence of 
the form 
$$
0\rightarrow H_{dR}^{p-1}(X,Y)/H_{dR\Bbb Z}^{p-1}(X,Y)
\rightarrow {\cal CS}^p_{X,Y} \rightarrow A_{\Bbb Z}^p(X,Y)
\rightarrow 0,
\eqno(3.2.1)
$$
where $A_{\Bbb Z}^p(X,Y)$ is the subset of the Cartesian product
$H_{s\Bbb Z}^p(X,Y)\times Z_{dR\Bbb Z}^p(X,Y)$ formed by  the pairs of a 
relative integer singular cohomology class in $H_{s\Bbb Z}^p(X,Y)$ and 
a representative of the image of such class  
in $H_{dR\Bbb Z}^p(X,Y)$. The relative de Rham 
cohomology torus $H_{dR}^{p-1}(X,Y)/H_{dR\Bbb Z}^{p-1}(X,Y)$ appears here. 
It parameterizes the group of all $\Phi\in {\cal CS}^p_{X,Y}$ characterized 
by the same pair of data in $A_{\Bbb Z}^p(X,Y)$. The sequence (3.2.1)
in the absolute case was found in ref. \ref{32}.

We note that the $\Bbb Z$ linear mappings $\Phi:Z^s_{p-1}(X,Y)\rightarrow 
\Bbb R/\Bbb Z$ constructed in sect. 2 all belong to ${\cal CS}^p_{X,Y}$
as they satisfy (3.1.1) on account of (2.3.6). Each such $\Phi$ is 
characterized by a relative integer singular cohomology class in 
$H_{s\Bbb Z}^p(X,Y)$ and a representative $(\Xi^p,\Upsilon^{p-1})\in 
Z_{dR\Bbb Z}^p(X,Y)$ of the image of such class in $H_{dR\Bbb Z}^p(X,Y)$.
As is easy to see, these relative data are precisely the ones defined 
abstractly in subsect. 3.1 above. Indeed, (3.1.4) is the statement 
in the language of singular cohomology that the sequence of
cochains $(\Xi^p,\Upsilon^{p-1};\bar\Phi;\Lambda^p,\Gamma^{p-1})$
is a differential cocycles (cfr. subsect. 1.3 and the discussion of 
subsect. 2.1). The set of the $\Phi$ compatible with a fixed choice of the 
relative data is parameterized by $H_{dR}^{p-1}(X,Y)/H_{dR\Bbb Z}^{p-1}(X,Y)$.
This justifies our claim that the family of $\Bbb Z$ linear mappings $\Phi$ 
of sect. 2 is precisely ${\cal CS}^p_{X,Y}$. 

\vskip.6cm
\item{\it 3.} {\it The second relative Cheeger--Simons exact sequence}
\vskip.4cm
\par
  
From the above discussion, there is another exact sequence of the form 
$$
0\rightarrow C_{dR}^{p-1}(X,Y)/Z_{dR\Bbb Z}^{p-1}(X,Y)
\rightarrow {\cal CS}^p_{X,Y} \rightarrow H_{s\Bbb Z}^p(X,Y)
\rightarrow 0,
\eqno(3.3.1)
$$
which is directly related to the first one. This sequence indicates that 
the group of all characters $\Phi\in {\cal CS}^p_{X,Y}$ characterized by
the same cohomology class in $H_{s\Bbb Z}^p(X,Y)$ is isomorphic to 
$C_{dR}^{p-1}(X,Y)/Z_{dR\Bbb Z}^{p-1}(X,Y)$. In the absolute case, 
the sequence was found in ref. \ref{11}. Its importance stems from the fact 
that it reveals the relation between the Cheeger--Simons differential
characters and the smooth Beilinson--Deligne cohomology \ref{12,13,14}.

The analysis of sect. 2 furnishes an expression of the depenedence 
of the Cheeger--Simons characters $\Phi\in {\cal CS}^p_{X,Y}$
on the cohomologically integer relative de Rham cocycle
$(\Xi^p,\Upsilon^{p-1})\in Z_{dR\Bbb Z}^p(X,Y)$ for a fixed class in 
$H_{s\Bbb Z}^p(X,Y)$. Indeed, from (2.3.6), if we shift 
$(\Xi^p,\Upsilon^{p-1})$ by a relative de Rham coboundary of the form 
(1.2.14), $\Phi$ varies of an amount 
$$
\Delta \Phi(S,T)=\langle S_{p-1},\xi^{p-1}\rangle
-\langle T_{p-2},\upsilon^{p-2} \rangle, 
\eqno(3.3.2)
$$
for $(S_{p-1},T_{p-2})\in Z_{p-1}^s(X,Y)$.

\eject
\par\vskip.6cm
\item{\bf 4.} {\bf Concluding remarks}
\vskip.4cm
\par

In this paper, we have shown that the proper treatment of the topological 
integrals appearing in many physical models such as gauge theory and 
string theory requires in an essential way relative (co)homology and 
leads to relative Cheeger--Simons differential characters. 
Instead of contenting ourselves with an abstract study of these matters, 
we have worked out a definition of relative Cheeger--Simons 
differential characters which is constructive, i. e. computable in principle,
and which contains the ordinary Cheeger--Simons differential characters
as a particular case. The resulting expressions are totally explicit 
and completely general and lend themselves also to a more formal type 
of study. 

Our method relies heavily on \v Cech (co)homological machinery.
This has its advantages and disadvantages. At any rate, it seems 
hardly avoidable when one has to deal with locally defined fields
on arbitrary topologically non trivial manifolds. A major part of the 
effort consisted in  showing independence from covering choices. 

We limited ourselves to the case where the quantization conditions can
be formulated in the framework of integral relative cohomology. This 
excludes interesting examples from D--brane theory, which require
more general cohomology theories such as K theory. 
It would be very interesting to generalize our constructions
to K theory. This is left for future work.

\par\vskip.6cm
\item{\bf A1.} {\bf Existence and cofinality of good open coverings}
\vskip.4cm
\par

Let $M$ be a manifold equipped with a Riemannian metric $g$.
For $m\in M$ and $u\in T_mM$, we set $|u|_{Mg}=g_m(u,u)^{1/2}$.
For $r>0$, we define $B_{Mg}(m,r)=\{u|u\in T_mM,~ |u|_{Mg}<r\}$.
The exponential $\exp_{Mg}$ is a map of an open neighborhood 
$N_{Mg}$ of the $0$ section of $TM$ into $M$. It has the basic property that,
for $m\in M$ and $u\in T_mM$, the curve $\gamma_{mu}(t)=\exp_{Mg}(tu)$,  
$0\leq t$, $tu\in N_{Mg}$, is the unique geodesic with initial 
condition $(m,u)$ \ref{40}.

The following theorem holds \ref{40}. For $m\in M$, there is $r_{Mg}(m)>0$ 
such that, for any $r$ with $0<r<r_{Mg}(m)$, $B_{Mg}(m,r)\subseteq N_{Mg}$
and there is an open neighborhood $U_{Mg}(m,r)$ of $m$ in $M$ such that 
$\exp_{Mg}:B_{Mg}(m,r)\rightarrow U_{Mg}(m,r)$ is a diffeomorphism.
Further, $U_{Mg}(m,r)$ is geodesically convex, that is every two points
$p,~q\in U_{Mg}(m,r)$ can be joined by a unique distance minimizing geodesic
contained in $U_{Mg}(m,r)$.
For $m\in M$, the family ${\cal U}_{Mg}(m)=\{U_{Mg}(m,r)|0<r<r_{Mg}(m)\}$
is a fundamental system of geodesically convex open neighborhoods of $m$. 
Since the intersection of any finite number of geodesically convex
open sets is geodesically convex, the open coverings $\cal O$
of $M$ made of sets $U_{Mg}(m,r)$ with varying $m$ and sufficiently 
small $r$ are good. Further, such good coverings are cofinal in the family of 
all open coverings (cfr. subsect. 1.1).

Let $X$ be a manifold equipped with a Riemannian metric $g$ and let $Y$ be 
a submanifold of $X$ with induced metric $i^*g$. Assume that $Y$ is totally 
geodesic \ref{40}. Then, every geodesic of $Y$ with respect to the metric  
$i^*g$ is a geodesic of $X$ with respect to $g$, so that
$\exp_{Yi^*g}=\exp_{Xg}|N_{Yi^*g}\cap N_{Xg}$. It follows that 
for $y\in Y\subseteq X$ and $0<r<r_{Yi^*g}(y)$, $U_{Yi^*g}(y,r)
=U_{Xg}(y,r)\cap Y$. Now, define 
${\cal U}'_{Xg}(x)=\{U_{Xg}(x,r)|0<r<r_{Xg}(x),~
U_{Xg}(x,r)\cap Y=\emptyset\}$, for $x\in X\setminus Y$,
${\cal U}'_{Xg}(y)=\{U_{Xg}(y,r)|0<r<r_{Yi^*g}(x)\}$, for 
$y\in Y$. Then, for any $x\in X$, ${\cal U}'_{Xg}(x)$ 
is a fundamental system of geodesically convex open neighborhoods of 
$x$ such that, for any $y\in Y$, ${\cal U}'_{Xg}(y)\cap Y
={\cal U}_{Yi^*g}(y)$. 
From the discussion of the previous paragraph, 
it follows that the open coverings $\cal O$ of $X$ made of sets 
$U_{Xg}(x,r)$ with varying $x$ and sufficiently small $r$ are good
for the pair $X$, $Y$ (cfr. subsect. 1.4) and that
such good coverings are cofinal in the family of 
all open coverings.

Therefore, given a manifold $X$ and a submanifold $Y$ of $X$, in order
a cofinal family of good open coverings of $X$, $Y$ to exist, it is 
sufficient that there is a Riemannian metric $g$ on  $X$ with respect 
to which $Y$ is totally geodesic.

\vskip.6cm
\par\noindent
{\bf Acknowledgements.} We are greatly indebted to R. Stora for providing 
his invaluable experience and relevant literature. 
\vskip.6cm
\centerline{\bf REFERENCES}
\vskip.6cm

\item{\ref{1}}
O.~Alvarez,
``Cohomology And Field Theory'', UCB-PTH-85/20
Plenary talk given at Symp. on Anomalies, Geometry and Topology, Argonne, 
IL, Mar 28-30, 1985.

\item{\ref{2}}
O.~Alvarez,
``Topological Quantization And Cohomology'',
Commun.\ Math.\ Phys.\  {\bf 100} (1985) 279.

\item{\ref{3}}
K.~Gawedzki,
``Topological Actions In Two-Dimensional Quantum Field Theories'',
in  Cargese 1987, proceedings, Nonperturbative Quantum Field Theory, 101.

\item{\ref{4}} 
D.~S.~Freed,
``Locality and integration in topological field theory'',
 hep-th/9209048.

\item{\ref{5}} 
D.~S.~Freed,
``Classical Chern-Simons theory. Part 1'',
Adv.\ Math.\  {\bf 113} (1995) 237
 hep-th/9206021.

\item{\ref{6}} 
D.~I.~Olive and M.~Alvarez,
``Spin and abelian electromagnetic duality on four-ma\-ni\-folds'',
Commun. \ Math. \ Phys. {\bf 217} (2001) 331,
 hep-th/0003155.

\item{\ref{7}} 
M.~Alvarez and D.~I.~Olive,
``The Dirac quantization condition for fluxes on four-manifolds'',
Commun.\ Math.\ Phys.\  {\bf 210} (2000) 13,
 hep-th/9906093.

\item{\ref{8}}
M.~Bauer, G.~ Girardi, R.~Stora, F.~Thuillier,
``A class of topological actions'',
incomplete draft, June 2000.

\item{\ref{9}}
J.~L.~ Koszul, 
``Travaux de S. S. Chern et J. Simons sur les classes caract\'eristiques''
Seminaire Bourbaki 26\`eme ann\'ee {\bf 440} (1973/74) 69.

\item{\ref{10}}
J.~Cheeger,
``Multiplication of Differential Characters'',
Convegno Geometrico INDAM, Roma maggio 1972, in 
Symposia Mathematica {\bf XI} Academic Press (1973) 441.

\item{\ref{11}}
J.~Cheeger and J,~Simons,
``Differential Characters and Geometric Invariants'',
Stony Brook preprint (1973) reprinted in 
Lecture Notes in Math. {\bf 1167} Sprin\-ger Verlag (1985) 50.

\item{\ref{12}}
P.~Deligne,
``Th\'eorie de Hodge'',
Inst. Hautes \'Etudes Sci. Publ. Math. {\bf 40} (1971) 5.

\item{\ref{13}}
A.~A.~Beilinson,
``Higher regulators and values of L--functions'',
J. Soviet Math. {\bf 30} (1985) 2036.

\item{\ref{14}}
A.~A.~Beilinson,
``Notes on absolute Hodge cohomology'',
in Contemp. Math. {\bf 55} Part I  Amer. Math. Soc. (1986).

\item{\ref{15}}
M.~J.~Hopkins and I.~M.~Singer,
``Quadratic Functions in Geometry, Topology, and M--Theory'',
incomplete draft, June 2000.

\item{\ref{16}}
D.~S.~Freed and E.~Witten,
``Anomalies in string theory with D-branes'',
 hep-th/ 9907189.

\item{\ref{17}}
A.~Kapustin,
``D-branes in a topologically nontrivial B-field'',
Adv.\ Theor.\ Math.\ Phys.\  {\bf 4} (2001) 127,
 hep-th/9909089.

\item{\ref{18}} 
C.~Bachas, M.~Douglas and C.~Schweigert,
``Flux stabilization of D-branes'',
JHEP {\bf 0005} (2000) 048,
 hep-th/0003037.

\item{\ref{19}}
J.~Polchinski, unpublished.

\item{\ref{20}} 
W.~Taylor,
``D2-branes in B fields'',
JHEP {\bf 0007} (2000) 039,
 hep-th/0004141.

\item{\ref{21}} 
A.~Alekseev, A.~Mironov and A.~Morozov,
``On B-independence of RR charges'',
Phys.\ Lett.\ B {\bf 532} (2002) 350,
 hep-th/0005244.

\item{\ref{22}} 
S.~Stanciu,
``A note on D-branes in group manifolds: Flux quantization and D0-charge'',
JHEP {\bf 0010} (2000) 015
 hep-th/0006145.

\item{\ref{23}} 
A.~Alekseev and V.~Schomerus,
``RR charges of D2-branes in the WZW model'',
 hep-th/0007096.

\item{\ref{24}} 
J.~M.~Figueroa-O'Farrill and S.~Stanciu,
``D-brane charge, flux quantization and relative (co)homology'',
JHEP {\bf 0101} (2001) 006,
 hep-th/0008038.

\item{\ref{25}} 
M.~R.~Douglas,
``Topics in D-geometry'',
Class.\ Quant.\ Grav.\  {\bf 17} (2000) 1057,
 hep-th/9910170.

\item{\ref{26}}
M.~B.~Green, J.~A.~Harvey and G.~Moore,
``I-brane inflow and anomalous couplings on D-branes'',
Class.\ Quant.\ Grav.\  {\bf 14} (1997) 47,
 hep-th/9605033.

\item{\ref{27}}
Y.~E.~Cheung and Z.~Yin,
Nucl.\ Phys.\  {\bf B517} (1998) 69,
 hep-th/9710206.

\item{\ref{28}}
R.~Minasian and G.~Moore,
``K-theory and Ramond-Ramond charge'',
JHEP {\bf 9711} (1997) 002,
 hep-th/9710230.

\item{\ref{29}}
E.~Witten,
``D-branes and K-theory'',
JHEP {\bf 9812} (1998) 019,
 hep-th/9810188.

\item{\ref{30}}
E.~Witten,
``Overview of K-theory applied to strings'',
Int.\ J.\ Mod.\ Phys.\ A {\bf 16} (2001) 693,
 hep-th/0007175.

\item{\ref{31}}
G.~Moore and E.~Witten,
``Self-duality, Ramond-Ramond fields, and K-theory'',
JHEP {\bf 0005} (2000) 032,
 hep-th/9912279.

\item{\ref{32}}
D.~S.~Freed and M.~J.~Hopkins,
``On Ramond-Ramond fields and K-theory'',
JHEP {\bf 0005} (2000) 044,
 hep-th/0002027.

\item{\ref{33}} 
R.~Gopakumar, S.~Minwalla, N.~Seiberg and A.~Strominger,
``OM theory in diverse dimensions'',
JHEP {\bf 0008} (2000) 008,
 hep-th/0006062.

\item{\ref{34}}
R.~Bott and L.~Tu,
``Differential Forms in Algebraic Topology'',
Springer Verlag, New York, 1982.

\item{\ref{35}}
M.~Karoubi and C.~Leruste,
``Algebraic Topology via Differential Geometry'',
Cambridge University Press, Cambridge, 1987.

\item{\ref{36}}
J.~W.~Vick,
``Homology Theory'',
Academic Press, New York and London, 1973.

\item{\ref{37}}
J.~-L.~Brylinski,
``Loop Spaces, Characteristic Classes and Geometric Quantization'',
Birkh\"auser, 1993.

\item{\ref{38}}
N.~Hitchin,
``Lectures on special Lagrangean submanifolds'',
math.DG/9907034.

\item{\ref{39}}
A.~Weil, 
``Sur les th\'eor\`emes de de Rham'',
Commen.\ Math.\ Helv. {\bf 26} (1952) 17.

\item{\ref{40}}
S. Kobayashi and K. Nomizu, ``Foundations of Differential Geometry'',
vols. I and II, J. Wiley \& Sons (1963).

\bye